\documentstyle{ar209-mod}
\newcommand\La{\Lambda}
\newcommand{\gaprox}{$ {\raisebox{-.6ex}{{$\stackrel{\textstyle >}{\sim}$}}} $}
\newcommand{\saprox}{$ {\raisebox{-.6ex}{{$\stackrel{\textstyle <}{\sim}$}}} $}

\newcommand{\beq}{\begin{equation}}
\newcommand{\eeq}{\end{equation}}
\newcommand{\beqa}{\begin{eqnarray}}
\newcommand{\eeqa}{\end{eqnarray}}
\newcommand{\boldpi}{\mbox{\boldmath $\pi$}}

\newcommand{\boldt}{\mbox{\boldmath $t$}}

\newcommand{\mpi}{m_\pi}
\newcommand{\mpis}{m_\pi^2}

\def\si{{}^1\kern-.14em S_0}
\def\siii{{}^3\kern-.14em S_1}
\def\piii{{}^3\kern-.14em P_1}
\def\diii{{}^3\kern-.14em D_1}

\begin{document}

\input epsf.tex    %<-If you need EPS figures to be
                   %  called in {figure} environment for PC
%\input epsf.def   %<-If you need EPS figures to be
                   %  called in {figure} environment for Macintosh

\input epsf.sty
\input psfig.sty

\jname{ARNPS} \jyear{2002} \jvol{52} \ARinfo{}

\title{Effective Field Theory for Few-Nucleon Systems}

\markboth{Bedaque \& van Kolck}{Effective Field Theory for
Few-Nucleon Systems}

\author{Paulo F. Bedaque
\affiliation{Nuclear Science Division,
Lawrence Berkeley National Laboratory,\\
Berkeley, CA 94720
}
 Ubirajara van Kolck \affiliation{Department of
Physics,
 University of Arizona,\\
Tucson, AZ  85721\\
~\\RIKEN BNL Research Center,
Brookhaven National Laboratory,\\
Upton, NY 11973
} }


\begin{abstract}

\vspace{0.3cm}
We review the effective field theories (EFTs) developed for
few-nucleon systems. These EFTs are controlled expansions in
momenta, where certain (leading-order) interactions are summed to
all orders. At low energies, an EFT with only contact interactions
allows a detailed analysis of renormalization in a
non-perturbative context and uncovers novel asymptotic behavior.
Manifestly model-independent calculations can be carried out to
high orders, leading to high precision. At higher energies, an EFT
that includes pion fields justifies and extends the traditional
framework of phenomenological potentials. The correct treatment of
QCD symmetries ensures a connection with lattice QCD. Several
tests and prospects of these EFTs are discussed.

\vspace{0.5cm}
With permission from the Annual Review of Nuclear and Particle Science. 
Final version of this material is scheduled to appear in the 
Annual Review of Nuclear and Particle Science Vol. 52, 
to be published in December 2002 by Annual Reviews 
(http:/AnnualReviews.org).

\end{abstract}

\maketitle

\section{INTRODUCTION}

\subsection{Why effective theories?}

Nuclear systems have often been described as pathologically
complicated. The forces between the constituent nucleons are
strong and non-central and the relatively small binding found in
nuclei results from detailed cancellations between much larger
contributions. Adding the fact that the basic interaction
among nucleons is not completely known, especially at short
distances, and the problems involved in the numerical solution of
the Schr\"odinger equation for systems with many fermions, one
can understand why nuclear structure remains an unsolved problem
after decades of intense effort. This seems even more frustrating
when one remembers that all nuclear processes are encoded in the
QCD Lagrangian and parameter-free predictions could, in principle,
be obtained.
 Despite all these difficulties enormous progress has been
made throughout the years by the use of models capturing different
aspects of nuclear phenomena. One dissatisfying aspect of these
 models however is their basically {\it ad hoc}
nature and the presence of uncontrolled approximations. These
models are not derived from any basic principle (and certainly not
from QCD) and contain information coming from decades of trial and
error hidden behind apparently arbitrary choices of some contributions
over others. Each new improvement involves the same process of
educated guesses and one is never sure of what a reasonable error
estimate would be. Effective field theories (EFTs) are useful
by providing a systematic expansion in a small parameter that organizes
and extends previous phenomenological knowledge about nuclear processes
and by providing a rigorous connection to QCD. They also help  with
more technical but important issues that have plagued nuclear physics 
in the past, like gauge invariance, ``off-shell'' effects and
relativistic corrections, by
borrowing heavily from the arsenal of field theory.

\subsection{What is an effective theory?}

Most of the uncertainty in nuclear processes comes from the 
short-distance interactions ($\saprox 1$ fm) between two or more nucleons 
(and photons, leptons).
Even when one is interested only in low-energy phenomena, the
short-distance contributions can be important. In perturbation
theory, for instance,
the influence of short-distance physics on low-energy observables appears 
in the existence of ultraviolet-divergent integrals, that is, 
in the dominance of high-momentum modes over the small-momentum ones.
 Sensitivity
of large-distance observables on short-distance physics is not an
unusual situation in physics, it is in fact pervasive in many
fields. One way of dealing with it is to model the short-distance
physics and solve the problem from a microscopic approach. In the
case of nuclear systems this would lead either to a calculation of
nuclear processes directly from QCD (which is currently impossible
and would be, even if possible, a highly inefficient way of
approaching the problem) or to the use of 
meson-exchange/quark/skyrmion/... models. 
Another approach is the use of effective
theories \cite{Manohar_review}\footnote{
The term ``effective theories"
has also other meanings besides the specific one assumed in this 
review. ``Effective Lagrangian" sometimes is meant as the one
including all quantum corrections. Other times the term 
``effective theory" is used to describe any model useful at low 
energies, whether or not there is a separation of scales and a
rigorous expansion in powers of the momentum.}. 
Before introducing the particular
case used in nuclear physics, let us
consider effective theories in general. 

Suppose we want to study
the low-energy behavior of a system described by some theory that
we will call the ``fundamental" theory. In the path-integral
formalism, we can imagine  integrating over  the high-momentum modes 
$k< \La$, where the scale $\La$ is chosen to be much larger than the
momentum scale we want to study. The result of this partial
integration over the high-momentum modes will be a complicated
Lagrangian containing an infinite number of terms. This
Lagrangian, called the effective Lagrangian, will also be, in
general, non-local, but this non-locality, arising from the
momenta $k< \La$ will be restricted to a spatial scale $\saprox 1/\La$.
One can thus expand those interaction terms in a Taylor series on
$\partial/\La$, where $\partial$ stands for a derivative of the
fields. The coefficients of this expansion do not depend on the
soft momenta carried by the field of the effective theory and
describe the hard physics within the scale $1/\La$. They are,
however, functions of $\La$ (the coupling constants ``run"). 
The soft ($k<\La$) and hard ($k>\La$) physics are factorized in
the effective Lagrangian. These effective-Lagrangian 
coefficients are usually called
``low-energy constants" (LECs) since they encode all we need to know
about the fundamental theory in order to compute low-energy observables.
Notice that, up to this point, no approximation was made and the
effective Lagrangian contains exactly the same information as the
fundamental one. Calculations of observables done using the
effective Lagrangian will contain two sources of $\La$ dependence. 
One is the implicit dependence contained in LECs, 
the other appears in the cutoff that should
be used in those computations. These two sources of $\La$
dependence, by construction, cancel each other.

One may wonder what the advantage is in separating the integration
over momentum modes in two steps. The answer to this question
depends on the particular situation in hand. In problems where the
integration over the high-momentum modes can be explicitly
accomplished the effective Lagrangian is  a bookkeeping device
that allows us to perform approximations in a very efficient way.
 That is the case, for instance, of
non-relativistic QED \cite{Lepage_NRQFT},
heavy-quark effective theory \cite{Georgi_heavyquark}, and 
high-density QCD chiral perturbation theory
\cite{Son_mesonmasses}. 
In other cases, as the nuclear systems considered here, we
will not be able to explicitly integrate the high-momentum modes.
We can however determine the effective Lagrangian by a combination
of self-consistency requirements and experimental data.
 We start  by considering the most general Lagrangian
consistent with the symmetries of the underlying theory. This
Lagrangian contains an infinite number of arbitrary constants. For
a fixed $\La$, different values of the LECs
describe different underlying theories. Just one set of these
values will make our low-energy theory  reproduce the same
observables as the fundamental theory.  We then resort to an
approximation scheme: we expand the low-energy observables in
powers of the small parameter $Q/\La \ll 1$, where $Q$ is a 
low-energy scale like the momenta of the external particles, light masses, 
{\it etc.}
Now the factorization between high- and low-momentum contributions
comes in handy.
 Instead of using the full effective Lagrangian
with its infinite number of terms we can argue that, {\it at a
given order in the $Q/\La$ expansion}, only a finite number of
them will contribute, since the remaining terms will include many
powers of $\partial/\La \sim Q/\La$. This way we are left with a
much simpler Lagrangian, with a finite (and hopefully small)
number of coefficients that can be determined from some
experimental data (or from the fundamental theory, if possible, or models)
and used to predict others. Increasing the order in $Q/\La$ of a
calculation will increase its precision but may also bring other
LECs that will have to be determined by experiment. The precise
argument connecting the order of the expansion in powers of $Q/\La$
and the terms in the effective Lagrangian that need to be included at 
that order (called ``power counting") varies case by case but always 
include two
steps. The first one is to estimate the size of diagrams, given
the size of the LECs appearing on the vertices, and it is simply done
by dimensional-analysis arguments. The second one is
to estimate the size of the LECs themselves. For that we first
determine their running, that is, their dependence on $\La$, by
requiring physical observables to be $\La$ independent (at the
order in $Q/\La$ we are working). The information about the
evolution of the LECs is not by itself enough to determine them
since we do not know their initial conditions. 
Although for some particular value of $\La$ one LEC 
might be passing through
zero, this is very unlikely.
We assume that a typical
size for a LEC $C(\La)$ is $ C(\La)\sim C(2\La)-C(\La)$,
that
is, the LECs should have the same order of magnitude as the size of
their running. In perturbative settings this principle amounts
to little more than dimensional analysis, and is known as 
``naive dimensional analysis'' \cite{Georgi_Manohar}. 
Strictly, this provides  only a reasonable lower bound, so
one should be aware of possible violations of this principle. This
estimate, of course, is used only in arguing that  some
terms in the effective Lagrangian will have a negligible effect and can
 be dropped. The values of the 
LECs actually kept in
the calculation are determined by experimental data.
Notice that,
for a given set of symmetries and low-energy degrees of freedom,
there is no guarantee that the effective Lagrangian can be
truncated at any order in $Q/\La$, that is, there is no guarantee
that a consistent power counting can be found.

The version of the EFT
method sketched above is sometimes called the ``Wilsonian"
effective theory. Another version of the same basic idea,
identical in spirit but differing in detail is given by the
``continuum" effective theory \cite{Georgi_heavyquark}. 
There, we construct the effective
theory in such a way as to reproduce the same  vertices and propagators as
the full theory at {\it low energy}. The two theories differ 
in the ultraviolet region but this difference can always
be absorbed in the values of the LECs. The technical advantage
over the Wilsonian approach
resides on being able to integrate over all momenta (used in
conjunction with dimensional regularization), and not only over
$k<\La$, which makes it simpler to maintain gauge, chiral and
spacetime symmetries, and to avoid power-law divergences that
sometimes complicate power counting.

\subsection{How?}

EFTs can be used in a few different ways in
nuclear physics. Historically, the first one was to set the
separation scale $\La$ around the $\rho$-meson mass and keep as 
low-energy degrees of freedom the pions and the 
nucleons\footnote{Since energies are
measured from the ground state {\it with a given baryon number},
slow nucleons, despite carrying large rest-mass energy, should be  
considered
low-energy degrees of freedom.} (and maybe
the $\Delta$ isobars), as well as photons and leptons 
\cite{inwei6, ciOvK, inwei5, ciOLvK, civK1, invk}. This approach builds on and
extends the success of Chiral Perturbation Theory (ChPT)
in the mesonic
and one-baryon sectors. It shares with nuclear potential models
the fact that it describes non-relativistic nucleons interacting
through a potential, but it also brings a number of ingredients of its own,
such as a small expansion parameter, consistency with the chiral
symmetry of QCD, and systematic and rigorous ways of including
relativistic corrections and meson-exchange currents. 

 Another way of applying EFT ideas in nuclear physics
 is made possible by the existence of shallow
bound states, that is, binding energies much below any reasonable
QCD scale 
\cite{Bedaque_vanKolck_quartet, vanKolck:1997ut, vanKolck_shortrange, 
seattle_pionless}. 
We can then set $\La$ around the pion mass and keep as
low-energy degrees of freedom only the nucleons (and photons, leptons).
 At
least in the case of two- and three-body systems the bound states
will be within the range of validity of this simpler theory. This
``pionless" effective theory can be considered as a  formalization
and extension of the old effective-range theory (ERT) 
\cite{Bethe}
and 
the work on ``model-independent results'' in three-body physics
\cite{Efimov_qualitative}. The new
features, besides the existence of a small parameter on which to
expand, appear in a  number of new short-distance contributions
describing exchange currents and three-body forces, as well as in
relativistic corrections, that are transparent in this approach. An
extra bonus is the possibility of deriving analytic, 
high-precision expressions for many observables that previously could
only be obtained after non-trivial numerical work.

In Sects. \ref{pionless} and \ref{pionful} we will review these two approaches
in few-nucleon systems,
emphasizing qualitative aspects of recent developments.
In Sect. \ref{outlook} we present an outlook,
including other approaches that are being developed
for larger nuclei. 
Some reviews have already appeared covering applications of EFT ideas
to nuclear physics, with different emphasis from the present one
\cite{vanKolck_review}.
Many developments of the last couple of years are described
in Ref. \cite{books}.

\section{EFT WITHOUT EXPLICIT PIONS}
\label{pionless}

\subsection{The two-nucleon system and the non-trivial fixed point}
\label{pionlesstwobod}

\subsubsection{Two-nucleon scattering}

Let us now apply the ideas outlined in the previous section to the 
specific case of two nucleons with momentum $k$ below the pion scale 
$k<m_\pi$. More details can be found in Ref. \cite{seattle_pionless}.
We start
by writing the most general Lagrangian involving only two nucleons
(electroweak external currents will be included later). A system
with two nucleons with zero angular momentum $L=0$ can exist in a
spin singlet ($^1S_0$) or spin triplet ($^3S_1$) state so there
are two independent interactions with no derivatives,
\begin{equation}\label{pionlesslag}
{\cal L}= N^\dagger (i \partial_0 + \frac{\vec{\nabla}^2}{2 M}+
\ldots)N - C_{0 t}(N^\dagger P_t N)^2  - C_{0 s}(N^\dagger P_s
N)^2 + \ldots,
\end{equation}
where
\begin{eqnarray}\label{projectors}
P_t^i&=&\frac{1}{\sqrt{8}} \sigma_2 \sigma^i \tau_2, \nonumber\\
P_s^A&=&\frac{1}{\sqrt{8}} \tau_2 \tau^A \sigma_2\
\end{eqnarray}
are the projectors in the triplet and singlet spin-isospin states
($\sigma$'s act on spin space, $\tau$'s on isospin space),
$M$ is the nucleon mass and $N$ the nucleon field. The
dots in Eq.~(\ref{pionlesslag}) stand for terms with more
derivatives that, as we will argue later, will be subdominant.

The nucleon-nucleon ($NN$) scattering amplitude can be written in terms of 
the phase shift $\delta$ as
\begin{eqnarray}\label{T_effectiverange}
T&=&\frac{4\pi}{M}\frac{1}{k \cot\delta - i k}\nonumber\\
&=&\frac{4\pi}{M}\frac{1}{-\frac{1}{a_s}+\frac{r_{0s}}{2}k^2+\ldots -
i k}
\end{eqnarray}
\noindent It can be shown that
for potentials of range $\sim R$ ($R\sim 1/m_\pi$ in our case), $k
\cot\delta$ is an analytic function around $k=0$ and that it has a
cut starting at $k^2\sim 1/R^2$, so it is well approximated by a
power series as shown in the last line of Eq.~(\ref{T_effectiverange}).
The parameter $a_s$ ($r_{0s}$) is called  the singlet scattering length 
(singlet effective range).
For notational simplicity we specialize for now on the spin singlet channel.
 
\begin{figure}[!t]
\centerline{\epsfysize=3cm\epsfbox{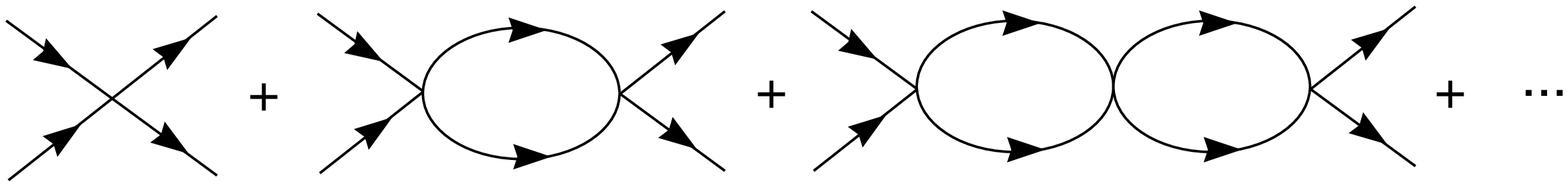}}
\caption{Graphs contributing to the LO $NN$ scattering amplitude.}
\label{bubbles}
\end{figure}

The graphs contributing to $NN$ scattering generated by
the
Lagrangian in Eq.~(\ref{pionlesslag}) are shown in Fig.~(\ref{bubbles}).
The $L$-loop graph factorizes into a power,
\begin{equation}\label{bubble}
L{\rm -loop\ graph}\sim (c \La - i k)^L,
\end{equation}\noindent each one containing a linearly divergent 
piece and the unitarity cut $i k$ (in the center-of-mass system
with total energy $k^2/M$). The loop integral
is linearly divergent and the coefficient $c$ is dependent on the
particular form of the regulator used, that is, the particular
form the high-momentum modes are separated from the low-momentum
ones. Using a sharp momentum cutoff, for instance, we have $c=2/\pi$,
using dimensional regularization (DR), $c=0$. 
The sum of all graphs in  Fig.~(\ref{bubbles}) is
a geometrical sum giving
 \begin{equation}\label{Tpionless}
T=\frac{4\pi}{M}\frac{1}{-\frac{4\pi}{M C_{0 s}}+ c\La-i k}.
\end{equation}
\noindent 
 We see then that terms  shown explicitly in
Eq.~(\ref{pionlesslag}) reproduce the first term of the effective
range expansion. The addition of terms with more derivatives will
reproduce further terms in the effective range expansion. 

We can
learn some important lessons from this simple calculation. Let us
consider two separate situations.

{\bf Natural case}: For a generic potential with range $R$, the
effective-range parameters typically have  similar size $a\sim
r_0 \sim R$. Using DR, $C_{0}$ can be chosen to be $C_{0}=4\pi a/M$
(this choice is called minimal subtraction). The effective theory
is valid for $k<1/R$ and, in this range, $T$ can be expanded as
\begin{equation}\label{T_natural}
T=\frac{4\pi}{M} \left(-a+i k a^2 + 
                       \left(\frac{a^2 r_0}{2} + a^3\right)k^2 + \ldots\right).
\end{equation}\noindent Since $C_{0}\sim a$, there is
a one-to-one correspondence between the order in the $k a$
expansion, the number of $C_{0s}$ vertices
 and the number of loops in a graph. The leading order
(LO) is given by one tree-level diagram, the next-to-leading order
(NLO) by the one-loop diagram, next-to-next-to-leading order
(N$^2$LO) by the two-loop diagram involving $C_0$ and one tree-level
diagram with a two-derivative vertex (not shown in Eq.~(\ref{pionlesslag})), 
and similarly for higher
orders. We have then a {\it perturbative} expansion, even though
the microscopic potential can be arbitrarily strong. If one uses a
cutoff regulator the situation is slightly more complicated.
Choosing $\La\sim 1/R \sim 1/a$ we note that the most divergent
piece of the multi-loop graphs is as large as the tree-level graph
and must be resummed to all orders, while the energy-dependent part
containing powers of $i k C_0$ is suppressed. The pieces
that need to be resummed at leading order merely renormalize the
constant $C_0$. The one-to-one correspondence between the order in
the $k a$ expansion and the number of loops is lost in any but the
DR with minimal subtraction renormalization/regularization scheme.
The technical advantages arising from the use of DR and renormalization 
theory in this perturbative setting was used in the study of 
dilute gases with short-range interactions in 
Ref. \cite{Braaten_dilute}.

{\bf Unnatural case}: In the nuclear case the scattering lengths of
the two S-wave channels are much larger than the range of the
potential. The $^1S_0$ (neutron-proton)
scattering length $a_s$ is $a_s=-23.714$ fm 
and the scattering length of the triplet
(deuteron) channel $^3S_1$ is $a_t=5.42$ fm, corresponding to
momentum scales of $1/a_s = 8.3$ and $1/a_t=36$ MeV, respectively.
Those scales are much smaller than the pion mass, $m_\pi\simeq 140$ MeV,
defining the range of the 
nuclear potential\footnote{Alkali atoms used in cold atomic traps 
frequently have
scattering lengths much larger than their sizes.
They can be made even larger by the use of a carefully tuned
 external magnetic field
(Feshbach resonances). All the ideas and formalisms developed to 
deal with this fact in the nuclear domain can and have already been 
used to study the physics of atomic traps 
\cite{stooges_3bosons, Bedaque_recombination}.}. Actually, we will see
later that the potential due to pion exchange is too weak to
describe  the low-energy phase shifts, and  the physics
corresponding to the large scattering lengths occurs at the QCD
scale $M_{QCD}\sim 1$ GeV, what makes the discrepancy between
nuclear and QCD scales even more startling . 
The origin of the
fine-tuned cancellations leading to the disparity between the underlying
scale and the $S$-wave scattering lengths (and deuteron binding
energy) is presently unknown. 
It does not appear in any known limit
of QCD like the chiral limit ($m_{q}\rightarrow 0$) or large number of
colors ($N_c\rightarrow\infty$). 
We will just assume that this
cancellation happens, track the dependence of observables on the
new soft scale $1/a_{s,t}$ and perform our low-energy expansion in
powers of $k R\ll 1$ while keeping the full dependence on $k a_{s,t}\sim
1 $. 
The singlet $NN$ scattering amplitude, for instance, will be
expanded as
\begin{equation}\label{Tlargea}
T=-\frac{4\pi}{M} \left(\frac{a_s}{1+i k a_s} + \frac{k^2 a_s^2 r_{0s}}{2}
\frac{1}{(1+i k a_s
)^2}+\ldots\right).
\end{equation}\noindent
 It is a little challenging to reproduce an expansion
of this form in the EFT. If one uses a momentum cutoff, for
instance, the constant $C_{0 s}$ has to be chosen to be $C_{0
s}=(4\pi/M) (1/a_{s}+c\La)$. The one-loop graph is then suppressed
compared to the tree-level one by a factor $\sim M k C_{0 s}\sim
k/\La$ and one would naively imagine that the leading-order
contribution is given solely by the tree-level graph. But there
are cancellations between the graphs in Fig.~(\ref{bubbles}) and
all these graphs need to be taken into account to reproduce the
expansion of $T$ above \cite{vanKolck:1997ut, vanKolck_shortrange}. 
In the $NN$ 
scattering case
considered here it is not difficult to see which graphs have to be
included at each order, but in more complex situations this can be
extremely tricky. A more convenient way to proceed is to use a
renormalization prescription that shifts contributions from 
high-momentum modes to the LECs in such a way as to
eliminate this ``accidental'' cancellations between different
diagrams. One can determine which diagrams contribute at each
order on a diagram-by-diagram basis (manifest power counting). One
way to do that is to use DR with a ``power divergence
subtraction'' (PDS) \cite{ksw_1}\footnote{Other schemes also
solve this problem \cite{Mehen_Stewart_momsub}.}.  
In this scheme, we add and
subtract to the denominator of the bubble sum in
Eq.~(\ref{Tpionless}) an amount $M\mu/4\pi$, where $\mu$ is an
arbitrary scale, and absorb the subtracted term in a redefinition
of the constant $C_{0 s}(\mu)$, that now is a function of $\mu$.
We have for the LO amplitude
\begin{equation}\label{TpionlessPDS}
T=-\frac{4\pi}{M}\frac{1}{\frac{4\pi}{MC_{0 s}(\mu) }+ik + \mu}.
\end{equation}
\noindent The constants $C_{0 s}(\mu)$ is now chosen to be
\begin{equation}\label{C0PDS}
C_{0 s}(\mu)=\frac{4\pi}{M}\frac{1}{\frac{1}{a_{s}}-\mu},
\end{equation}
\noindent in order to reproduce the LO piece of the expansion
in Eq.~(\ref{Tlargea}). The explicit dependence on $\mu$ cancels against
the implicit dependence contained in $C_{0 s}(\mu)$.
 The point of this rearrangement is that
if $\mu$ is chosen so that 
$\mu\sim 1/a_{s}$, $C_{0 s}(\mu)\sim 4\pi/M\mu$ and the
contribution of all diagrams in the bubble sum are of the same
order, justifying the need to resum them. Let us see how this works
in some detail. Denoting the soft scales $1/a_{s}$, $\mu$ and
$k$ collectively by $Q$, the tree-level diagram is of the order
$C_{0 s}\sim 4\pi/M\mu$. The one-loop graph contains two
powers of $C_{0 s}$, two nucleon propagators, each one counting
as $1/(k^2/M)\sim M/Q^2$, and a loop integral with three powers of
momentum ($\sim Q^3$), one of energy ($\sim Q^2/M$) and the usual
factor of $1/4\pi$ from the loop integration, for a total of
$(4\pi/MQ)^2 (M/Q^2)^2 Q^5/M\sim 4\pi/MQ$. Thus the one-loop
diagram has the same size of the tree-level graph. The same occurs
for the remaining diagrams and they all have to be resummed. It is
interesting to note that this reshuffling of contributions between
the divergent loop and the LECs amounts to
subtracting the poles $1/(D-2)$, where $D$ is the number of 
spatial dimensions, that would exist in two space dimensions. 
One can easily go to higher orders and include terms
with derivatives in the Lagrangian. A simple calculation (again,
subtracting the pole occurring in two spatial dimensions)
leads to expressions for all the LECs in terms of
the effective-range parameters (and of the arbitrary scale $\mu$).
For instance, denoting by $C_{2n}$ the 
coefficient of operators with $2n$ derivatives,
\begin{eqnarray}\label{C2PDS}
C_{2 s}&=&\frac{4\pi}{M} \frac{r_{0  s}}{2}\left(\frac{1}{\frac{1}{a_{s}}
-\mu}\right)^2,\\
C_{4 s}&=&\frac{4\pi}{M}\left( \frac{r_{0  s}^2}{4}\left(
\frac{1}{\frac{1}{a_{s}}-\mu}\right)^3+ \frac{r_{1
s}^3}{2}\left( \frac{1}{\frac{1}{a_{s}}-\mu}\right)^2\right),
\end{eqnarray}
\noindent where $r_{1 s}$ is the coefficient of the third term
of the effective-range expansion (``shape parameter'').

The $\beta$-function describing the evolution of the dimensionless
coupling $\hat{c}_{0 s}\equiv -M\mu C_{0 s}/4\pi$ is
\begin{equation}\label{betafunction}
  \mu \frac{\partial}{\partial\mu}\hat{c}_{0 s}(\mu)=\hat{c}_{0
  s}(\mu)\left(1-\hat{c}_{0 s}(\mu)\right)
\end{equation}
\noindent Note the existence of two fixed points \cite{ksw_1, 
Birse_fixedpoints}, the trivial
(perturbative) one at $\hat{c}_{0 s}=0$ and a non-trivial one at
$\hat{c}_{0 s}=1$. For $\mu\ll 1/|a_s|$, as appropriate to the
natural case discussed above,  $\hat{c}_{0 s}$ is close to the
trivial fixed point. Diagrams involving more $C_{0 s}$ vertices
are suppressed by powers of $\hat{c}_{0 s}\ll 1$ and the system
is perturbative. The value $\hat{c}_{0 s}\sim \mu a_s$
corresponds to the naive-dimensional-analysis one and the effects
of the $C_{0 s}$ operator become smaller at lower energies (the
operator is {\it irrelevant}). On the other hand, for values of
$\mu\sim 1/|a_s|$ or larger, as adequate to the fine-tuned case
discussed here, the flow is close to the non-trivial fixed point.
Since $\hat{c}_{0 s}\sim 1$, the addition of more $C_{0 s}$
vertices is not suppressed and all graphs containing only this
vertex should be resummed. The dimensionless coupling $\hat{c}_{0
s}$ goes from the naive-dimensional-analysis value $\hat{c}_{0
s}\sim \mu a_s$ to $\hat{c}_{0 s}\sim 1$ and its effects do not go
away in the infrared ({\it marginal} operator).

Since the $^3S_1$ scattering length is also 
unnaturally large 
(and consequently the deuteron is unnaturally shallow),
 the same power counting used in the singlet channel applies also to 
the triplet channel.
$NN$ scattering in this channel is more complicated  because
 nuclear forces, being non-central,
 mix it with the $^3D_1$ channel. There are new operators, starting
with two derivatives, describing this mixing.
 Their coefficients
are determined from an expansion of mixing angle analog to
Eq.~(\ref{Tlargea}). Also, the LECs are usually determined
by matching to an effective range expansion centered around the deuteron
pole, as opposed to the one centered around $k=0$ as is done in
the singlet channel. The $^3S_1$ $NN$ amplitude is
parameterized as
\begin{equation}\label{deuteroneffectiverangeexp}
T=\frac{4\pi}{M}\frac{1}{-\gamma+\frac{\rho
(k^2+\gamma^2)}{2}+\ldots-ik}
\end{equation}
\noindent where $\gamma^2/M$ is the deuteron binding energy
and $\rho$ the effective range parameter.
 Explicit expressions for the leading terms in the
Lagrangian and numerical values for the LECs can be found in
Ref. \cite{seattle_pionless}.

The inclusion of external currents (photons, neutrinos, ...) is
straightforward. All terms involving nucleons and the new fields
or currents should be included, as long as they satisfy the
symmetries of the underlying theory. In the case of photons, some
of these terms are just those required by gauge invariance and are
determined by minimally coupling the photon to the nucleon Lagrangian. Their
coefficients are thus fixed by $NN$ scattering data and
gauge invariance. There are also terms that are gauge invariant by
themselves and  whose coefficients are {\it not} determined by
$NN$ scattering data alone. They represent the physics
of exchange currents, quark effects, {\it etc.}, and need to be
determined through some extra piece of experimental data. To
perform the low-energy expansion though, it is necessary to have
an {\it a priori} estimate of their size. This estimate is
obtained by using the fact that observables should be independent
of the cutoff (or $\mu$ if using DR), at least up to the order one
is computing. Consider some two-nucleon operator of the form $X=C^X_{2n}
N^\dagger N^\dagger\Gamma_X \vec{\partial}^{2n} N N$, where
$\Gamma_X$ is some tensor in spin-isospin space. Its matrix
element on two-nucleon states is given by the diagrams
involving the operator $X$ ``sandwiched'' between two two-nucleon scattering
amplitudes and by one-loop one-body diagrams that do not involve $X$.
 Typically the one-body diagram is
not divergent and does not introduce any $\mu$
dependence\footnote{One exception is the two-nucleon, 
no-external-current $C_4$ operator whose renormalization is driven by $C_2$.
This explains the apparent discrepancy between Eq.~(\ref{C2PDS})
and Eq.~(\ref{CXestimate}).} so the remaining graphs have to
be $\mu$ independent by themselves. We have to make a distinction
now between the cases where the operator $X$ connects two $S$-wave
states, two non-$S$-wave states, or one $S$-wave and one non-$S$-wave state.
In the first case renormalization-group invariance of the two-nucleon
matrix element of $X$ implies
\begin{equation}\label{CXmuindependent}
\mu\frac{\partial}{\partial\mu}C^X_{2n}(\mu)
\left(\frac{T}{C_0(\mu)}\right)^2=0,
\end{equation}
\noindent where $T$ is the LO $NN$ scattering matrix, which is $\mu$
independent. From that it follows that $C^X_{2n}(\mu)$ scales as
$\sim (\mu-1/a)^{-2}$. Similarly, for the case where $X$ connects one 
$S$-wave or no
$S$-wave states
$C^X_{2n}(\mu)$ scales as $\sim (\mu-1/a)^{-1}$ and $\sim (\mu-1/a)^{0}$, 
respectively.
Using dimensional analysis to fix the powers of $\La$ we then have
\begin{equation}\label{CXestimate}
C^X_{2n}(\mu)\sim\frac{1}{M(1/a-\mu)^\alpha}\frac{1}{\Lambda^{2n+1-\alpha}},
\end{equation}
\noindent where $\alpha$ is the number of $S$-wave states the operator $X$ 
can connect
(either $0,1$ or $2$).

In a nutshell, the power counting rules valid for the two-nucleon system 
are
\cite{vanKolck:1997ut, vanKolck_shortrange, ksw_1, seattle_pionless}:
\begin{eqnarray}\label{lowQrules}
{\rm fermion\ line}&\rightarrow& M/Q^2\nonumber\\
{\rm loop}&\rightarrow& \frac{Q^5}{4\pi M}\nonumber\\
\vec{\partial}&\rightarrow& Q\nonumber\\
\partial_0&\rightarrow& Q^2/M\nonumber\\
C_{2n}&\rightarrow& \frac{4\pi}{M \Lambda^n Q^{n+1}}\nonumber\\
C^X_{2n}&\rightarrow&\frac{4\pi}{M \Lambda^{2n+1-\alpha} Q^\alpha},
\end{eqnarray}
\noindent where $C_{2n}$ is the coefficient of the two-nucleon interaction 
with $2n$ derivatives, 
$C^X_{2n}$ is the coefficient of a two-nucleon operator 
with external current $X$ and $2n$ derivatives, 
and $\Lambda$ is the high-energy scale $\La\sim m_\pi$. 

\begin{figure}[!t]
\centerline{\epsfysize=5cm\epsfbox{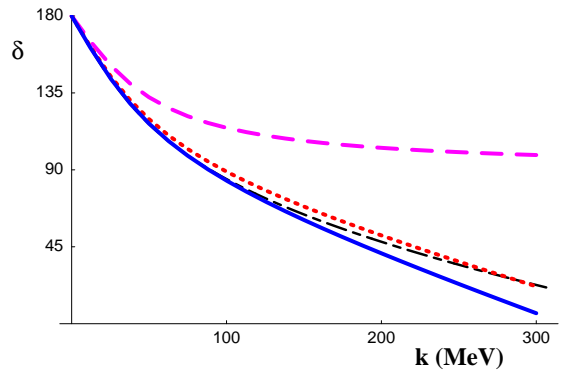}}
\caption{$^3S_1$ $NN$ phase shift (in degrees)
as function of the center-of-mass momentum. 
The LO result is the dashed (purple) line, 
the N$^2$LO
the dotted (red) line and N$^4$LO the thick (blue) solid curve. 
The dot-dashed line is the Nijmegen PSA. 
{}From Ref. \cite{seattle_pionless}, courtesy of M. Savage.}
\label{3S1_nopions}
\end{figure}

Using this rule 
we can determine the contributions to $NN$ scattering at any given order. 
At LO, for instance, we have the
series of diagrams shown in Fig. (\ref{bubbles}), with
all the vertices containing no derivative. That is the only 
non-perturbative resummation
necessary. At NLO we have the insertion of one
$C_2$ operator in a chain of $C_0$ operators. At N$^2$LO
we have two insertions of $C_2$ and one insertion of $C_4$, and so on.
The resulting $^3S_1$
phase shift, for example, is shown in Fig. (\ref{3S1_nopions}),  
and compared 
to the Nijmegen phase-shift analysis (PSA) \cite{nijmanal}.
Analytic expressions for the phase shifts
 can be found in Ref. \cite{seattle_pionless}.
They suggest convergence for momenta
$k\saprox 100$ MeV, 
as it is reasonable for an EFT without explicit pions.
Electromagnetic effects in proton-proton scattering
were considered in the EFT approach in Ref. \cite{Ravndal}.

Up to this point we have considered only $NN$ scattering, 
where the predictive power of the pionless EFT is very small.
We were able however to determine many LECs using scattering data
and understand the effects of the fine-tuning on the $S$-wave channels.
We now apply the formalism developed above 
%(and the values of the
%LECs determined through $NN$ scattering and deuteron
%properties) 
to the computation of form factors and processes
involving external currents. We will omit the diagrams needed to
be computed and the explicit analytic expressions that are always
available in the two-nucleon sector. They can be found in the
literature cited.

\subsubsection{Electromagnetic form factors of the deuteron}

The matrix element of the electromagnetic current on the deuteron
has the non-relativistic parameterization
\begin{eqnarray}\label{emformfactors}
\langle p',\epsilon^j| J^0_{em}|p,\epsilon^i\rangle  &=& e \left[  F_C(q^2)
\delta_{ij} + {1\over 2 M_d^2}F_{\cal Q} (q^2)
\left( q_i q_j-{1\over
3}q^2 \delta_{ij}\right)\right] \left({E+E^\prime\over 2 M_d}\right),
\nonumber\\
\langle p',\epsilon^j| \vec{J}^{\; k}_{em}|p,\epsilon^i\rangle &=& {e\over 2 M_d}
\left[ F_C(q^2) \delta_{ij}(p+p')^k + F_M(q^2)\left(\delta_j^k q_i -
\delta_i^k q_j\right)
\right.\nonumber\\
&&\qquad\quad \left.+{1\over 2M_d^2} F_{\cal Q}(q^2) \left( q_i
q_j-{1\over 3} q^2\delta_{ij}\right)(p+p')^k\right],
\end{eqnarray}
\noindent where $|p,\epsilon^i\rangle $ is the deuteron state with
momentum $p$ and polarization $\epsilon^i$, $M_d$ is the deuteron
mass, $q=p'-p$ and the form
factors are normalized such that $F_C(0)=1$ (deuteron charge), $e
F_M(0)/2M_d=\mu_D$ (deuteron magnetic moment) and
$F_Q(0)/M_d^2=\mu_Q$ (deuteron quadrupole moment).

At LO and NLO the computation of $F_C(q^2)$ involves only the
constants $C_{0 t}$ and $C_{2 t}$ and is identical to the 
ERT
calculation. At N$^2$LO a one-body term describing the nucleon
charge mean square radius ($\langle r^2\rangle_N$) appears, which is the first
deviation from ERT \cite{seattle_pionless}. 
Formally there are also relativistic
corrections, but they are suppressed by powers of $Q/M$ as opposed
to $Q/m_\pi$, and are numerically small. Still they can be readily
computed in EFT. Defining the deuteron charge mean square radius by 
$\langle r^2\rangle_d \equiv 6 (dF_C/dq^2)$, 
one  finds
\begin{equation}\label{r2electric}
\langle r^2\rangle_d =
\langle r^2\rangle_N +\frac{1}{1-\gamma\rho}\frac{1}{8\gamma^2}+\frac{1}{32M^2}
=4.565\ {\rm fm}^4,
\end{equation}
\noindent to be compared with the experimental value 
$\langle r^2\rangle_d =4.538$ fm$^4$.

The magnetic form factor $F_M(q^2)$ at LO and NLO is simply the
electric form factor $F_C(q^2)$ multiplied by the isoscalar nucleon
magnetic moment $\kappa_n+\kappa_p$, except for a new two-body
term appearing at NLO without an ERT analog,
\begin{equation}\label{L2}
  {\cal L}=-e L_2 i \epsilon^{ijk} (NP_i N)^\dagger (NP_j N) B_k + {\rm
  h.c.}
\end{equation}
\noindent $L_2$ can be determined through the experimental value
of the deuteron magnetic moment and, using this value,
 the momentum dependence of
$F_M(q^2)$ is then predicted.

The $F_Q(q^2)$ form factor involves a transition between the $S$-
and $D$-wave components of the deuteron. At LO its value is
determined by a $S$- to $D$-wave transition operator whose
coefficient is extracted from the asymptotic $D/S$ ratio of the
deuteron, $\eta_{D/S}$. At NLO there is a new two-body term
\begin{equation}\label{CQ}
  {\cal L}=-e C_Q  (NP_i N)^\dagger (NP_j N)
  \left(\nabla^i\nabla^j-\frac{1}{3}\nabla\delta^{ij}\right)A^0,
\end{equation}
\noindent whose coefficient $C_Q$ can be fitted to the experimental 
deuteron quadrupole moment. At N$^2$LO the only contribution comes 
from the finite size
of the nucleon charge distribution $\langle r^2\rangle_N$.
 The value of $F_Q(0)$ is  then a  fit, but the momentum dependence is 
an EFT prediction. The presence of a counterterm not determined by 
$NN$ scattering at NLO indicates 
that $\mu_Q$ is sensitive at the $\sim 10\%$ level to short-distance physics 
not determined by $NN$ scattering. 
That is probably the reason different potential-model calculations 
underpredict $\mu_Q$ by $\simeq 5\%$.

\subsubsection{Deuteron polarizabilities and Compton scattering}
The deuteron is a very loose bound state so it is no surprise that its electric
polarizability depends mostly on the large-distance part of its wave-function.
Consequently, a model-independent prediction can be made with a 
high degree of accuracy.
The electric (magnetic) scalar deuteron polarizabilities
$\alpha_{E 0}$ ($\alpha_{M 0}$)  are defined by
\begin{equation}\label{polarizabilities_definition}
{\cal L_D}=2\pi \alpha_{E 0} d^{i \dagger}d^i \vec{E}^{\, 2}+
           2\pi \alpha_{M 0} d^{i \dagger}d^i \vec{B}^{\, 2},
%&=& 2\pi a\lpha_{E 2}\left[d^{i \dagger}d^j+d^{j \dagger}d^i-\frac{2}{3}
%\delta^{ij}d^{k \dagger}d^k    \right]\vec{E}^i\vec{E}^j+
%2\pi a\lpha_{M 2}\left[d^{i \dagger}d^j+d^{j \dagger}d^i-\frac{2}{3}
%\delta^{ij}d^{k \dagger}d^k    \right]\vec{B}^i\vec{B}^j
\end{equation}
\noindent  where $d^i$ is a canonically normalized deuteron field.

The EFT  result for  $\alpha_{E 0}$ up to N$^3$LO is \cite{seattle_pionless}
\begin{equation}\label{electric_polarizability}
\alpha_{E 0}=\frac{\alpha M}{32 \gamma^4}\frac{1}{1-\gamma\rho}
\left( 1+\frac{2\gamma^2}{3M^2}
+\frac{M \gamma^3}{3\pi}D_P+\ldots \right)=  0.6325\pm 0.002\ {\rm fm}^3,
\end{equation}
\noindent where $D_P=-1.51$ fm$^3$ is a combination of constants
describing $P$-wave two-nucleon interactions and in the last line
we used the fact that higher-order computations always build the
deuteron wave-function renormalization factor $Z=1/(1-\gamma\rho)$
\footnote{A modified method counting the effective range parameters as 
$r_{0s}, \rho\sim 1/Q$ formalizes this observation \cite{range_large}.}.
The first two-nucleon current operator appears only at N$^5$LO.
Since all the inputs were determined
from $NN$ scattering, we expect that any model describing
the phase shifts well should make predictions within the estimated error.
Indeed a number of models predict $\alpha_{E 0}=0.6328\pm 0.0017$ fm$^3$.

%%%*** Precisamos mesmo de duas linhas na formula acima?
%%%nao
Magnetic, vector and tensor polarizabilities, and their momentum-dependent 
analogs were not yet analyzed in the pionless EFT (see below for results
using the pionful EFT). However, these polarizabilities determine 
the amplitude
for low-energy Compton scattering off the deuteron, that was analyzed in
Ref. \cite{Rupak_Griesshammer_compton}. 
The {\it nucleon} polarizabilities appear 
already
at NLO and Compton scattering can be used to extract the hard-to-measure
neutron polarizabilities.

\subsubsection{Radiative capture of neutrons by protons}
\begin{figure}[!t]
\centerline{\epsfysize=5cm\epsfbox{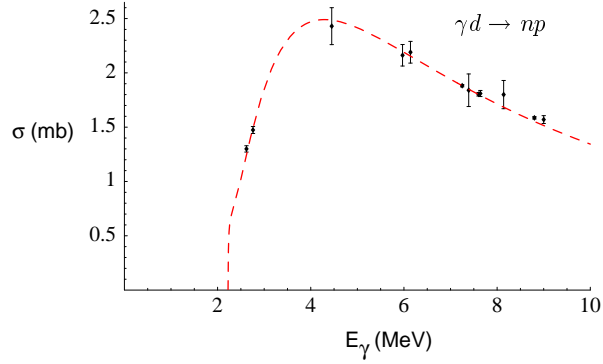}}
\caption{Cross section for $\gamma+d\rightarrow n+p$ 
as function of the photon energy
at N$^4$LO, compared to data.
{}From Ref. \cite{Rupak_BBN}, courtesy of G. Rupak.
}
\label{photodissociation}
\end{figure}

The $n+p\rightarrow d+ \gamma$ reaction at low energies is a key
ingredient in Big-Bang nucleosynthesis calculations. The amplitude for
this process can be expanded in multipoles as
\begin{eqnarray}\label{np_to_dgamma_amplitude}
{\cal A}&=&i e X_{M1V} \epsilon^{ijk} \epsilon^{* i} k^j \varepsilon^{ k} \;
n P^3 p
+ e  X_{E1V} \; n \tau_2\tau_3\sigma_2 \vec{\sigma} \cdot\vec{\epsilon}^* p P^i\varepsilon^{* i} \nonumber \\
&& + \frac{e X_{M1S}}{\sqrt{2}} \; n P^i
[k^i\vec{\epsilon}^* \cdot\vec{\varepsilon}^*-\vec{\epsilon}^* \cdot \vec{k}\varepsilon^{* i}    ]p \nonumber \\
&& +\frac{e X_{E2S}}{\sqrt{2}} \; n P^i
[k^i\vec{\epsilon}^* \cdot\vec{\varepsilon}^*+\vec{\epsilon}^* \cdot \vec{k}\varepsilon^{* i} -
\frac{2}{3}\epsilon^{* i} \vec{k}\cdot \vec{\varepsilon}^*]p+\ldots, 
\end{eqnarray}
\noindent where $n$ and $p$ are the neutron and proton Pauli spinors,
$k$ is the photon momentum and $\epsilon$ ($\varepsilon$) is the polarization
of the deuteron (photon). At low energies the form factor $X_{M1V}$ dominates
the cross section by a few orders of magnitude. 
Its computation at LO is the same
as the ERT one and underpredicts the experimental value for thermal neutron
capture by $10\%$. This discrepancy was explained long ago as due to a
pion-exchange current contribution \cite{Riska_Brown}. In EFT
\cite{seattle_pionless} the same effect
is encapsulated in the two-nucleon current
\begin{equation}\label{L_1M}
{\cal L}= e L_{M1V} (N P^i N)^\dagger (N P^A N)\delta_{A3}B^i + {\rm h.c.},
\end{equation}
\noindent whose coefficient $L_{M1V}$ can determined by the cold-capture
cross section. The momentum dependence is then predicted. However for photon
energies larger than a few MeV the $X_{E1V} $ form factor dominates the
cross section. Up to N$^3$LO the computation of $X_{E1V}$ involves only
$C_{0 t}$, $C_{2 t}$, the $P$-wave interaction combination $D_P$ encountered in
the polarizability calculation, and $L_{M1V}$ fitted at threshold. 
At N$^4$LO some
relativistic effects and a new
term appears \cite{Rupak_BBN},
\begin{equation}\label{L_1E}
{\cal L}= e L_{E1V} (N P^i N)^\dagger (N
(\vec{\nabla}_i\tilde{P}_{jA}
-\overleftarrow{\nabla}_j \tilde{P}_{iA}) N)\delta_{A3}E^j,
\end{equation}
\noindent where $\tilde{P}_{iA}=\sigma_2\sigma_i \tau_2\tau_A/\sqrt{8}$, 
which is fitted using data for the inverse reaction
$d+\gamma\rightarrow n+p$. The resulting cross section for the
photodissociation of the deuteron is shown in
Fig. (\ref{photodissociation}),
and compared to data \cite{data_BBN}. The estimated error is of the
order $(Q/m_\pi)^5\sim 1\%$. These precise, analytical computations
are particularly useful for Big-Bang nucleosynthesis
codes.

A set of polarization observables in the $ n+p\rightarrow\gamma+d $
reaction was analysed in Ref. \cite{seattle_suppressed}.

\subsubsection{Neutrino-deuteron scattering and proton-proton fusion}

A complete set of reactions involving (anti-)neutrino 
breakup of the deuteron
were computed in the pionless EFT approach 
\cite{Chen_neutrino_1}.
Proton-proton fusion and the interplay of Coulomb interactions
 were analysed in 
Ref. \cite{ravndal1}.
 These
reactions are essential for the understanding of solar
neutrino physics since they are relevant for both the production
($p+p\rightarrow d+e^++\nu_e$) and detection in heavy water
detectors (through the neutral current $\nu+d\rightarrow \nu+n+p$ and the
charged current $\bar{\nu}+d\rightarrow e^++n+n$  reactions).

The weak interactions are described by the familiar neutral and
charged current pieces
\begin{equation}\label{EWlagrangian}
{\cal L}=-\frac{G_F}{\sqrt{2}}(l_{Z \mu}J_Z^\mu+l_{+
\mu}J_-^\mu+{\rm h.c.}),
\end{equation}\noindent with the leptonic currents $l_{Z
\mu}=\bar{\nu}(1-\gamma_5)\gamma_\mu\nu$, 
$l_{+ \mu}=\bar{\nu}(1-\gamma_5)\gamma_\mu e$
and hadronic currents $J^\mu_-=V^\mu_--A^\mu_-$ and $J_Z^\mu=-2
\sin^2\theta_W V^\mu_0 +
(1-2\sin^2\theta_W)V^\mu_3-A^\mu_0-A^\mu_3$. The isosinglet vector
(axial) current $V^\mu_0$ ($A^\mu_0$) and isotriplet vector
(axial) current $V^\mu_A$ ($A^\mu_A$), written in terms of the
nucleon fields, have contributions in the form of one- and two-nucleon
operators. The one-nucleon operators are determined by the axial
coupling constant $g_A=1.26$, the neutron and proton magnetic
moments $\kappa_n$ and $\kappa_p$, the strange contribution to the
proton spin $\langle \bar{s}\gamma^\mu\gamma_5 s\rangle$ and the strange
magnetic moment of the proton $\mu_s$. The two-nucleon currents
contributing up to N$^2$LO are
\begin{eqnarray}\label{neutrino_counterterms}
{\cal A}^i_a&=&L_{1 A}(NP^iN)^\dagger(NP_AN),\nonumber\\
{\cal A}^i_0&=&-2i L_{2A}
\epsilon_{ijk} (NP^iN)^\dagger(NP^jN)+{\rm h.c.},\nonumber\\
{\cal V}^i_a&=&2i L_{1}\epsilon_{ijk}
(NP^jN)^\dagger(\vec{\nabla}^k+\overleftarrow{\nabla}^k)(NP_aN),
\nonumber\\
{\cal V}^i_0&=&2i L_{2}
(NP^jN)^\dagger(\vec{\nabla}_j+\overleftarrow{\nabla}_j)(NP^iN).
\end{eqnarray}\noindent $L_1=L_{M1V}$ was
 determined through the neutron-proton cold-capture cross section and  
$L_{2}$  was determined
by the value of the deuteron magnetic moment. The parameters $L_{2
A}, \langle \bar{s}\gamma^\mu\gamma_5 s\rangle$ and $\mu_s$, 
which are not well
determined experimentally,  have a negligible impact on the
cross section ($<1\%$) due to the almost orthogonality between
initial and final states in the triplet channel. The only relevant
unknown in a N$^2$LO calculation is then the value of $L_{1A}$.
Using the estimate in Eq.~(\ref{CXestimate}) we find
$L_{1A}\sim 4\pi/M \mu^2 \sim 5$  fm$^3$.
It is found that two potential-model results, 
one with and another without exchange-current
terms, and differing by about $5\%$, can be reproduced by varying
the value of $L_{1A}$ within this range. This shows that the
difference between these calculations comes from different
assumptions about the short-distance physics. To fix this
indeterminacy and, consequently, have predictions for the
$\nu d$ reactions at the percent level the value of
$L_{1A}$ needs to be determined experimentally. One possibility is
the measurement of one of these reactions at one energy. The other
is to extract $L_{1A}$ through another process sensitive to this
term, like tritium $\beta$-decay or muon capture on the deuteron.
The challenges involved in this extraction from the well-measured
value of the tritium lifetime will be discussed below.

\subsection{The three-body system and the limit cycle}
\label{greatness}

The study of three-nucleon systems using EFT is still in its infancy compared
to its mature status in the two-nucleon sector. Only now calculations
accurate enough are appearing for the triton-$^3$He channel that open
the door for precision calculations of processes involving external currents.
Those processes may turn out to be a very important way of fixing the value of
two-body LECs that are hard to measure in the deuteron.
  The new ingredients in going from two- to three-body systems
are three-body interactions. In the absence of fine-tuning their
typical size is determined by dimensional analysis. Since they subsume
physics contained within the range $1/\Lambda\sim 1/m_\pi$, a 
three-body force
with $2n$ derivatives would have the typical size
\begin{equation}\label{3force_naive}
{\cal L}_3= D_{2n} \vec{\partial}^{\, 2n} (N^\dagger N)^3\rightarrow
D_{2n}\sim \frac{(4\pi)^2}{M \Lambda^{4+2n}}.
\end{equation}
\noindent Just as it happens in the two-body force, the fine-tuning
in the two-body $S$-wave channels introduces a new scale 
$ \gamma\sim 1/a_s$
that invalidates the estimate in Eq.~(\ref{3force_naive}). We will resort
to the same argument used before to estimate the size of these contributions.
We will demand  observables to be cutoff independent order by order in the
low-energy expansion, which determines the running of the three-body forces
with the assumption that their typical size is set by the size of their 
running,
$D_{2n}(\Lambda)\sim D_{2n}(2\Lambda)-D_{2n}(\Lambda)$. 
As mentioned before, it is unlikely that 
$D_{2n}(\Lambda)$ is much smaller than this estimate 
for a particular value of the regulator $\Lambda$,
or that it contains a large $\Lambda$-independent piece.
Another way of looking at this
way of estimating the LECs is to remember that if the inclusion  of a
particular LEC is necessary in order to have cutoff-independent
results (at a particular order in the expansion), this LEC
needs to be large enough to appear at that same order of the expansion.

The doublet $S$-wave channel (where the triton and $^3$He are) has
a very different behavior from the other channels. The physical reason 
is that this is the only channel
where all three nucleons can occupy the same point in space (two spin
and two isospin states allow for a maximum of four nucleons in the same state).
A system of three bosons also displays this property and is qualitatively 
similar to three nucleons in the doublet $S$-wave channel.
In the remaining channels either the angular-momentum barrier or the 
Pauli exclusion principle forbids the three nucleons to touch.
One would then expect  that the doublet $S$-wave channel 
(and systems of three bosons) are much more sensitive to 
short-distance physics than the other channels, an 
expectation that we will see confirmed by further analysis.

To avoid unnecessary complications we will present explicit expressions
only for the case of the $S$-wave three-boson system.
The formulae for the nucleon
cases in the different channels can be deduced in an analogous way and can
be found, for instance, in
Refs. \cite{stooges_dependence, stooges_triton, Bedaque_highL}. 
A convenient first step is to rewrite the action
in terms of an auxiliary ``dimeron'' field $d$ \cite{Kaplan:1996nv},
\begin{eqnarray}\label{transvestite_lagrangian}
{\cal L}_{bosons}&=&N^\dagger(i\partial_0+\frac{\vec{\nabla}^2}{2M}+\ldots)N
-C_0 (NN)^\dagger NN -D_0 (NNN)^\dagger NNN+ \ldots \nonumber\\
&\rightarrow&N^\dagger(i\partial_0+\frac{\vec{\nabla}^2}{2M}+\ldots)N
+ \Delta d^\dagger d \nonumber\\
&&+ y(d^\dagger NN + d N^\dagger N^\dagger)
 +gd^\dagger d N^\dagger N+\ldots,
\end{eqnarray}
\noindent where $C_0=y^2/\Delta$ and
$D_0=-g y^2/\Delta^2$. The equivalence between the two Lagrangians
 above can
be shown by simply performing the Gaussian integral over the auxiliary field
$d$. Since $d$ has the quantum numbers of
a two-particle bound state, it can be used as an interpolating field for the
bound state.
Note that the normalization of the field $d$ (and the value of $\Delta$)
 is arbitrary and all the
observables should be a function only of the combinations 
$y^2/\Delta$ and $g y^2/\Delta^2 $,
not of $y$, $g$ and $\Delta$ separately.

\begin{figure}[!t]
\centerline{\epsfysize=5cm\epsfbox{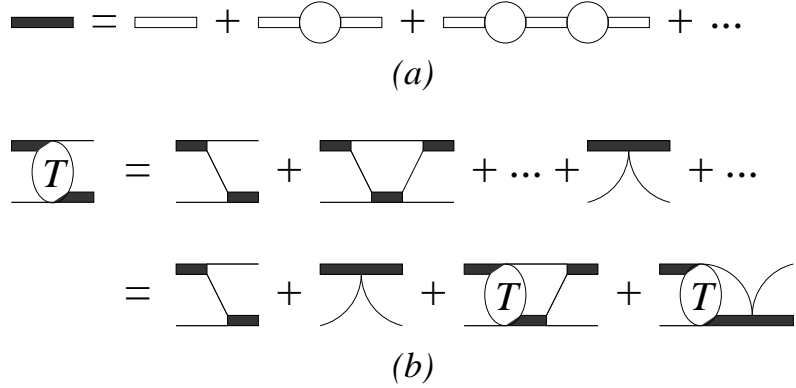}}
\caption{LO graphs contributing to the dressed propagator of the
dimeron {\it (a)} and to the particle/dimeron amplitude {\it (b)}.}
\label{three}
\end{figure}

The propagator ${\mathbf\Delta}(p)$ for the dimeron field of momentum $p$
is given by the sum of bubble graphs shown at the top of
Fig. (\ref{three}).
The power counting leading to the necessity
of summing all these graphs when the scattering length is large is identical
to the one discussed in connection to $NN$ scattering.
Taking the arbitrary constant $\Delta$ to scale as $\Delta\sim 1$, we have
$y^2\sim C_0\sim 4\pi/MQ$ and thus ${\mathbf\Delta(p)}$ scales as
 ${\mathbf \Delta}(p)\sim (4\pi/My^2) (1/Q)\sim 1$.
Summing all graphs, we can write
\begin{equation}
{\mathbf\Delta}(p) =\frac{1}{-\gamma+\sqrt{\frac{3p^2}{4}-ME}}.
\end{equation}

Let us now consider the graphs describing the scattering of one particle
off the bound state of other two,
shown at the bottom of  Fig.~(\ref{three}). We can determine the impact
of each graph by power counting. For each additional loop we have two
particle propagators, one $d$ propagator, two powers of $y$ and one loop
integration, for a total of $(M/Q^2)^2 (1) (4\pi/MQ) (Q^5/4\pi M)\sim 1$.
There is no suppression for additional loops, and all diagrams involving
an arbitrary number of particle exchanges contribute at LO. Graphs including
the two-body $C_2$ operator are suppressed and start appearing at NLO. 
Graphs
including the three-body force may or may not be leading, depending
of the size of the three-body forces. Since at this point we do not know
how large they typically are, we will provisionally include them.
The
chain of diagrams to be summed, contrary to the two-particle case,
does not form a simple geometrical series and cannot be summed
analytically. We can however consider the second line in
Fig.~(\ref{three}) as an equation determining this sum. For the
bosonic case, $\lambda=1$, we have
\begin{equation}\label{faddeev}
t_k(p)=K(p,k) +\frac{2H}{\Lambda^2}+
\frac{2\lambda}{\pi}\int_0^\Lambda\ dq q^2
{\mathbf\Delta}(q)\left(K(p,q)+ \frac{2H}{\Lambda^2}\right)t_k(q),
\end{equation}\noindent with
\begin{eqnarray}\label{kernel}
K(p,q)&=&\frac{1}{pq}\ln
\left(\frac{p^2+pq+q^2-ME}{p^2-pq+q^2-ME}\right),\nonumber\\
H(\Lambda)&=&\frac{g\Lambda^2}{4My^2},
\end{eqnarray}\noindent where
$t_k(p)$ is the scattering amplitude with all but
the outgoing single-particle line on-shell
(``half-off-shell amplitude''), 
$k$  is the incoming momentum in the center-of-mass system,
$p$ is the outgoing momentum, 
and $E=3k^4/4m - \gamma^2$ is the total energy. 
The on-shell point is at $p=k$. 
In the case of nucleons in the quartet channel the same
equation is obtained, but with $\lambda=-1/2$ 
and some additional momentum dependence in the three-body force
(since the simple momentum-independent
three-body force does not contribute to this channel).
Because all spins are parallel in the quartet channel, only triplet two-body
interactions occur and the value of $\gamma$
is determined by the deuteron pole. 
The auxiliary field $d$ carries 
the quantum numbers of the deuteron.
 In the doublet case, singlet and triplet two-body interactions contribute.
The analog of Eq.~(\ref{faddeev}) is   a pair of coupled 
integral equations that, 
in the $SU(4)$ limit\footnote{See Ref. \cite{Mehen:1999qs} for a discussion
of the $SU(4)$ limit in the two-nucleon sector.}
where the singlet and triplet scattering lengths are equal (or in the
ultraviolet where $1/a_{s}$,  $\gamma$ can be discarded) decouple
into a pair of equations like
Eq.~(\ref{faddeev}), one with $\lambda=1$, another with $\lambda=-1/2$.
Two auxiliary fields appear, one with the quantum numbers of the deuteron, 
another with the $^1S_0$ quantum numbers.
In all spin channels, equations for the higher partial waves are obtained
by substituting the logarithm in the kernel by a Legendre function
$Q_l(pq/(p^2 + q^2 - ME))$.
Eq.~(\ref{faddeev}) is the version of the Faddeev equation
(see Ref. \cite{Amado_review} for a nice introduction)
% that is valid for an arbitrary potential, 
 appropriate for contact interactions.
It was first derived, by very different methods, in Ref. \cite{skorny}.
It is only for {\it separable} potentials, like the contact 
interactions considered
here,  that the Faddeev equation reduces to a one-dimensional integral 
equation. This simplification reduces the numerical work involved
by many orders of magnitude.

\subsubsection{Most channels are easy}
\begin{figure}[!t]
\centerline{\epsfysize=7cm\epsfbox{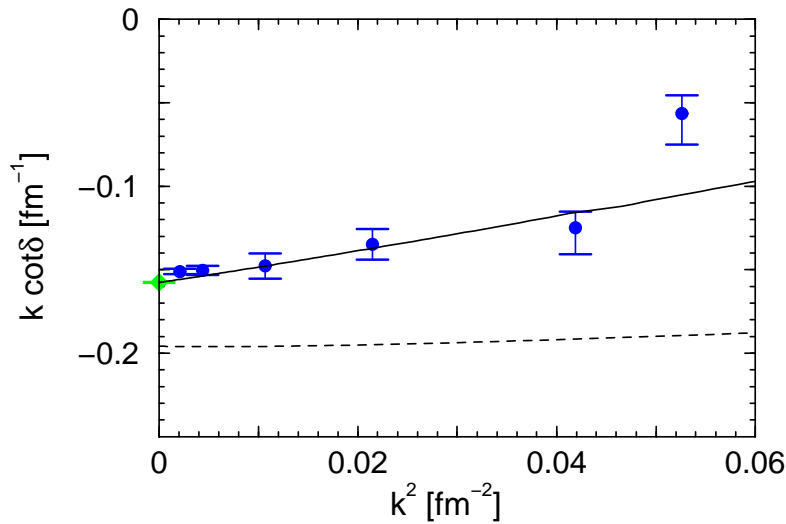}}
\caption{$k \cot\delta$ for $nd$ scattering in the quartet  $S$-wave channel
as function of the energy. 
The dashed line is the LO and the full line the N$^2$LO result
\cite{stooges_dependence}. 
Points are from a PSA \cite{data_quartet_phaseshifts} 
and a near-threshold measurement \cite{data_quartet}.
Figure courtesy of H.-W. Hammer.}
\label{quartet}
\end{figure}

Let us first consider the channels in the second group, that is, all
channels but the doublet $S$ wave.
The diagrams summed by Eq.~(\ref{faddeev}) are all ultraviolet finite.
That would suggest that there is no need to include  three-body forces to
absorb the cutoff dependence since this dependence is a subleading
$1/\Lambda$ effect. The (numerical) solution confirms this.
Even in
the absence of a three-body force the phase shifts are only weakly
dependent on
$\Lambda$ and $t_k(k)$ has  a finite limit when
$\Lambda\rightarrow\infty$. Higher-order corrections can be 
included either
perturbatively
(as was done in the two-body sector) or non-perturbatively
through
the denominator in the dimeron propagator. This
last resummation amounts
to including some (but not all!) higher-order effects.
It can be automatically
computed by solving a modified version of Eq.~(\ref{faddeev}), which is easier
than performing perturbative calculations at high orders.
Calculations of the neutron-deuteron ($nd$)
phase shifts are presently available up to
N$^2$LO \cite{Bedaque_vanKolck_quartet, stooges_dependence, Bedaque_highL}. 
At this order the only inputs are $\gamma$ and $r_{0t}$ 
(for the three-body quartet channels), and 
$1/a_s$ and $r_{0s}$ (needed only
for the doublet channels). There are no unknown LECs
appearing at N$^3$LO,
so this approach can be easily pushed to higher orders (and precision).
For a flavor of the results, we show the predicted
quartet $S$-wave  phase shift
in Fig.~(\ref{quartet}),
compared to a PSA \cite{data_quartet_phaseshifts} 
(at finite $k$) and a scattering length measurement \cite{data_quartet}
(essentially at $k=0$).

Most of the data is not precise enough to provide a strict test for the
convergence of
the low-energy expansion, but the zero-energy point is much better measured.
The EFT calculation gives for the quartet $S$-wave scattering length 
$a_{3/2}=5.09+0.89+0.35+\ldots=6.33\pm 0.05$ fm 
\cite{skorny, Efimov_range, Bedaque_vanKolck_quartet},
while the
experimental value is $a_{3/2}^{exp}=6.35\pm 0.02$ fm \cite{data_quartet}. 
The EFT  error
is probably overestimated since the N$^2$LO calculation resummed some of
the N$^3$LO pieces and the remaining ones are known to be smaller
than the naive estimate (like the
effect of the two-body shape parameter).
Since the whole input in these calculations were the threshold parameters of
$NN$ scattering, these results are universal (a ``low-energy theorem'').
Any model
with the correct scattering lengths and effective ranges
 (and not wildly wrong phase shifts above threshold) should reproduce them,
within the estimated error. The small discrepancy with 
``first-generation'' $NN$ potentials can be explained by the imprecise values
of $a_{s,t}$ that those models predicted. 
``Second-generation'' (or ``realistic'') potentials
fixed this discrepancy \cite{Friar_quartet}.

Coulomb interactions are important in proton-deuteron scattering at small
energies. In Ref. \cite{Rupak_coulomb} it was shown how they can be easily
incorporated in the quartet channels by a simple change in the kernel of
Eq.~(\ref{kernel}).

\subsubsection{Triton-$^3$He channel and the limit cycle}
\label{tritonsect}

\begin{figure}[!t]
\epsfysize=8cm\epsfbox[70 520 370 715]{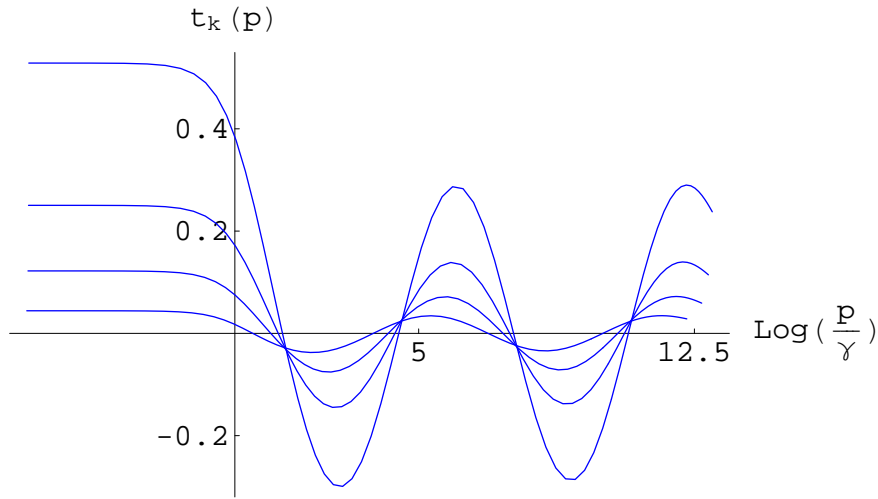}
\caption{Zero-energy half-off-shell
amplitude for boson/two-boson scattering as a function of 
the outgoing momentum $p$, from a numerical
solution of Eq.~(\ref{faddeev}) with no three-body force, 
for several different cutoffs.}
\label{oscillations}
\end{figure}

The change of sign  of $\lambda$ between the bosonic and quartet
equations qualitatively changes  the behavior of respective solutions.
The most striking feature is that the solution of the bosonic (and
of the $S$-wave doublet) equation in the absence of
a three-body force
depends sensitively on the
value of cutoff used. This happens despite the fact that, as mentioned
above,
 there is no ultraviolet divergence in the graphs summed. To illustrate
this we show in Fig.~(\ref{oscillations}) the $k=0$ solutions of
Eq.~(\ref{faddeev}) corresponding to various cutoffs (with
$\lambda=1$, $H=0$). Notice that the three-body scattering
length, determined by the asymptotic plateau
on the left, $t_{k=0}(k=0)$, can take any value by just varying the cutoff
within a reasonable range.

\noindent 

It has been known for a long time that Eq.~(\ref{faddeev}) is
not well defined in the limit $\Lambda\rightarrow\infty$. This
disease can be shown in a variety of ways
\cite{Faddeev_Minlos, Danilov, Efimov_effect, stooges_3bosons}. 
Consider the intermediate-momentum regime $Q\ll p \ll \Lambda$ (remember
that $Q$ stands for any infrared scale $Q\sim \gamma\sim k$). Up
to terms suppressed by powers of $Q/\Lambda$, and assuming $H\sim
1$, Eq.~(\ref{faddeev}) reduces to \cite{Faddeev_Minlos}
\begin{equation}\label{asymptotic_equation}
t_k(p)=\frac{4}{\sqrt{3}\pi}\int_0^\infty dq\ \frac{1}{q} \ln
\left(\frac{p^2+pq+q^2}{p^2-pq+q^2}\right) t_k(q).
\end{equation}
\noindent 
Eq.~(\ref{asymptotic_equation}) has two
symmetries\footnote{These symmetries suggest that in the limit
$\gamma,1/a_s\rightarrow 0$  the full conformal symmetry holds.
 In the two-body sector it was shown that this is indeed the
case, even for off-shell amplitudes \cite{conformal}. 
It is not known whether the three-body amplitude is conformal in this limit.
The three-body forces break scale invariance at the order they first appear.}
\begin{eqnarray}\label{conformal}
t_k( p)&\rightarrow&t_k(\alpha p)\ \ \ ({\rm scale\ invariance})\nonumber\\
t_k( p)&\rightarrow&t_k (1/p)\ \ \ ({\rm inversion\
symmetry)},
\end{eqnarray}
\noindent 
that suggest the power-law solution
$t_k(p)\sim p^{s}$. The allowed values of $s$ are determined by
plugging this {\it ansatz} back into Eq.~(\ref{asymptotic_equation}). 
For values of
$s$ such that ${\mathbf Re}(s)< 0$ we find
\begin{equation}\label{s}
\frac{8 \lambda}{\sqrt{3}s}\ \frac{\sin \frac{\pi s}{6}}{\cos\frac{\pi
s}{2}}=1.
\end{equation}
\noindent  
For values of $\lambda$ smaller than
$\lambda_c\equiv 3\sqrt{3}/4\pi=0.4135\ldots$, the roots 
of Eq.~(\ref{s})
have ${\mathbf Re}(s)< -1$ and the 
half-off-shell amplitude goes rapidly to zero as 
$p\rightarrow\infty$. In the quartet channel, for instance, 
$t_k(p)\sim 1/p^{3.17}$, which is a softer ultraviolet behavior than 
the one expected in perturbation theory, $t_k(p)\sim 1/p^{2}$. For the
$S$-wave doublet case (or the bosonic case) though, $\lambda=1$ and there is
a pair of imaginary solutions $s=\pm i s_0$, with $s_0= 1.006\ldots$ 
The asymptotic 
behavior of the half-off-shell amplitude is then
\begin{equation}\label{asymptotics}
t_k(Q\ll p \ll \Lambda)= A \sin (s_0 \ln p + \delta)
\end{equation}
\noindent 
where $A$ and $\delta$ are $p$-independent numbers.
 This oscillatory 
behavior can be seen in Fig.~(\ref{oscillations}).

The normalization $A$
is fixed by the inhomogeneous term in Eq.~(\ref{faddeev}).
Because that is important
only in the infrared region $p\sim Q$, $A$ can be determined only by 
matching Eq.~(\ref{asymptotics}) to the solution in the infrared
region. The phase $\delta$ is determined by matching Eq.~(\ref{asymptotics})
to the solution in the ultraviolet region and will depend on the ultraviolet
 physics (like the value of  the regulator $\Lambda$ and 
the three-body force 
$H(\Lambda))$. Actually, by dimensional analysis,
the $\Lambda$ dependence of $\delta$ has to be the form $\delta(\Lambda)=
-s_0 \ln \Lambda + \bar{\delta}$.
 If $\Lambda$ is varied while keeping $H$ constant, 
as we have done so far,
the dependence
 on the asymptotic phase $\delta$ will ``spill over''
 the infrared region and 
change the on-shell amplitude by a factor of ${\cal O}(1)$. 
On the other hand, 
$\delta$ does not depend on the infrared scales ($k$, $\gamma$, $1/a_s$)
and so $H(\Lambda)$ 
can be adjusted in such a way that 
$\delta=-s_0 \ln \Lambda+\bar{\delta}(H(\Lambda))\equiv -s_0 \ln\bar{\Lambda}$
is cutoff independent (for any value of $k$), with $\bar{\Lambda}$ being
 a constant. 
Thus, the requirement of cutoff invariance means that $H(\Lambda)$
runs at LO and, for generic values of $\La$,  $H(\Lambda)\sim 1$.
 The typical size of the three-body force is then 
\begin{equation}
D_0\sim \frac{(4\pi)^2 H}{M \Lambda^2 Q^2}
   \sim \frac{(4\pi)^2}{M \Lambda^2 Q^2},
\end{equation}
as opposed to the naive estimate in Eq.~(\ref{3force_naive}). 
This means
that $D_0$ is 
enhanced in the presence of fine-tuning in the two-body sector by a factor
of $(\Lambda/\gamma)^2$. The arbitrary parameter $\bar{\Lambda}$ has to 
be fixed by one piece of experimental input coming from a three-body 
observable. The two-body phase shifts are {\it not} enough to fix the
three-body physics already at LO. Another way of looking at the 
renormalization of the three-body system that is more easily generalizable to
higher-order calculations is to consider  two amplitudes $t_k^\Lambda(p)$
and $t_k^{\Lambda'}(p)$ obtained by solving Eq.~(\ref{faddeev}) with two 
cutoffs $\Lambda$ and $\Lambda'$ and 
the corresponding three-body forces $H(\Lambda)$ and 
$H(\Lambda')$. The integral equation determining  $t_k^{\Lambda'}(p)$ can
 be written as
\begin{eqnarray}\label{3bosonsLambdaprime}
&&  \int_0^\Lambda dq \left(\delta(p-q)- \frac{2}{\pi} q^2
  {\mathbf\Delta}(q)
   K(p,q)\right)
  t_k^{\Lambda'}(q) \nonumber\\
&=& K(p,k) +\frac{2H(\Lambda)}{\Lambda^2}
 + \frac{2}{\pi}\frac{2H(\La)}{\La^2}\int_0^\Lambda dq
  q^2 {\mathbf\Delta}(q)t_k^{\Lambda'}(q)\nonumber\\
&&+\frac{2H(\Lambda')}{\Lambda^{'2}}
-\frac{2H(\Lambda)}{\Lambda^{2}}
 +\frac{2}{\pi}\left(\frac{2H(\Lambda')}{\Lambda^{'2}}
-\frac{2H(\Lambda)}{\Lambda^{2}}\right)
 \int_0^\Lambda dq q^2
  {\mathbf\Delta}(q) 
t_k^{\Lambda'}(q)\nonumber\\
&& 
  +\frac{2}{\pi}
 \int_\Lambda^{\Lambda'} dq q^2
  {\mathbf\Delta}(q)
 \left( K(p,q)+  \frac{2H(\Lambda)}{\Lambda^2}   \right)
  t_k^{\Lambda'}(q).
\end{eqnarray}
\noindent The first two lines in Eq.~(\ref{3bosonsLambdaprime})
are identical to the equation determining $t_k^{\Lambda}(p)$. 
If the effect of the remaining terms are small 
(up to terms of ${\mathcal O}(Q/\La)$),
 $t_k^{\Lambda}(p)= t_k^{\Lambda'}(p) $ 
(again, up to terms of ${\mathcal O}(Q/\La)$).
These terms are indeed small, suppressed by a factor of $Q/\La$ compared
to the leading ones. However, their effect on $t_k^{\Lambda}(p)$ is 
not suppressed. That is because the operator acting on $t_k^{\Lambda}(p)$
on the left-hand side of Eq.~(\ref{3bosonsLambdaprime}) is nearly singular,
that is, it has a small eigenvalue of order $Q/\Lambda$. 
\begin{equation}\label{eigenvalue_equation}
\int_0^\Lambda\left(\delta(p-q)-\frac{2}{\pi} q^2
{\mathbf\Delta}(q)K(p,q)\right) a(q) = {\mathcal O}(Q/\Lambda)
a(p)
\end{equation}
\noindent The eigenfunction $a(p)$ has the same asymptotics shown in 
Eq.~(\ref{asymptotics}) as $t_k^{\Lambda}(p)$. In our determination of
the asymptotics, Eq.~(\ref{asymptotics}), we have already shown 
Eq.~(\ref{eigenvalue_equation}) in the limit
$\La\rightarrow\infty$ and the total energy $E\rightarrow 0$. 
We now see that the operator 
in the  left-hand side of Eq.~(\ref{3bosonsLambdaprime}) is almost
non-invertible and that the projection of $t_k^{\Lambda}(p)$ in the $a(p)$ 
direction is sensitive to  the  ${\mathcal O}(Q/\Lambda)$ terms in
the right side of Eq.~(\ref{3bosonsLambdaprime}). The amplitudes computed
with different $\La$'s shown in Fig. (\ref{oscillations}) indeed differ
in the intermediate regime by a term of the form $\sin(s_0 \ln p+\delta)$.
The solution
$t_k^{\Lambda}(p)$ has an increased sensitivity to the ultraviolet physics
and changes by a factor of 
${\mathcal O}(1)$ 
if the right-hand 
side of Eq.~(\ref{3bosonsLambdaprime}) changes by a factor 
of ${\mathcal O}(Q/\La)$.
That explains the apparent contradiction between the cutoff sensitivity 
and the absence of ultraviolet divergences.

\begin{figure}[!t]
\epsfysize=7.8cm\epsfbox[50 510 390 710]{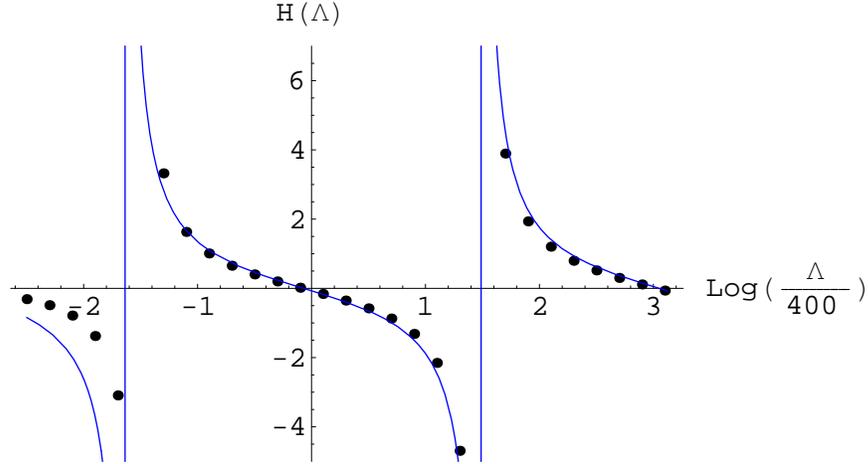}
\caption{Three-body force coefficient $H(\La)$ computed analytically
(blue line) and numerically (black dots) as a function of 
$\log(\La/400\ {\rm MeV})$.}
\label{hansplot}
\end{figure}

Using the asymptotic form of $t_k^{\Lambda}(p)$,
Eq.~(\ref{asymptotics}),
even in the region $p\sim \La$ where it is unjustified,
and dropping the terms suppressed by $(Q/\La)^2$ we can derive 
an approximate analytical expression for 
 $H(\La)$ needed to cancel the $Q/\La$ terms in  
Eq.~(\ref{3bosonsLambdaprime}) (and to guarantee $t_k^{\Lambda}(p)$ 
is cutoff independent up to order $Q/\La$):
\begin{equation}\label{H}
H(\La)=\frac{\cos\left(s_0 \ln\frac{\La}{\bar{\La}}+\arctan s_0\right)}
            {\cos\left(s_0 \ln\frac{\La}{\bar{\La}}-\arctan s_0\right)},
\end{equation}
\noindent 
where $\bar{\La}$ is the parameter
 determining the asymptotic phase, to be fixed by experiment.
We show $H(\La)$ in Fig.~(\ref{hansplot}). 
The points there 
 were determined
by numerically finding values of $H(\La)$ that, when 
inserted in Eq. (\ref{faddeev}), reproduce the same three-body 
phase shifts for different values of $\La$. 
The solid line is Eq.~(\ref{H}).
The agreement between 
Eq.~(\ref{H}) and the numerical points,
as well as the independence of the phase shifts with $\La$ after $H(\La)$
is properly changed, confirms our understanding of this somewhat unusual 
renormalization.
For further discussion of renormalization-group invariance
at this order, see Ref \cite{Hammer:2000nf}.

The asymptotic running of the three-body coupling can be interpreted
as a limit cycle.
The possibility of limit cycles in addition to fixed points
was suggested in Ref. \cite{wilson1},
but apparently never before seen in a simple physical system.
Limit cycles are now being further studied \cite{wilson2}.

\begin{figure}[!t]
\centerline{\epsfysize=7cm\epsfbox[75 462 493 709]{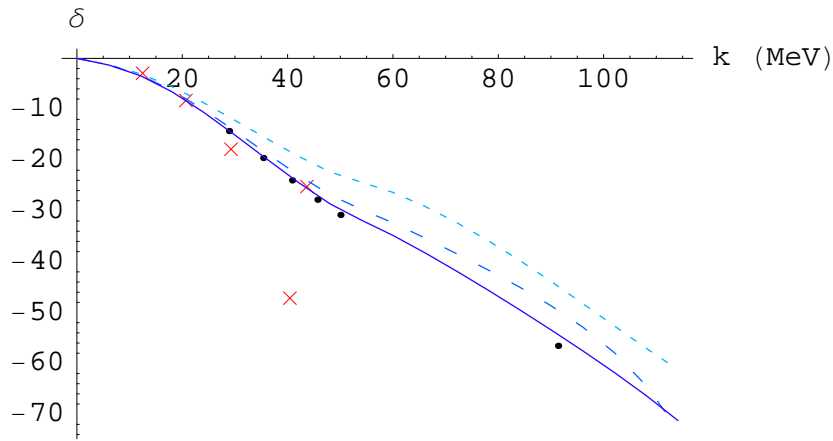}}
\caption{$nd$ phase shift (in degrees) in the 
doublet $S$-wave channel as function of the center-of-mass momentum: 
EFT at LO (light-blue dotted line), 
at NLO (blue dashed line), and at N$^2$LO (purple solid line),
a PSA (red crosses),
and a potential model (black dots).
{}From Ref. \cite{toappear}. 
}
\label{doublet_phaseshifts}
\end{figure}

NLO calculations of phase shifts were performed in Ref. \cite{Hammer_range}
(scattering lengths were computed by different, but equivalent methods
in Refs. \cite{Danilov, Efimov_range}). 
No new three-body force
is needed at this order (although $H(\La)$ has a correction proportional 
to the two-body effective range). As in other channels, it is easier to
compute higher-order corrections by computing the kernel at the order desired
and iterating it. 
That is, one numerically solves Eq.~(\ref{faddeev}) with
a kernel that includes higher-order effects.
At NLO we demand $t_k^{\Lambda}(p)$ to be independent of $\La$ up to
terms of ${\cal O}(Q/\La)^2$. Due to the almost singular nature of the
integral equation, the  right-hand 
side of Eq.~(\ref{3bosonsLambdaprime}) has to be $\La$ independent 
up to
terms of order $(Q/\La)^3$. 
This can be accomplished with the same no-derivative three-body force
because the terms in the  right-hand 
side of Eq.~(\ref{3bosonsLambdaprime}) of ${\cal O}(Q/\La)^2$
are $k$, $p$ independent. 
(However, $H(\La)$ will have a different form than in
Eq. (\ref{H})). Similarly, at  N$^2$LO we need to cancel the  
${\cal O}(Q/\La)^3$
terms in the  right-hand 
side. Those terms, however, are proportional
to $k^2$, $p^2$ and can be absorbed only by a three-body force with two 
derivatives. So, at N$^2$LO, a new three-body 
force appears, and we need yet another piece of three-body data
to fix this new LEC. 
The cutoff sensitivity is dramatically reduced by going to higher 
orders, as it should \cite{toappear}.
In Fig.~(\ref{doublet_phaseshifts}) we show
the phase shifts in the $S$-wave doublet
channel computed at the first three orders \cite{toappear}.
As inputs in these calculations we have, besides 
the two-body 
interactions,   a no-derivative three-body
force fitted to the experimental binding energy of the triton at LO and a 
two-derivative three-body force fitted to the experimental value of the
doublet $nd$ scattering length at NLO.
The EFT results are compared to a PSA  \cite{data_quartet_phaseshifts}
and to  results from the
Argonne V18 potential plus the Urbana three-body force adjusted to
reproduce the correct triton binding energy \cite{kievsky}.

\begin{figure}[!t]
\vspace{0.5cm}
\centerline{\epsfysize=7cm\epsfbox[73 456 530 700]{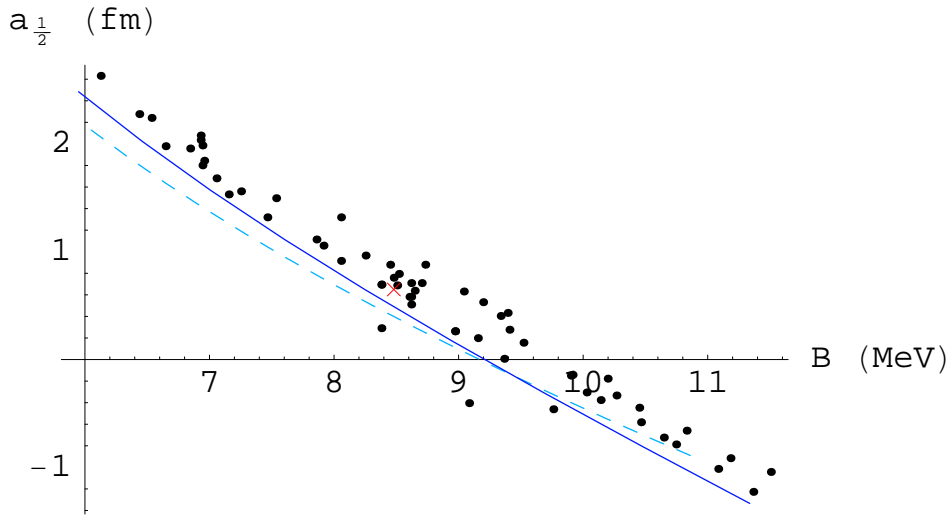}}
\caption{Correlation between the doublet $S$-wave $Nd$ scattering
length and the triton binding energy (Phillips line):
predictions of different models (black points),
EFT in LO (light dashed line)
and NLO (dark solid line), 
and experimental value (red cross). }
\label{phillips}
\end{figure}

The existence, already at LO, of a parameter not determined by 
$NN$ scattering means that models tuned to reproduce the low-energy
$NN$ phase shifts may differ widely on their predictions for
three-body properties. However, since up to NLO there is only a 
one-parameter
arbitrariness in the three-body predictions, there must be correlations 
among
these predictions. This was noted empirically in Ref. \cite{Phillips}.
Fig.~(\ref{phillips}) shows results for the doublet $S$-wave scattering 
length and the triton binding energy from a number of models, all reproducing 
the same low-energy $NN$ scattering \cite{tkachenko}. 
The predictions cover a wide 
range but show a clear correlation (``Phillips line''). 
Also shown are the LO and NLO predictions
for this correlation, obtained by varying the value of the three-body 
force at fixed cutoff. An equivalent explanation  for the 
existence of the Phillips line was first found in 
Ref. \cite{Efimov_phillips}.
Analogous results can be obtained in the EFT for the
hypertriton \cite{Hammer:2001ng}.

In the $3N$ system, the pionless EFT seems to converge
over a range of momenta that is large enough to
include the interesting physics associated with the
bound states.
For example, if the three-body force is fitted to
the scattering length, binding energy of triton is
$B_3= 8.08 + 0.23+ \ldots=(8.31 + \ldots)$ MeV \cite{stooges_triton, toappear},
to be compared to the experimental result 
$B_3^{exp}= 8.48$ MeV.
The success of these EFT calculations opens the way
for the study of low-energy reactions involving 
triton and $^3$He.

%%%%%%%%%%%%%%%%%%%%%%%%%%%%%%%%%%%%%%%%%%%%%%%%%%%%%%%%%%%%%%%%%%

\section{EFT WITH EXPLICIT PIONS}
\label{pionful}

As the typical momentum $Q$ approaches 
the pion mass $m_\pi$, it becomes increasingly difficult
to account for pion exchange as a short-range effect.
As we further increase momenta past $Q\sim m_\pi$, 
we have to include in the EFT an explicit pion field
and build up all its interactions allowed by symmetries.
Because {\it numerically} the mass difference between the delta isobar
and the nucleon,
$\delta m=m_\Delta - m_N$, is $\sim 2 m_\pi$,
convergence of the ``pionful'' EFT is optimized by
the concomitant inclusion of an explicit delta degree of freedom.
The delta can be included without additional problems, since
at these momenta the delta is, like the nucleon,
a non-relativistic object. 
All other degrees of freedom can be considered
heavy. Their effects are still subsumed in contact interactions, 
as they were in the pionless EFT. 
What we are doing is to
remove part of the pion (and possibly the isobar) contributions
from the contact interactions.
One hopes the new EFT works for momenta up to $M_{QCD}\sim 1$ GeV,
the mass scale of the heavier particles.

Adding explicit pions to the theory
will generate all sorts of non-analytic contributions
to nuclear amplitudes.
We want to devise a rationale for a controlled
expansion in the presence of pions.

\subsection{Chiral symmetry and Chiral Perturbation Theory}

Fortunately, pion interactions are not arbitrary.
Once explicit pion fields are considered, approximate chiral symmetry
imposes important restrictions on the way pions couple among themselves
and to other degrees of freedom.

In the limit where we neglect the masses $m_u$ and $m_d$ of the up and down
quarks (``chiral limit''), QCD has a chiral $SU(2)_L \times SU(2)_R$ symmetry.
Since this symmetry is not apparent in the QCD spectrum,
it is reasonable to assume that it is broken
spontaneously down to its diagonal subgroup, the
$SU(2)_{L+R}$ of isospin.
Goldstone's theorem \cite{inGSW} tells us that 
massless Goldstone bosons, naturally 
identified as pions,
are associated with the three broken generators, 
and their fields $\boldpi$ live on the 
``chiral circle'' (actually a three-sphere $S^{3}$)
that represents the set of possible vacua. 
We call the radius of this sphere
$f_{\pi}$; it is directly related to 
the chiral-symmetry-breaking scale, 
$\Lambda_{\chi SB} \sim 4\pi f_{\pi} \sim M_{QCD}$, but the
precise factor can only be obtained from the 
(so far elusive) mechanism of 
dynamical symmetry breaking in QCD.
It turns out that this diameter can be 
determined from pion decay,
and is called the pion decay constant, $f_{\pi}\simeq 92.6$ MeV.

The ``chiral circle'' can be parameterized in different ways but it is
 convenient to  choose fields for which
chiral symmetry is respected term-by-term in the effective Lagrangian. 
Since the interactions of
pions have to be invariant under chiral rotations,
it is possible to choose fields where an infinitesimal 
rotation is represented as
$\boldpi \rightarrow \boldpi +\epsilon$.
Interactions of these fields 
always involve derivatives on the sphere, that is,
derivatives together with certain non-linear self-interactions.

As long as quark masses are small enough, their only
effect is to change this picture slightly.
A common quark mass breaks chiral symmetry
explicitly down to the diagonal subgroup.
Points on the chiral circle are no longer degenerate in energy,
and a particular minimum is selected, in
a direction given by the quark-mass terms that we define as the
fourth direction.
The quark mass difference further breaks isospin explicitly. 
In the low-energy EFT, the effect of quark-mass terms
can be reproduced if
we construct all terms that break chiral symmetry in the same way.
These interactions can involve $\boldpi$ without derivatives,
but are always accompanied by powers of $m_u+m_d$
or $m_u-m_d$.
One example is a pion mass term, $m_\pi^2 \propto (m_u+m_d)$.
Likewise, 
electromagnetic and weak interactions can be constructed as well.

The well-studied theory of non-linear representations of symmetries
\cite{inCCWZ}
provides the tools to write down the appropriate interactions
between pions and other fields. 
The resulting chiral Lagrangian ${\cal L}$
has an infinite number of terms
that can be grouped 
according to the 
index $\Delta$:
\begin{equation}
{\cal L}= \sum_{\Delta=0}^{\infty} {\cal L}^{(\Delta)}, 
\; \; \; {\Delta}\equiv {d}+{f}/2-2,
\label{index}
\end{equation}
where $d$ is the number of 
derivatives, powers of $m_\pi$ and/or powers of $\delta m$,
and $f$ is the number of fermion fields.
Because of chiral symmetry, pion interactions stemming
from QCD
bring derivatives and/or powers of the pion mass.  
As a consequence, $\Delta\ge 0$.
The explicit form of ${\cal L}^{(\Delta)}$ for the lower values of
$\Delta$ can be found in the literature
\cite{bkm,vanKolck_review}.

As in the pionless EFT, the only hope of any predictive power rests on
finding some ordering of contributions to amplitudes.
This can be done for processes where all the external three-momenta
are $Q\sim m_\pi$.
Powers of $Q$ of any particular Feynman diagram can be counted
as done for the superficial degree of
divergence.
Each space derivative in an interaction produces a three-momentum
in a vertex and therefore counts as $Q$.
A complication in the counting of energies
stems from the presence of heavy particles such as the nucleon
together with light particles such as the pion.
In any loop, integration over the zeroth component of the 
four-momentum
will involve {\it two} types of poles, according to the scales appearing
in the propagators:
{\it (i)} standard poles at $\sim Q$ corresponding to
external three-momenta and to 
the mass of the pion;
and {\it (ii)} shallow poles at $\sim Q^2/2m_N$
corresponding to external nucleon energies.

Processes that involve at most one heavy particle
($A=0,1$) are the simplest
because the contour of integration can always be closed 
so as to avoid shallow poles. 
In this case all energies are $\sim Q$.
As a consequence,
each time derivative counts as $Q$ and 
four-momentum integration brings a factor $Q^{4}$.
A pion propagator is  $Q^{-2}$ and
a nucleon (or delta) propagator is $Q^{-1}$ from its
static term; kinetic terms are of relative ${\cal O}(Q/m_N)$ 
and thus can be treated as corrections.
With these ingredients one can write the 
contribution of any diagram to the amplitude as
\begin{equation}
T \propto Q^\nu {\cal F}(Q/\Lambda), 
\label{powercount}
\end{equation}
where $\Lambda$ is a renormalization scale, ${\cal F}$ is a calculable
function of LECs,
and $\nu$ is a counting index. 
For strong interactions \cite{inwei2},
\begin{equation}
\nu=4 -2C - A + 2L+{\sum _i}{\Delta _i}, 
\label{nu}
\end{equation}
where $C=1$ is the number of connected pieces,
$L$ is the number of loops, and the sum runs over all vertices.
In addition, electroweak interactions can be considered
through a simultaneous expansion in 
$\alpha=e^2/4\pi$ and $G_F f_\pi^2$.
Since $L\ge 0$ and $\Delta\ge 0$,
$\nu\ge 2-A\equiv \nu_{min}$.
Assuming that all LECs have ``natural'' size (given by 
powers of $M_{QCD}$
once the lower scales have been identified explicitly),
an expansion in  $Q/M_{QCD}$ results.
Its first two orders are equivalent
to the current algebra of the 60's, but at higher orders 
unitarity corrections can be accounted for systematically.
In the sector of $A=0,1$, the EFT is called Chiral Perturbation Theory
(ChPT).

In processes that involve more than one stable heavy particle ($A\ge 2$),
on the other hand,
a failure of perturbation theory can
lead to bound states \cite{inwei6}.
The shallow poles cannot be avoided:
they represent ``reducible'' intermediate states 
that, 
as in the pionless EFT, 
differ from initial states only
by nucleon kinetic energies, which are $\sim Q^2/m_N$.
This ${\cal O}(m_N/Q)$ infrared enhancement 
over intermediate states where energies are $\sim Q$
invalidates Eq. (\ref{nu})
for reducible diagrams.
Fermion lines, loops and derivatives then scale
with $Q$ as in Eq. (\ref{lowQrules}).
A pion propagator still counts as $Q^{-2}$,
but the pion can be taken in first approximation as
static, and it is sometimes referred to as a ``potential'' pion.
Contributions that come from
standard poles naively scale as in processes with no more than 
one heavy particle.
Pions there are non-static or ``radiative''.

The issue now is how to estimate the
size of pion contributions. 
One needs to find the importance of 
{\it (i)} pion exchange relative to short-range
interactions;
{\it (ii)} multi-pion exchange  relative to one-pion exchange (OPE).
Both issues are related, via renormalization,
to the size of the contact interactions.
How large are they in the pionful EFT?
What are the contributions that {\it must} be resummed in nuclear amplitudes? 

\subsection{The two-nucleon system}

As we have discussed, the $NN$ system
is characterized by scattering lengths $a_s$, $a_t$
that are much larger
than $1/M_{QCD}$.
In the pionless theory, this fine-tuning cannot be explained,
but it can be accommodated in the power counting 
by assigning to the contact interactions the scaling 
given in Eq. (\ref{lowQrules}).

A new scale 
appears naturally in the pionful theory.
The leading ($\Delta=0$) coupling of the pion to the nucleon 
is derivative with a coupling constant $g_A/f_\pi$,
where $g_A \simeq 1.26$ is a parameter not fixed by symmetry
but determined in $\beta$-decay.
The OPE contribution to the $NN$ amplitude is,
schematically, $g_A^2 Q^2/f_\pi^2 (Q^2+ m_\pi^2)$.
Because a reducible intermediate state
contributes $m_N Q/4\pi$,
relative to OPE, once-iterated OPE  
can be estimated to give a contribution
\begin{equation}
\frac{(g_A^2 Q^2/f_\pi^2 (Q^2+m_\pi^2))^2 (m_N Q/4\pi)}
     {g_A^2 Q^2/f_\pi^2 (Q^2+ m_\pi^2)}
\sim \frac{Q}{M_{NN}}
\label{TtoOPE}
\end{equation}
(as long as $Q \gaprox m_\pi$).
Here we introduced the scale
\begin{equation}
M_{NN} \equiv \frac{4\pi f_\pi^2}{g_A^2 m_N},
\end{equation}
which sets the relative strength
of multi-pion exchange.
Numerically (for $N_c=3$) $m_N\sim 4\pi f_\pi$ 
and $g_A \sim 1$, so
$M_{NN}\sim f_\pi$.
This naive dimensional analysis cannot, however, 
capture the numerical factors that actually set the relative
size of pion contributions.
A more accurate estimate requires concrete calculations.

\subsubsection{Perturbative pions}

For $Q \saprox M_{NN}$, 
iteration of OPE should be suppressed
with respect to OPE according to Eq. (\ref{TtoOPE}).
Moreover,
if we assume that the
leading short-range effects are $\sim 4\pi a/m_N$
(as in Eq. (\ref{lowQrules}) with $Q \to 1/a$),
then 
OPE is suppressed by ${\cal O}(1/a M_{NN})$ compared 
to the leading contact interaction.
With such estimates, 
if $M_{NN}$ is sufficiently large (compared to $1/a$ and $m_\pi$)
and $Q$ sufficiently small, pions may be treated perturbatively.
The suggestion that this can be profitably done
was made in Refs. \cite{lutz,ksw_1}.

A simple power counting, which became known as KSW counting,
follows by taking
$Q\sim 1/a \sim m_\pi < M_{NN}$, and counting powers
of the light scale $Q$.
This is a direct extension of the power counting in Eq. (\ref{lowQrules}).
In particular, the scaling  of the
contact operators is assumed to be the same as before 
with $\Lambda \to M_{NN}$, and thus their ordering is unchanged.
Because of chiral symmetry, each insertion of a pion exchange brings
a factor of $Q/M_{NN}$.
Electroweak interactions can be treated in much 
the same way as before.
One can show that renormalization can be carried out consistently
within this power counting \cite{ksw_1}.

However, for this power counting to be relevant to nuclear physics
$M_{NN}$ has to be sufficiently large.
If $M_{NN}$ is not larger than $m_\pi$, the domain of perturbative pions
is no larger than that of the simpler pionless theory.
The issue of the range of validity of 
the EFT with perturbative pions can only be settled by explicit calculation
of dimensionless factors and comparison with precise
observed quantities, this being done to sufficiently
high order so that a significant number of pion effects be tested. 

With this power counting,
the LO $NN$ amplitude
coincides with that in the pionless EFT,
see Fig. (\ref{bubbles}).
At this order there are contributions only in the
two $S$-wave channels from chirally-symmetric, non-derivative
contact interactions (the $C_0$ terms).
Subleading terms, of ${\cal O}(Q/M_{NN})$ relative to leading,
are constructed in a direct
extension of subleading terms of the pionless EFT.
Besides two two-derivative contact interactions ($C_2$ terms),
we also insert OPE and 
two non-derivative contact interactions that break chiral symmetry explicitly
($m_\pi^2 C_{2}^{qm}$ terms).
Both $C_2$ and $C_{2}^{qm}$ contact interactions only contribute to
$S$ waves. The tensor operator from OPE, on the other hand,
introduces mixing between $^3S_1$ and $^3D_1$ waves.
To this order, all but the $S$-wave phases
are
{\it predicted}
in terms of pion parameters.
A calculation of the $NN$ system to NLO was carried out
in Refs. \cite{ksw_1,cohenhansen1}.
When the results were compared to the Nijmegen
PSA \cite{nijmanal},
it was suggested that $M_{NN}\simeq 300$ MeV \cite{ksw_1}.
The relative size of OPE and contact interactions 
was extensively discussed 
\cite{variouspert,mehenstewart,furnsteele}, 
but the issue is clouded by the details of fitting
procedures. 

\begin{figure}[!t]
\vskip 0.15in
\centerline{\epsfxsize=8.truecm \epsfbox{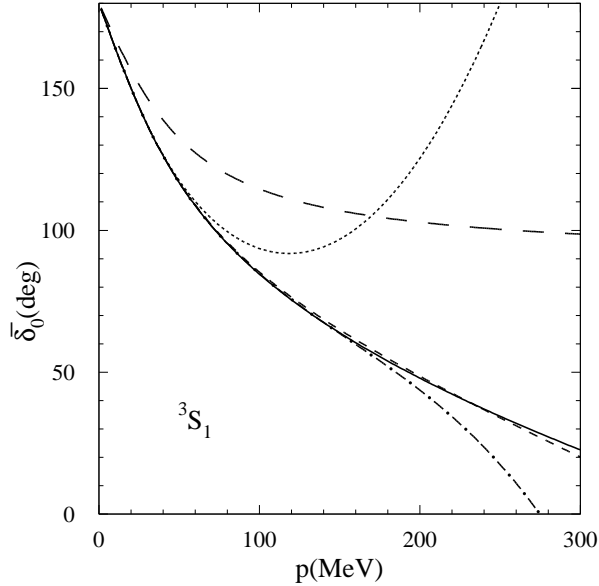}  } %\vspace{.5cm}
\noindent
\caption{
$^3S_1$ $NN$ phase shift in the EFT with perturbative
pions, as function of the center-of-mass momentum.
The long-dashed, short-dashed, and dotted 
lines are, respectively,  the EFT results in LO, NLO, and 
N$^2$LO. 
The dash-dotted line is the N$^2$LO result
with a higher-order contact interaction added.
The solid line is the Nijmegen PSA.
{}From Ref. \cite{fms}, courtesy of T. Mehen.
}
\label{fig:fms3S1}
\vskip .2in
\end{figure}

At N$^2$LO we encounter new pion exchanges:
both non-static (or radiative) OPE and once-iterated OPE.
A calculation at this order was carried out in all $S$, $P$, and $D$
waves \cite{fms,rupakshoresh}.
Results were found to depend on the channel. 
While in singlet channels, like $^1S_0$, there seems to be
good agreement with the Nijmegen PSA,
in triplet channels, such as $^3S_1$, $^3D_1$, and $^3P_{0,2}$,
the N$^2$LO corrections are big at momenta $\sim 100$ MeV
and lead to large disagreement.
One example is shown in Fig. (\ref{fig:fms3S1}).
The effects of perturbative pions (and of delta isobars) is milder in the
higher partial waves \cite{twokaiserpapers}.

The problem can be traced to the iteration of the tensor part of OPE.
These results suggest that pions, or more explicitly, 
the Yukawa part of 
potential and radiation pions, when treated perturbatively give rise to a
converging expansion for the $\si$ scattering amplitude
up to fairly large momenta.
However, 
OPE in the $\siii-\diii$ coupled channels is
{\it not} perturbatively convergent
for momenta around 100 MeV, because the tensor force,
which survives in the chiral limit, is too large. 
This, in turn, suggests that the naive estimate $M_{NN}\sim f_\pi$
is not entirely unreasonable.

\subsubsection{Renormalization of the pion ladder and power counting} 
\label{piladder}

If indeed $M_{NN} \ll M_{QCD}$, we might
hope to improve on the expansion of the previous section 
by a controlled resummation
of terms that go as $Q/M_{NN}$.
If $M_{NN} \sim 100$ MeV, the lack of other known
particle thresholds there suggests that the
resummation could involve {\it only} pions.
Indeed, Weinberg, who first attempted the use of EFT in 
the derivation of nuclear forces \cite{inwei6},
had already suggested a power counting
in which pions appear in leading order, and should therefore
be iterated.
(Some elements of this power counting had been
anticipated
by Friar \cite{friarcoon}).

Weinberg's original proposal \cite{inwei6} for an EFT describing
multi-nucleon systems was to split the full amplitudes
into reducible and irreducible parts.
Irreducible diagrams,
in which typical energies
superficially resemble those in ordinary ChPT,
should satisfy the power counting, Eq. (\ref{nu}).
We call the sum of irreducible diagrams the potential $V$. 
Note that the potential, being a set of subgraphs,
can be defined in alternative ways.
All definitions that exclude the infrared-enhanced
contributions but differ by a smaller amount are equally good,
as long as no double-counting or omissions are made.
A field redefinition might change the potential but not
the full amplitude. 
The important point is that the only
scale appearing explicitly in the potential is $Q$, 
so that the power counting 
proceeds as in the case of diagrams with at most one heavy particle.
Reducible diagrams can be obtained by sewing together irreducible diagrams
via intermediate states 
that contains the propagation 
of only the initial particles.
The full amplitude  $T$ for an $A$-nucleon system
is thus a sum of the potential and its 
iterations; schematically, 
\begin{equation}
T = V+ VG_0V+ VG_0VG_0V+\ldots = V+ VG_0T,
\label{T}
\end{equation}
\noindent
where $G_0$ is the $A$-nucleon free (Schr\"odinger) Green's function.
This is just the Lippmann-Schwinger equation,
which is {\it formally} equivalent to the Schr\"{o}dinger equation
with the potential $V$, from which wave-functions $|\psi \rangle$ 
can be constructed.

Unfortunately, just counting powers of $Q$ is not in itself sufficient 
for an ordering of interactions. 
We need to find which other scales accompany $Q$. 
Given that we do not yet possess a full solution of QCD, it is
obvious that some assumptions have to be made about the LECs.
Because OPE has short-range components,
it is natural to assume that they set the scale for 
the short-range interactions.
In this case, since OPE is ${\cal O}(g_A^2/f_\pi^2)$ for $Q\sim m_\pi$,
we expect the leading two-nucleon contact interactions
to be ${\cal O}(4\pi/m_N M_{NN})$.
This assumption naturally explains why nuclear bound states
are much shallower than naively expected \cite{inwei6,ciOLvK}.
The series in Eq. (\ref{T})
is roughly
\begin{equation}
T\sim \frac{4\pi}{m_N M_{NN}} 
   \left[ 1+ {\cal O}\left(\frac{Q}{M_{NN}}\right)
                            + \ldots \right],
\end{equation}
which requires resummation and 
exhibits a (real or virtual) bound state
at $Q\sim M_{NN}$.
In other words, the natural scales for the $NN$ scattering length 
and for the binding energy of a nucleus are
\begin{equation}
|a|\sim \frac{1}{M_{NN}}, \ \ B\sim \frac{M_{NN}^2}{m_N}, 
\label{natbinding}
\end{equation}
respectively.
If we use again that $m_N \sim 4\pi f_\pi$ and $g_A\sim 1$,
then $B\sim f_\pi/4\pi \sim 10$ MeV.
So we find that it is not $M_{QCD}$ by itself that sets
the scale for binding energies, but a certain ratio
of powers of $f_\pi$ and $m_N$. 
Now, as we have seen, this is not the whole story.
It remains mysterious why the $NN$ (real and virtual)
bound states are even shallower, or equivalently,
why $|a|> 1/M_{NN}$, by factors of a few.
This still has to be accommodated by fine-tuning the contact interactions. 

Because they serve as counterterms
to pion loops in the potential, which are expected to be suppressed
(as in ChPT) by powers of $Q/4\pi f_\pi$,
Weinberg implicitly assumed that 
LECs related to more derivatives and powers of the pion mass
 contain
inverse powers of $M_{QCD}$.
That is,
a (renormalized) contact operator with index $\Delta$ would
scale as 
\begin{equation}
C_{\Delta}\sim \frac{4\pi}{m_N M_{NN} M_{QCD}^{\Delta}}, 
\label{Wscaling}
\end{equation}
as in naive dimensional analysis.

With this assumption and disregarding the fine-tuning,
a simple power counting results from
taking $Q\sim M_{NN}\sim 1/a \sim m_\pi$.
The potential obeys Eqs. (\ref{powercount},\ref{nu}).
The leading potential consists of   
no-derivative, chirally-symmetric contact interactions plus static OPE,
\begin{equation}
 V^{(0)} =  C_{0}^{(S)}+C_{0}^{(T)}\vec{\sigma}_{1}\cdot\vec{\sigma}_{2}
     -\left(\frac{g_{A}}{f_{\pi}}\right)^{2}\boldt_{1}\cdot\boldt_{2}
                \frac{\vec{\sigma}_{1}\cdot\vec{q}\,\vec{\sigma}_{2}\cdot
                \vec{q}}{\vec{q}\,^{2}+m_{\pi}^{2}}, 
\label{leadingpot}
\end{equation}
where $\vec{\sigma}_i$ ($2\boldt_{i}$) are the Pauli matrices
in spin (isospin) space and $\vec{q}$ is the momentum transferred.
All contributions to nuclear forces other than Eq. (\ref{leadingpot})
would come as corrections in powers of $Q/M_{QCD}$,
starting at $(Q/M_{QCD})^2$.
The structure of the potential rapidly becomes more complex
with increasing order \cite{ciOvK}.
The leading potential has to be resummed,
while corrections can be treated in perturbation theory.
If the corrections are truly small, resumming them
should cause no major harm.
This method requires numerical solution of the
Schr\"odinger equation and is similar in spirit to the 
traditional 
potential-model approach.
As we will see below,
Weinberg's power counting has been extensively and successfully 
developed during the past
decade to study processes involving few-nucleon systems. 

However, there is a subtlety not present in ChPT.
Loops in reducible diagrams probe, as other loops do, high energies
(when nucleons are far off shell).
As in the pionless EFT, the potential 
does not vanish at
large momenta: it is singular.
In addition to the delta function and its derivatives
already present in the pionless theory, 
pion exchange generates potentials
that behave as $1/r^n$ with $n\ge 3$, as the radial coordinate
$r \rightarrow 0$.
The large-momentum or short-distance behavior
is the same as in the chiral limit $m_\pi^2 \rightarrow 0$.
Already in leading order the tensor force
goes as $1/r^3$ in the chiral limit.
As a consequence of ultraviolet divergences generated 
by the iteration of the potential,
the infrared enhancement of $M_{QCD}/M_{NN}$ might contaminate
LECs, possibly invalidating
Eq. (\ref{Wscaling}).

The crucial issue is whether, at any given order, all divergences
generated by iteration can be absorbed
in the parameters of the potential truncated at that order.
There is some indication that 
equally good fits can be achieved 
in leading orders with various cutoffs
\cite{ciOLvK,nonppions,taesunpions,furnsteele,FTT,epelfit},
as required of a sensible EFT,
but the numerical nature of the results makes a definite
answer difficult.

Unfortunately, there seem to be formal inconsistencies in Weinberg's
counting \cite{ksw_1,mehenstewart}.
Divergences that arise in the iteration
of leading-order interactions
apparently
cannot be absorbed by the leading-order operators themselves. 
Two examples are two-loop diagrams where:
{\it i)} OPE happens between
two contact interactions, having a divergence proportional to
the square of the pion mass \cite{ksw_1};
{\it ii)} OPE is iterated three times, having a divergence proportional to
the square of the momentum \cite{mehenstewart}.
Although these two particular cases could be resolved
by the promotion to leading order of two counterterms,
it is likely that a similar problem would show up
at higher orders in the expansion.
The correspondence between
divergences and counterterms appears to be lost,
a fundamental problem with the chiral expansion {\it and} the
momentum expansion resulting from Weinberg power counting.

This argument is not decisive, though.
It has been known that in the context of the
Schr\"odinger equation  perturbative arguments
are not in general reliable for singular
potentials \cite{singpotrev}.
The perturbative expansion might have a cut starting at $g_A^2/f_\pi^2=0$;
insistence on a $g_A^2/f_\pi^2$ expansion
would then reflect itself on different orders offering
correlated contributions to counterterms,
each bringing powers of $M_{QCD}/M_{NN}$
yet resulting in a much better behaved sum.
How can the resummation be consistent?
There is a mapping between the singular two-body $1/r^2$ central potential
and the three-body problem with
short-range interactions.
In Sect. \ref{tritonsect} we saw that for the latter,
the renormalization
of the nonperturbative equation is very different
from the renormalization of individual terms in the
associated perturbative series \cite{stooges_3bosons}.
In particular,
the relevant counterterm exhibits a limit-cycle behavior.

\begin{figure}[!t]
\centerline{{\epsfxsize=3.6in \epsfbox{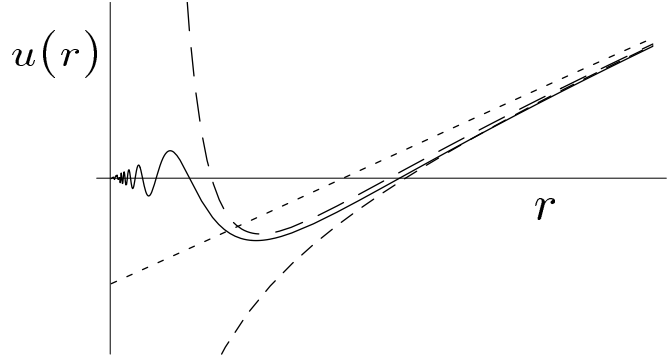}} }
\noindent
\caption{Zero-energy radial wave-functions of the $1/r^4$ 
potential:
exact (solid line) and 
perturbation theory to
LO (small dashes), 
NLO (medium dashes), and  
N$^2$LO (large dashes). 
{}From Ref. \cite{kiddies}.}
\label{perturb}
\vskip .2in
\end{figure}

It turns out that 
the correct renormalization of
singular potentials is, indeed, in general intrinsically nonperturbative
\cite{singpotrev,kiddies}.
In contrast to regular potentials, both solutions of the Schr\"odinger equation
for an atractive $1/r^n$ central potential are equally acceptable:
the radial wave-function $u(r)$, whose large-distance behavior
determines low-energy observables,
oscillates rapidly as $r$ decreases.
Perturbative approximations to the wave-function fail
at distances comparable to the intrinsic scale $r_0$ present
in the potential, as illustrated in Fig. (\ref{perturb}).
The problem can be rendered 
essentially cutoff independent with a single counterterm
associated with the short-distance physics.
For example, short-distance physics can be represented by a square well
potential of radius $R\ll r_0$ whose depth $V_0=V_0(R)$ 
can be adjusted so that physics at $r\gaprox r_0$ be
independent of $R$ \cite{kiddies}.
(The advantage of this coordinate-space
regulator is that one can do an analytic matching
of the outer and inner solutions of the Schr\"odinger equation, 
thus finding the desired $V_0(R)$.
For another technique to deal with a singular potential,
see, for example, Ref. \cite{Camblong:2001zi}.)
While for a repulsive potential
there exist only fixed points \cite{Barford:2001sx},
the situation in the attractive case 
is similar to the three-body system.

These results were extended 
to the $NN$ potential in the $^1S_0$ and $\siii-\diii$ channels
in Ref. \cite{towards},
where the long-range part of the potential
was taken as that of OPE for $r>R$.
The asymptotic behavior is that of the chiral limit,
where the relevant scale is $r_0 \sim 1/M_{NN}$.
The depths of the short-range part of the potential
can be different in the singlet and triplet channels,
as there are two parameters ($C_0^{(S)}$ and $C_0^{(T)}$)
in Eq. (\ref{leadingpot}). 

\begin{figure}[!t]
\vskip 0.15in
\centerline{{\epsfxsize=2.5in \epsfbox{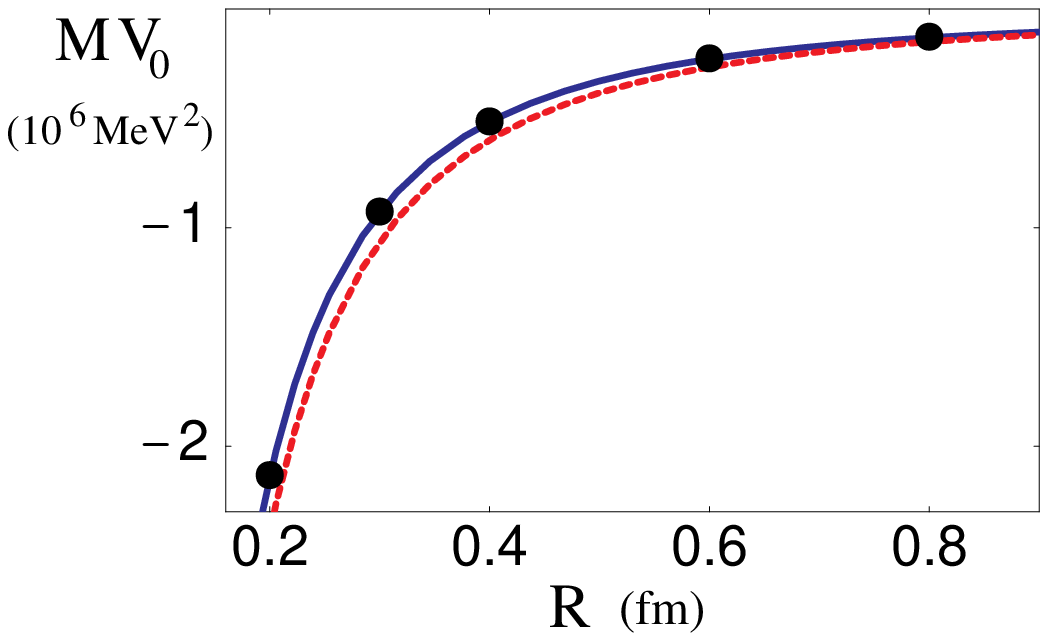}}\hskip0.1in{\epsfxsize=2.5in\epsfbox{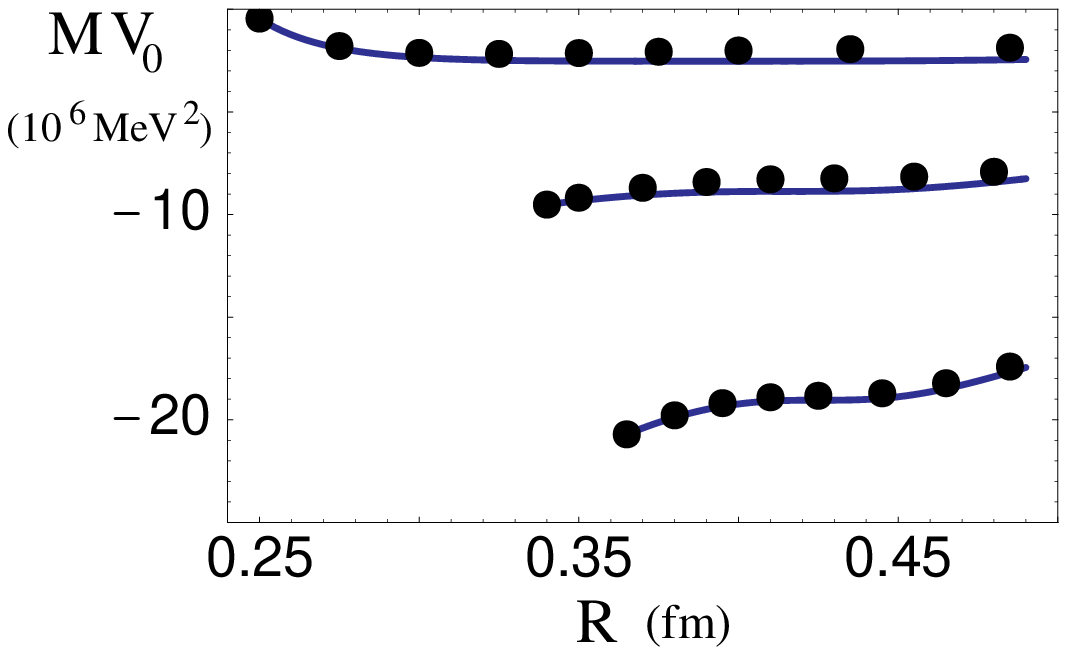}}} 
%\vskip 0.05in
\noindent
\caption{
Running of $m_N V_0$ as a 
function of the cutoff $R$. 
Singlet channel (left panel):
Eq. (\ref{eq:logderivativeVsingletapp}) for $n=1$ (blue solid line),
and same but with $R^{-1}$ part neglected (red dotted line).
Triplet channel (right panel):
numerical solution of an analytic matching equation (solid lines).
Dots are extracted directly from a numerical solution of the 
Schr\"odinger equation in the respective channel.
{}From Ref. \cite{towards}.}
\label{fig:vrunning}
\vskip .2in
\end{figure}

In the $^1S_0$ channel the calculation is straightforward.
The pion potential is simply a Yukawa, and the explicit solution
is 
\beq
V_0(R;n)=-(2n+1)^2{{\pi^2}\over {4 m_N R^2}} -
{{g_A^2 m_\pi^2}\over{8 \pi f_\pi^2 R}} \log \left({R\over{R_*}}\right)
+{\cal O}(R^0),
\label{eq:logderivativeVsingletapp}
\eeq
where $R_*$ is an intrinsic length scale to be determined numerically
from a fit to low-energy data,
and $n$ labels the branch of a cotangent. In 
the left panel of Fig. (\ref{fig:vrunning})
the $R$ dependence of $V_0$, as  given by
Eq.~(\ref{eq:logderivativeVsingletapp}), 
is compared to the numerical solution of 
the Schr\"odinger equation with the observed
singlet scattering length. 
The presence of a
multi-branch structure is related to the accumulation of bound states
inside the square well. Of course the presence of unphysical bound
states is innocuous as long as the binding energies of such states 
are near the cutoff of the EFT.  
One sees that the formal problem with the chiral expansion in
Weinberg's counting survives the resummation.
While the first cutoff-dependent term 
in Eq. (\ref{eq:logderivativeVsingletapp})
can be represented by a chiral-symmetric contact interaction,
the second would require a chiral-breaking one.
In momentum-space notation where the cutoff is denoted by $\Lambda$,
\beq
C_0(\Lambda)+\mpis C_{2}^{(qm)}(\Lambda )=
\frac{4\pi}{6 m_N \Lambda} \left[
(2n+1)^2 {{\pi^2}\over {2}} +
{{m_\pi^2}\over{M_{NN} \Lambda}}\log\left({{\Lambda_*}\over\Lambda}
\right) \right].
\label{eq:logderivativeVsingletappczero}
\eeq 
Although the logarithmic divergence is suppressed
by a power of $\Lambda$ compared to the first term, 
it is a true divergence in physical
quantities that must be renormalized at leading order in 
Weinberg power counting. 
The $C_{2}^{(qm)}$ operator, which is formally subleading, 
must be promoted to leading order if the full
OPE is iterated, in agreement with the perturbative
argument of Ref. \cite{ksw_1}.
On the other hand, 
the $C_{2}^{(qm)}$ contribution is numerically small, as demonstrated by the
dotted curve in the left panel of Fig. (\ref{fig:vrunning}) 
which neglects the $\Lambda^{-2}$
corrections to the running.  This smallness explains why Weinberg's 
power counting has
been found to work well in this channel over a moderate range of 
cutoffs \cite{ciOLvK,nonppions,taesunpions,FTT,furnsteele,epelfit}.

A possible conclusion is that OPE
and the ${\cal O}\left(m_q\right)$ LECs
contribute to any amplitude at the same order in the expansion, and this is
what leads to KSW power counting.  However, a more general conclusion to draw
is that the difference between the OPE contribution for $m_q\ne 0$ and the OPE
contribution in the chiral limit must occur at the same order as the 
${\cal O}\left( m_q\right)$ counterterms.  In many cases these two conclusions
yield identical amplitudes, however, in the $\siii-\diii$ channel they do not.

In the $\siii-\diii$ channel,
in addition to the long-distance Yukawa interaction and the 
contact interaction,
there is a strong tensor component of OPE
that couples $S$ and $D$ waves. 
At distances $r\ll 1/\mpi$ the central potential is negligible,
while 
in the region $R<r\ll 1/M_{NN}$ we can
neglect the angular-momentum barrier.  Moreover, for 
$\sqrt{m_N E}\ll M_{NN}$ the total energy $E$ can be
treated as a perturbation.  In this short-distance limit, 
we can keep only the chiral limit of the
tensor force, and
the Schr\"odinger equation can be diagonalized and
solved exactly.  In
the diagonal basis the Schr\"odinger equation decouples into an attractive
singular potential and a repulsive potential. The solution for the attractive
singular potential  is a linear combination of Bessel
functions \cite{Sprung,kiddies}, and the wave-function at this order is
\beq
u(r)= A\ r^{3/4}\cos\left(\sqrt{{{6}\over {M_{NN} r}}}
+\phi_0 \right),
\eeq
where $A$ is a dimensionful normalization constant
and $\phi_0$ is the asymptotic phase which
determines the triplet scattering length.
This solution oscillates ever faster as it approaches the
origin, just as in Fig. (\ref{perturb}). 
As before, the issue is whether a $V_0(R)$ can be found
in such a way that the asymptotic phase $\phi_0$ is made $R$ independent.  
Matching logarithmic derivatives of the interior
square-well and exterior attractive solutions at
$r=R$ yields an equation whose solution is 
shown in the right panel of Fig. (\ref{fig:vrunning}).
The renormalization-group flow is multibranched and non-analytic 
in $g_A^2/f_\pi^2$.

\begin{figure}[!t]
\centerline{{\epsfxsize=3.5in \epsfbox{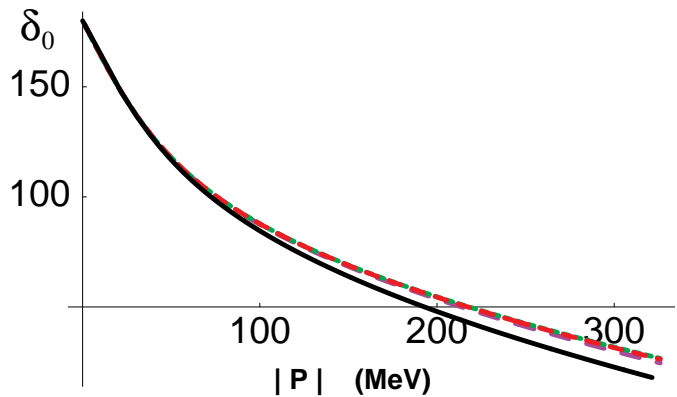}}} 
%\vskip 0.15in
\noindent
\caption{ 
$^3S_1$ $NN$ phase shift in leading order in the EFT with non-perturbative
pions, as function of the center-of-mass momentum.
The (purple) long-dashed, (red) medium-dashed, and (red) short-dashed   
lines have, respectively,  
$R=0.45~{\rm fm}$ ($\Lambda=438~{\rm MeV}$),
$R=0.21~{\rm fm}$ ($\Lambda=938~{\rm MeV}$),
and
$R=0.10~{\rm fm}$ ($\Lambda=1970~{\rm MeV}$).
The (black) solid line is the Nijmegen PSA.
{}From Ref. \cite{towards}.
}
\label{fig:delzeroW}
\vskip .2in
\end{figure}

The leading-order $^3S_1$ phase shift
is clearly cutoff independent as shown in
Fig. (\ref{fig:delzeroW}).
The situation is similar for the $^3D_1$ phase shift,
while the mixing parameter $\varepsilon_1$ exhibits some $R$
dependence. However, an error plot of $\varepsilon_1$ suggests
that the $R$
dependence and the deviations from the Nijmegen PSA \cite{nijmanal} are higher
order in the momentum expansion. 
These results \cite{towards} are in agreement with the numerical
analyses of Refs. \cite{Sprung,FTT}.
In these channels, contrary
to the perturbative argument \cite{mehenstewart}, 
Weinberg's power counting does not seem
to be formally inconsistent. 

It was conjectured
in Ref. \cite{towards} that 
a formally consistent expansion in the pionful EFT
is an expansion around the chiral limit.
This expansion is equivalent to KSW power
counting in the $\si$ channel and equivalent to Weinberg 
power counting in the
$\siii-\diii$ coupled channels, {\it i.e.} 
it selects only the desirable features of
both power countings.  
The leading-order potential $V({r}; 0)$, to be treated exactly,
consists of the chirally-symmetric component 
of OPE and non-derivative contact interactions.
Deviations from the chiral limit 
$V({r}; m_\pi)-V({r}; 0)$ can be treated perturbatively in all channels, 
and in fact,
such a perturbative expansion is required in the $\si$ channel but not in the
$\siii - \diii$ channel.
Evidence was presented 
that this expansion converges, even though convergence is slow
due not to
the long-range pion physics itself, but
to the fine-tuned short-distance physics 
(as argued previously \cite{Kaplan:1999qa}).

Although existing obstacles to a 
derivative and pion-mass expansion were removed,
higher orders must be studied before
the issue can be considered settled.
For example, in Ref. \cite{scaldeferri} an incomplete sub-leading
calculation with non-perturbative pions
has found limits in fitting the effective range.
Note also that
alternative views of the renormalization of the pion ladder exist
\cite{gegeliasoto}. 
Finally,
there is an interesting suggestion of expanding the $NN$ amplitude
in the energy region where the $S$-wave phase shifts vanish \cite{newlutz}.
The connection between this expansion and the low-energy expansion
described here
has not been fully analyzed.

\subsubsection{Potentials and fits to $NN$ data} 
\label{potfits}

The picture that emerges from the previous section
is close, conceptually and numerically, to Weinberg's original proposal.
To be formally consistent, we should expand
in the pion mass. Yet,
if we resum the effects of the pion mass in pion exchange,
which can be done with higher-order error,
the leading-order potential becomes the same as in Weinberg's power counting.
Corrections to the leading potential do not need to be iterated to all
orders. Yet, as has been shown explicitly in the pionless EFT
\cite{vanKolck_shortrange}, they {\it can}
be iterated with small error, as long as one uses a regularization with a
cutoff $\Lambda \sim M_{QCD}$.
Clearly, a potential which is correct up to a certain
order ensures that the amplitude is correct to the same
order.

Much work has been done in developing an EFT potential
based on Weinberg's power counting.
Traditionally, potential models have been 
plagued by problems of principle, such as 
the form of meson-nucleon interactions
(for example pseudoscalar {\it vs.} pseudovector pion coupling),
renormalization issues, absence of a small expansion parameter, 
{\it etc.} 
Because the EFT potential includes explicitly the exchange of only pions,
all these problems can be resolved.
For any given choice of pion field, the form of interactions
is fixed by the pattern of chiral symmetry breaking.  
Renormalization can be performed because {\it all} interactions
consistent with symmetries are included.
And the power counting Eq. (\ref{nu}) 
for the EFT potential implies that diagrams with an increasing
number of loops $L$ ---and, in particular, with increasing number
of exchanged pions--- should be progressively less important.
The EFT potential can be thought of as a low-energy
approximation to standard potential models,
although this can only be taken in an average sense.

\begin{figure}[!t]
\centerline{\epsfysize=8cm \epsfbox{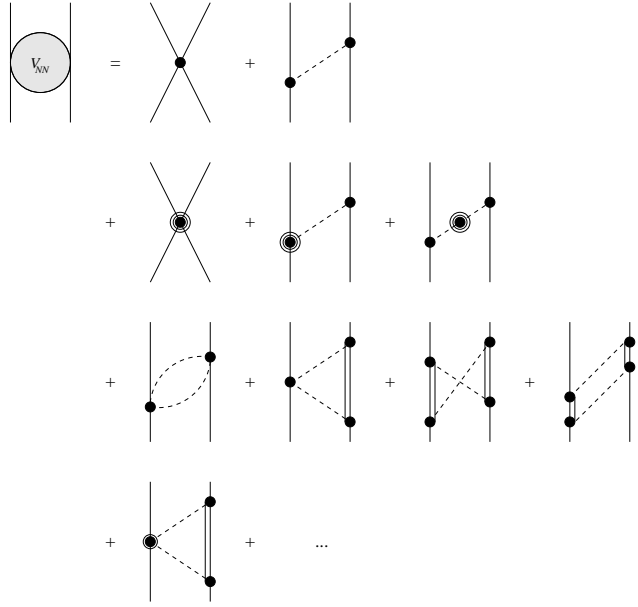}}  
%\vspace{0.5cm}
\caption{Some time-ordered diagrams contributing to the $NN$ potential
in the pionful EFT.
(Double) solid lines represent nucleons (and/or deltas), 
dashed lines pions, a heavy dot an interaction in ${\cal L}^{(0)}$,
a dot within a circle an interaction in ${\cal L}^{(1)}$,
and 
a dot within two circles an interaction in ${\cal L}^{(2)}$.
The first line corresponds to $\nu=\nu_{min}$,
second and third lines to $\nu=\nu_{min}+2$, 
fourth line to $\nu=\nu_{min}+3$, and 
``$\ldots$'' denote $\nu\geq \nu_{min}+4$. 
All orderings with at least one pion
(or delta) in intermediate states are included.
Not shown are 
diagrams contributing only to renormalization of parameters.
\label{F:vkolck:V2N}}
\end{figure}

In leading order, $\nu=\nu_{min}=0$, 
the $NN$ potential is, as we saw in the previous
section, simply static OPE
and momentum-independent contact terms. 
This is obviously a very crude approximation to the $NN$ potential:
it is known that the nuclear force 
has other sizable components, like a spin-orbit force, a strong short-range
repulsion and an intermediate range attraction.
These are all generated in the next orders:
$\nu=\nu_{min}+1$ corrections vanish due to parity and 
time-reversal invariance, but
$\nu=\nu_{min}+2$ corrections are several.
First, there are short-range corrections;
they come from one-loop pion dressing of the lowest-order contact
interactions, and from
four-nucleon contact interactions with two derivatives
or two powers of the pion mass.
It is easy to show that the result of loop diagrams amount
to a simple shift of the contact parameters.
Second, there are corrections to OPE; these come from vertex dressing
and from recoil upon pion emission.
Third, there are two-pion exchange (TPE) diagrams built out 
of lowest-order $\pi NN$ (and $\pi N\Delta$) interaction.
At $\nu=\nu_{min}+3$ a few more TPE diagrams appear,
which involve the $\pi\pi NN$ seagull vertices from 
the $\Delta=1$ Lagrangian.
To this order there are also small some relativistic corrections. 
At $\nu=\nu_{min}+4$ a host of two-loop diagrams and 
new contact interactions emerge, and so on. 
Some diagrams are shown in Fig. (\ref{F:vkolck:V2N}).

A calculation of all contributions to the
$NN$ potential up to $\nu=\nu_{min}+3$ was carried out
in Refs. \cite{ciOvK,ciOLvK}
using time-ordered perturbation theory. 
This EFT potential is energy dependent, but
equivalent
potentials can be obtained through unitary transformations.
An energy-independent potential is more convenient
in many situations, and the corresponding version was derived
in Ref. \cite{jimandsid}.

The potential to this order
has all the spin-isospin structure of phenomenological
models, but its profile is determined by explicit degrees of freedom,
symmetries, and power counting. 
The power counting suggests a hierarchy of short-range effects:
$S$ waves should depend strongly on the short-range parameters $C_0^{(S,T)}$;
contact interactions affect $P$-wave phase shifts 
only in subleading order,
so their effect should be smaller and approximately linear;
and $D$ and higher waves are directly affected by contact interactions at
higher orders, being thus essentially determined by pion exchange.
While phenomenological potentials such as that in Ref. \cite{argonne}
have similar short- and long-range structure,
it is on TPE that
chiral symmetry is particular influential. 
TPE here includes a particular form of terms previously
considered \cite{ciBwciSHciSO}, plus a few new terms.
These new terms come from nucleon-structure properties,
such as the axial polarizability and the sigma term,
and they provide a correlation between the exchanged pions
that is important in the isoscalar central force.
(Even though graphs where pions interact in flight appear
only in next order and should thus be relatively small.)
Not surprisingly, in the chiral limit these components of the potential
behave at large separations as van der Waals forces.
The components of the TPE potential were studied
in detail in Ref. \cite{twokaiserpapers}. 
In particular, it is shown explicitly that 
{\it i)} relativistic corrections are mostly small;
{\it ii)} both isoscalar central and spin-orbit
potentials are numerically similar to $\sigma$ and $\omega$ exchange
in models;
{\it iii)} the OPE isovector tensor potential is reduced by the
TPE contribution.
More recently, (the tail of the energy-independent version of) the EFT
TPE potential in the limit of a heavy
delta was substituted for 
(the tail of) the one-boson exchange in a 
Nijmegen phase-shift reanalysis of $pp$ data below 350 MeV 
\cite{robetal}.
A drop in 
$\chi^2$ was observed. When $\pi\pi NN$ seagull LECs
are fitted to the $NN$ data, they come out close to values
extracted from $\pi N$ scattering.
This 
confirms unequivocally the validity of chiral TPE.
A new full Nijmegen PSA is in the works, in which
chiral TPE is used in the long-range potential \cite{robetal}.

\begin{figure}[!t]
\centerline{{\epsfxsize=3.5in \epsfbox{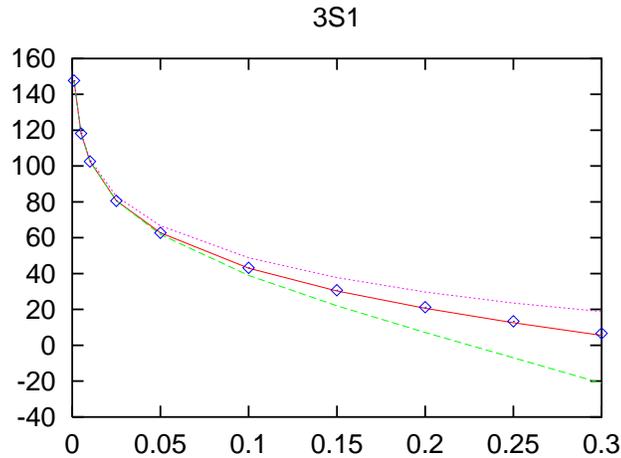}}} 
%\vskip 0.15in
\noindent
\caption{ 
$^3S_1$ $NN$ phase shift (in degrees) in the EFT with Weinberg counting, 
as function of the lab energy (in GeV).
The (purple) dotted, (green) dashed and (red) solid lines 
are the results at orders
$\nu=\nu_{min}$,
$\nu=\nu_{min}+2$,
and
$\nu=\nu_{min}+3$, 
respectively.
The squares are the Nijmegen PSA.
{}From Ref. \cite{epelfit}, courtesy of U.-G. Mei{\ss}ner.
}
\label{fig:delzeroMei}
\vskip .2in
\end{figure}

Fits to $NN$ phase shifts were done to this order
with \cite{ciOLvK}
and without \cite{epelfit}
explicit delta degrees of freedom.
For numerical convenience, smooth cutoffs are used to regulate the loops
generated by the Schr\"odinger equation.
For each cutoff value a set of nine independent 
(bare) parameters stemming from contact
interactions were fitted to phase shifts
below 100 MeV laboratory energies.
However,
the fits differed in the details.
In Ref. \cite{ciOLvK}, 
a Gaussian cutoff was used in all loops, 
and calculations performed with the cutoff parameter
$\Lambda$ taking values 500, 780 and 1000 MeV. 
In contrast, Ref. \cite{epelfit} used a mixed regularization:
while the cutoff parameter in the smooth regulator 
was varied between 500 and 600 MeV,
loops in the potential were treated by a subtraction procedure
equivalent to DR.
Although
OPE and TPE diagrams are completely determined by
LECs accessible in $\pi N$ reactions,
most of these LECs were not known at the time of Ref. \cite{ciOLvK},
and were also searched in the fit.
In Ref. \cite{epelfit}, these parameters were taken
from a (DR) fit to $\pi N$ scattering,
which allowed for a simpler, partial-wave-by-partial-wave fit
of $NN$ phase shifts.
Reasonable agreement with existing PSAs and deuteron
properties was found,
especially in Ref. \cite{epelfit}.
As an example,
the $^3S_1$ phase shift at various orders
is shown in Fig. (\ref{fig:delzeroMei})
and compared to the Nijmegen PSA \cite{nijmanal}.
(Cf. Figs. (\ref{3S1_nopions}, \ref{fig:fms3S1}, \ref{fig:delzeroW}).)
The contributions
from the short-range parameters in this fit
turn out to be comparable to those from heavier resonances
in phenomenological models \cite{ulfressat}.

Although the fits to $\nu=\nu_{min}+3$ are good,
they are inferior to the so-called ``realistic''
potentials that use $40-50$ parameters
to fit data up to 300 MeV with a $\chi^2$ near 1.
First steps are being made to extend the EFT potential
to  $\nu=\nu_{min}+4$.
In Ref. \cite{mach1}, contact interactions of this order were added
to the complete deltaless $\nu=\nu_{min}+3$ potential,
bringing the number of adjustable parameters to the
level of realistic potentials and the Nijmegen PSA.
The resulting fit is of quality comparable to realistic potentials.
(Indeed, the difference among realistic potentials
arise from high-momentum intermediate states that probe scales
which cannot be uniquely fixed by low-energy fits
\cite{Bogner:2001gq}.)
In order to achieve a complete $\nu=\nu_{min}+4$ potential,
two-loop diagrams and corrections to one-loop diagrams need to be included.
Classes of these diagrams that are invariant under
pion-field redefinitions have been calculated in Ref. \cite{kaiserloops}.
They are compared to peripheral partial waves in Ref. \cite{mach2},
where they are found to be relatively small.
We expect to soon witness a complete $\nu=\nu_{min}+4$
fit of high accuracy.

\subsubsection{Isospin violation}

The mass difference between $u$ and $d$ quarks breaks isospin symmetry
explicitly. 
Indication from the meson masses is that
the ratio $\varepsilon \equiv (m_u-m_d)/(m_u+m_d) \sim 1/3$. Naively,
this suggests that isospin might not be a much better symmetry than
the rest of the chiral group. On the other hand, a cursory
look at hadron masses and a more complete analysis of
dynamical amplitudes show that isospin is typically broken
only at the few percent level.

Why is isospin such a good symmetry
at low energies? The answer can be found in the 
pattern of chiral-symmetry breaking incorporated in 
the chiral Lagrangian \cite{invk}.
While explicit chiral-symmetry-breaking effects are present
already at index $\Delta=0$ through the pion mass term,
operators generated by the quark mass difference appear only at 
$\Delta=1$ through a term that
contributes to the nucleon mass splitting 
and,
due to chiral symmetry, to certain pion-nucleon interactions. 
As a consequence, in most quantities
isospin breaking competes with isospin-conserving
operators of lower order,
and its relative size is not $\varepsilon$
but $\varepsilon (Q/M_{QCD})^n$, where $n$ is a positive integer. 
In other words, isospin is an accidental symmetry \cite{invk}:
a symmetry of the lowest order EFT which is not a symmetry
of the underlying theory.
The only known exception to this rule is in the isoscalar
$t$ channel in $\pi N$ scattering at threshold. 
There, there is no contribution from the $\Delta=0$ Lagrangian, and
both the isospin-conserving and -violating amplitudes start
at the same order. The isospin-violating
piece comes from the pion-nucleon interactions
linked to the nucleon mass splitting.
Unfortunately, this is hard to see experimentally.
%finish

Along these lines, one can study the expected size of
isospin breaking in the nuclear potential.
We follow  the standard nomenclature and refer to an 
isospin-symmetric potential as  ``class I'',
to a potential that breaks charge dependence but not charge symmetry
---defined as a rotation of $\pi$ around the 2-axis in isospin space---
as  ``class II'',
to one that breaks charge symmetry but vanishes in the $np$ system
as  ``class III'',
and to one that breaks charge symmetry but causes mixing
in the $np$ system
as  ``class IV''.

At $Q\sim M_{NN}$,
photon exchanges are perturbative. 
These standard electromagnetic effects from ``soft'' photons can
be obtained straightforwardly from
gauge-invariant operators involving the photon field.
In addition, isospin violation
arises from the quark masses,
from indirect electromagnetic effects, and from simultaneous
pion-photon exchange. 
In order to compare the various sources of isospin breaking,
we note that the size of electromagnetic effects in loops
is typically $\sim \alpha/\pi$ which, numerically, is 
$\sim \varepsilon (Q/M_{QCD})^3$.

The leading isospin-breaking interactions in Weinberg's power counting
have been derived in 
Ref. \cite{invk}.
(The necessary modification of Weinberg's counting discussed
in Sect. \ref{piladder}
shows that chiral-symmetry-breaking terms are even more suppressed 
{\it vis-a-vis} chiral-symmetric ones, and
is not expected to affect the relative sizes among  
isospin-breaking interactions.)
No isospin-violating effects appear at
leading order, $\nu = \nu_{min}$, so the 
leading potential is class I.
The first isospin-breaking effect (in addition to Coulomb exchange)
appears at 
$\nu = \nu_{min} +1$ in the form of a class II
isospin violation from the pion mass splitting 
($\Delta m_\pi^2= {\cal O}(\alpha M_{QCD}^{2}/\pi)$) in OPE.
One order down, $\nu = \nu_{min} +2$,
a class III force appears
mainly from the quark mass difference
through breaking in the $\pi NN$ coupling 
($\beta_1={\cal O}(\varepsilon m_{\pi}^{2}/M_{QCD}^{2})$)
in OPE ,
from contact terms
($\gamma_{s,\sigma}={\cal O}(\varepsilon m_{\pi}^{2}/M_{QCD}^{4})$),
and from the nucleon mass difference
($\Delta m_N={\cal O}(\varepsilon m_{\pi}^{2}/M_{QCD})$). 
To this order the isospin-violating 
nuclear potential is a two-nucleon potential of the form
\begin{equation}
 V_{ib}=V_{\rm II}\left[ (t_1)_{3}(t_2)_{3}-\boldt_1\cdot\boldt_2 \right] +
        V_{\rm{III}}\left[ (t_1)_{3}+(t_2)_{3} \right] ,  \label{iv78}
\end{equation}
\noindent
where 
\begin{equation}
 V_{\rm II}=
   -\left(\frac{g_{A}}{f_{\pi}}\right)^{2}
     \frac{\vec{q}\cdot\vec{\sigma}_1 \vec{q}\cdot\vec{\sigma}_2}
          {(\vec{q}\,^{2}+m_{\pi^{0}}^{2})
  (\vec{q}\,^{2}+m_{\pi^{\pm}}^{2})}(\Delta m_{\pi}^{2}+\Delta m_{N}^{2}),
                                                     \label{iv79}
\end{equation}
\begin{equation}
 V_{\rm III}=
    \frac{g_{A}\beta_{1}}{2f_{\pi}^{2}}
    \frac{\vec{q}\cdot\vec{\sigma}_1\vec{q}\cdot\vec{\sigma}_2}
         {\vec{q}\,^{2}+m_{\pi}^{2}}
    -(\gamma_{s}+\gamma_{\sigma}\vec{\sigma}_{1}\cdot\vec{\sigma}_{2}). 
\label{iv80}
\end{equation}
\noindent
Finally, class IV forces
appear only at order $\nu = \nu_{min} +3$.

We conclude that the pattern of symmetry breaking in QCD 
naturally suggests a hierarchy of classes in the nuclear 
potential \cite{invk}:
\begin{equation}
\frac{\langle V_{\rm M+I}\rangle}{\langle V_{\rm M}\rangle}
\sim {\cal O}\left(\frac{Q}{M_{QCD}}\right),
\end{equation}
where $\langle V_{\rm M}\rangle$ denotes the average contribution of 
the leading class ${\rm M}$ potential. 
This qualitatively explains not only why isospin is a good 
approximate symmetry at
low energies, but also why 
charge symmetry is an even better symmetry.
It gives, for example, the observed isospin structure 
of the Coulomb-corrected scattering lengths \cite{jerry}, 
$a_{np}\simeq 4 \times ((a_{nn}+a_{pp})/2 - a_{np}) 
       \simeq 4^2 \times (a_{pp}-a_{nn})$. 

One can use the above formalism to do consistent and systematic
calculations of isospin violation.
For example, the  
isospin-violating potential of ranges $\sim 1/m_{\pi}$
and $\sim 1/2m_{\pi}$
up to $\nu = \nu_{min} +3$ were computed in 
Refs. \cite{prlwithnijm,mejim,niskanenpot}.
In contrast to previous attempts lacking an EFT framework,
the EFT results are invariant under both gauge transformation and
pion-field redefinition and have simple forms.
The component of range $\sim 1/m_{\pi}$
comes from diagrams with all possible one-photon
dressings of OPE, plus the relevant counterterms
\cite{prlwithnijm}.
Its isospin structure allows
only charged-pion exchange and therefore 
is class II. 
This $\pi\gamma$ potential has been incorporated in a
Nijmegen phase shift reanalysis of 
$np$ data below 350 MeV \cite{prlwithnijm}.
We can use the values for the $\pi NN$ coupling constants
determined by the analysis to find that their isospin breaking 
($\beta_1$) is consistent with zero,
with an uncertainty comparable to our expectation
from 
dimensional
analysis and from 
$\pi-\eta-\eta'$ mixing \cite{mejimandterry}.
Similarly, the two contact interactions ($\gamma_{s,t}$)
might be viewed 
as originating in $\rho-\omega$ mixing and 
pseudovector-meson exchange 
(in particular close-lying doublets such as $a_{1}-f_{1}$)
\cite{mejimandterry,coon}. 
The components of range $\sim 1/2m_{\pi}$ come from
two sources.
One is the pion mass difference ($\Delta m_\pi^2$) in TPE that
generates a class II potential \cite{mejim};
the other is a $\pi\pi NN$ seagull 
that arises as a chiral partner of the nucleon mass difference
($\Delta m_N$),
and produces a class III TPE potential \cite{niskanenpot}.
All these effects are
relatively small, with estimated
contributions to the scattering lengths of $\sim \pm 0.5$ fm.
%added
These long-range components could be included
in the new Nijmegen PSA \cite{robetal}.

A fit to $NN$ phase shifts including various isospin-breaking interactions
was carried out in Ref. \cite{waltz},
improving on an earlier analysis with perturbative
pions \cite{Epelbaum:1999zn}.
A slightly different counting is used that enhances by one order
the electromagnetic effects,
electromagnetic corrections being counted
as $\alpha \sim (Q/M_{QCD})^2$.
The EFT potential to order $\nu=\nu_{min}+2$ (including
Coulomb) is then fitted to low-energy $S$ and $P$, $np$ and $pp$
phase shifts,
using cutoffs in the range $300-500$ MeV.
With modest cutoff
dependence, isospin breaking in the scattering lengths can be 
accommodated, and 
higher energies and partial waves predicted.
A next-order calculation should achieve the level of precision
of modern phenomenological potentials.

\subsection{Three- and four-nucleon systems}
\label{3and4}

When few-body systems are considered, one needs to address the issue
of few-body forces, which, being in general not forbidden by symmetries,
will at some level contribute to observables. 
One of the advantages of a field-theoretical
framework is the possibility of the derivation of consistent few-body
forces, free of off-shell ambiguities. 
In the standard nuclear-physics approach, few-nucleon
forces are 
either inspired by arguments that are independent
of the assumptions invoked in the $NN$ potential,
or simply guessed
on phenomenological terms.

The pionful EFT offers some insight into few-nucleon forces.
In addition to contact interactions as in the pionless theory, one has further
pion-exchange components.
The potential, being defined as a sub-amplitude, includes (for $A>2$)
diagrams that have $C\ge 1$ separately connected pieces.
An $n$-nucleon force is a contribution
to the {\it potential} that connects $n$ nucleons.

Weinberg's power counting embodied
in Eq. (\ref{nu}) suggests a hierarchy of few-nucleon forces.
As in the two-nucleon case, this power counting relies
on an implicit assumption about the scale appearing
in contact interactions.
We have seen in Sect. \ref{tritonsect} 
that in the pionless
EFT the running of the renormalization group toward
low energies 
enhances the size of three-body forces.
The latter get contaminated by the fine-tunning present in the two-body sector.
Whether the same happens at the higher energies relevant to the
pionful theory is not clear.
Part of the $3N$ forces in the pionless theory
match onto diagrams of the pionful theory 
which are the iteration of the $NN$ potential
(through $NN$ intermediate states where at least
one nucleon has momentum ${\cal O}(m_\pi)$).
It is conceivable that the enhancement is removed from contact interactions
once the pion is introduced explicitly in the EFT. 

The new forces that appear in systems with more than two 
nucleons have been derived in 
Refs. \cite{ciOvK,inwei5,civK1}.
The dominant potential, at $\nu=6-3A=\nu_{min}$, 
is simply the $NN$ potential of lowest order that 
we have already encountered in Sect. \ref{potfits}.
We can easily verify that, if the delta is kept
as an explicit degree of freedom, a $3N$ potential will arise at 
$\nu=\nu_{min}+2$,
a $4N$ potential at $\nu=\nu_{min}+4$, and so on. 
It is (approximate) chiral symmetry therefore that
implies that $n$-nucleon forces $V_{nN}$ obey a hierarchy
of the type
\begin{equation}
\frac{\langle V_{(n+1)N}\rangle}{\langle V_{nN}\rangle}
\sim {\cal O}\left( \frac{Q}{M_{QCD}} \right)^2,
\end{equation}
with $\langle V_{nN}\rangle$ denoting the contribution per $n$-plet.
(This hierarchy is a non-trivial consequence of chiral symmetry,
as there exist non-chiral models that produce large
three-body forces.) 
As we discussed in Sect. \ref{piladder},
we expect $|\langle V_{2N}\rangle | \sim 10$ MeV.
Using $m_\pi$ and $ m_\rho$ for $Q$ and $M_{QCD}$ respectively,
the suppression factor is $\sim 0.05$, give or take a factor
of 2 or 3.
These estimates are in accord with detailed few-nucleon phenomenology
based on potentials that include small $3N$ and no $4N$ forces.
This is shown in Table \ref{table:sizes}
in the case of the AV18/IL2 potential \cite{Pieper:2001mp}.

\begin{table}[!b]
\caption{Contributions of the two-, three- and four-nucleon potentials
(per doublet, per triplet, and per quadruplet, respectively):
Weinberg power counting (W pc) and calculations \cite{Pieper:2001mp}
with the AV18/IL2 potential
for the ground states of various light nuclei ($^2$H, $^3$H, {\it etc.}).
\label{table:sizes}}
\vspace{0.25cm}
\begin{center}
\begin{tabular}{||c||c||c|c|c|c|c|c|c|c||} 
\hline
(MeV)  & W pc & $^2$H &  $^3$H & $^4$He & $^6$He & $^7$Li & $^8$Be & $^9$Be 
       & $^{10}$B \\
\hline
\hline
$|\langle V_{2N}\rangle |$& $\sim 10$ & 22 & 20 & 23 & 13 & 11 & 11 &  9.4 
                          & 8.9 \\
$|\langle V_{3N}\rangle |$& $\sim 0.5$ & -- & 1.5 & 2.1 & 0.55 & 0.43 & 0.38 & 0.29 
                          & 0.30 \\
$|\langle V_{4N}\rangle |$& $\sim 0.02$& -- & -- & ? & ? & ? & ? & ? 
                          & ?\\
\hline
{\Large $\frac{|\langle V_{3N}\rangle |}{|\langle V_{2N}\rangle |}$}
& $\sim 0.05$ & -- &0.075 & 0.091 & 0.042 & 0.039 & 0.035 &  0.031 
& 0.034\\

{\Large $\frac{|\langle V_{4N}\rangle |}{|\langle V_{3N}\rangle |}$}
& $\sim 0.05$ & -- & -- & ? & ? & ? & ? & ? 
& ? \\
\hline
\end{tabular}
\end{center}
\end{table}
%till here

\begin{figure}[!t]
\centerline{\epsfysize=10cm \epsfbox{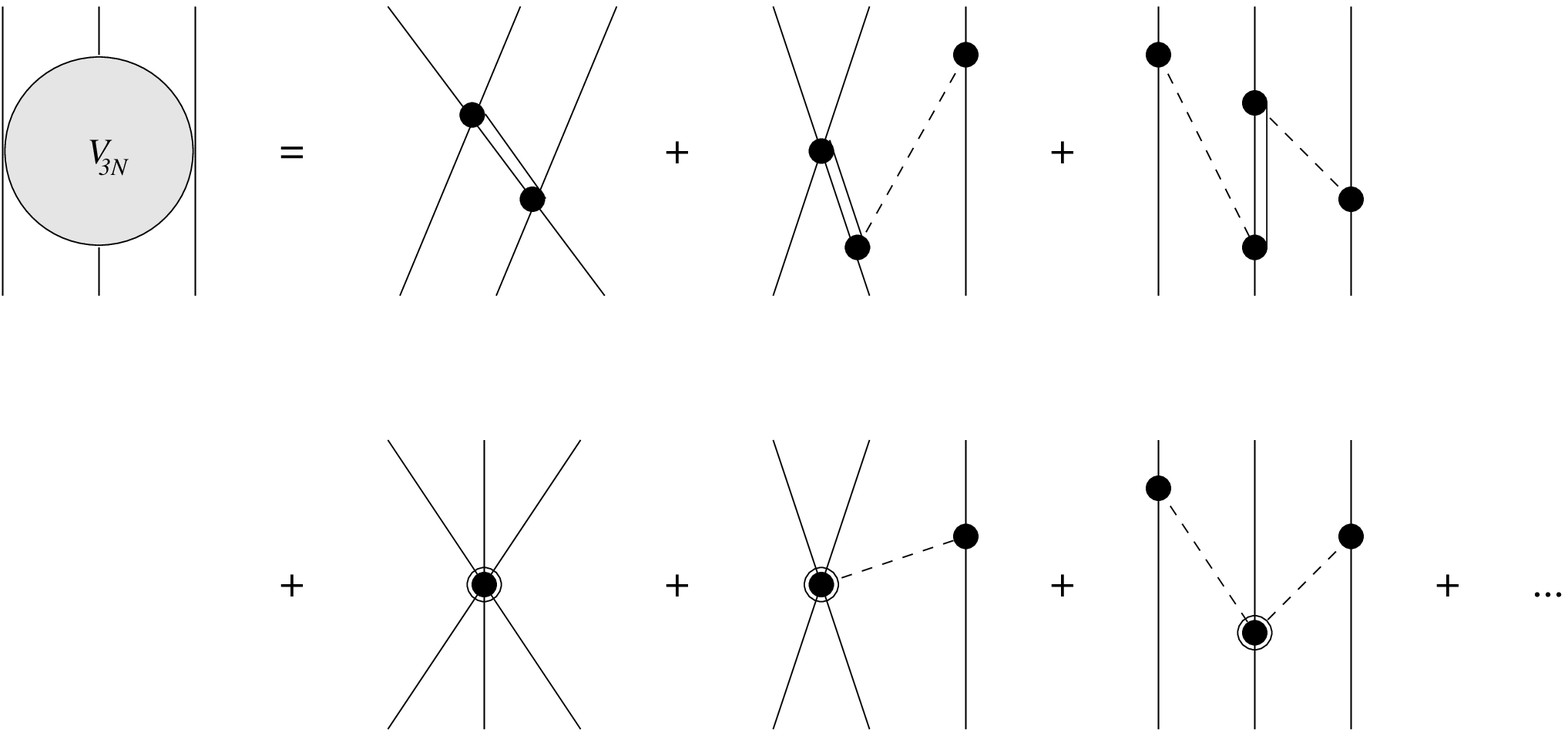}}
\vspace{-5.50cm}
\caption{Some time-ordered diagrams contributing to the $3N$ potential
in the pionful EFT.
(Double) solid lines represent nucleons (and/or deltas), 
dashed lines pions, a heavy dot an interaction in ${\cal L}^{(0)}$,
and a dot within a circle an interaction in ${\cal L}^{(1)}$.
First line corresponds to $\nu=\nu_{min}+2$,
second line to $\nu=\nu_{min}+3$, 
and ``$\ldots$'' denote $\nu\geq \nu_{min}+4$. 
All nucleon permutations and orderings with at least one pion
or delta in intermediate states are included.}
\label{F:vkolck:V3N}
\end{figure}

It proves instructive to look at the 
form of the first few terms in the
few-nucleon potential.
At $\nu=\nu_{min}+2$, in addition to corrections to the $NN$
potential, one also finds diagrams
involving either three nucleons or two pairs of nucleons
connected via leading contact interactions and static pions.
One finds \cite{inwei6,inwei5,civK1} that the various
orderings of these diagrams cancel among themselves
and against contributions from the energy-dependent piece of the 
iterated $NN$
potential that appears at the same order.
Alternatively, redefining the potential
to eliminate energy dependence leads to
no $3N$ TPE forces of this type at all \cite{jimandsid}.
Remaining to this order are only $3N$ forces generated by
the delta isobar, if the delta is kept explicit in the EFT.
At $\nu_{min}+3$ further terms with similar structure
arise \cite{invk,civK1} ---see
Fig. (\ref{F:vkolck:V3N}).

If, instead, the delta is integrated out, 
its contributions appear through the parameters of 
the potential at 
$\nu=\nu_{min}+3$. 
In this case,
there are no $3N$ forces up to $\nu=\nu_{min}+3$.
In Ref. \cite{epel34NLO}
the $\nu=\nu_{min}+2$ deltaless potential of Ref. \cite{epelfit}
was used to predict properties of the $3N$ and $4N$ systems.
When the cutoff was varied between 540 and 600 MeV,
the low-energy $3N$ scattering observables showed relatively
little cutoff dependence, and came out similar to
conventional-potential results.
The 
$3N$ and $4N$ binding energies also fell into conventional range, 
but they changed by about 10\%.
This nearly-linear dependence on the cutoff could
be indication of inconsistent renormalization
(presumably curable by a short-range $3N$ force),
although the small range of cutoff variation does not allow firm conclusions.
It is interesting that
when certain $\nu=\nu_{min}+3$ $NN$ contributions
are included, the cutoff dependence decreases.
In the process
one loses the agreement for the elastic neutron
analyzing power, $A_y$.
An ``$A_y$ puzzle'' plagues conventional models:
potentials that fit $NN$ data well all fall short of
reproducing $A_y$ at energies as low as 3 MeV,
even when existing $3N$ potentials are added.
This seems to be the case with chiral potentials as well
\cite{AywithchiralNN}.

The leading $3N$ potential \cite{civK1}
has components with three different ranges:
TPE; OPE/short-range; and purely
short ranged.
(Relativistic corrections neglected below 
are discussed in Ref. \cite{friarcoon}.)
The TPE part of the potential is determined
in terms of $\pi N$ scattering observables \cite{civK1}. 
It is similar to the Tucson-Melbourne (TM) 
and Brazil potentials \cite{TMbrazil},
but it corrects a deficiency of the TM potential 
in regard to chiral symmetry.
The TM potential
was built from a $\pi N$ scattering amplitude
that corresponds to a particular
choice of pion fields 
for which chiral symmetry is not respected term by term \cite{vk:huber}.
Although the on-shell TM $\pi N$ amplitude agrees
with that obtained in EFT to sub-leading order, they differ off shell.
Of course, in a field theory
the corresponding $3N$  potentials involve other interactions 
that enforce chiral symmetry and invariance under pion-field
redefinitions.
Unfortunately the original TM potential was not
derived within field theory, and the required extra terms were not 
included.
After this is done, the EFT result is obtained \cite{vk:huber}.
The resulting TM' potential is similar to the 
Brazil potential,
and has been studied in detail in
Refs. \cite{huber2,coonrework}.
In Ref. \cite{huber2} it was shown that one of the 
components of the force is dominant in $3N$ elastic scattering
observables. This explains why all existing TPE $3N$ forces
give essentially the same results for the $3N$ continuum
after being fitted to the triton energy. 
This type of $3N$ force does not improve agreement for
$A_y$ much \cite{huber2}.

The novel
OPE/short-range components of the potential 
involve two $\pi (\bar{N}N)^2$ interactions of strengths
not determined by chiral symmetry alone \cite{civK1}.
These parameters can be thought of as representing short-range effects 
such as $\sigma$ and $\omega$ exchanges with an intermediate
$N(1440)$ resonance, and $\rho$ exchange from a $\pi\rho$ Kroll-Ruderman
term \cite{newformod}.
It can be shown that due to the antisymmetry
of the wave-function only one combination of parameters contributes
\cite{stewartepelbaumPC}.
This combination can be determined
from reactions involving only two nucleons,
as discussed in Sect. \ref{pionfulprobes} below.
In Ref. \cite{huber2} the effect
of these novel terms was estimated assuming that their LECs have
natural size. In conjunction with TM', this force can 
bridge a significant part
of the discrepancy between ``realistic'' $NN$ potentials
and $A_y$ data.
Although not a consistent EFT calculation,
this estimate serves to assess the size
of $3N$ forces in the EFT.
Clearly, agreement with $A_y$ data at  $\nu=\nu_{min}+2$ in the deltaless EFT
is purely accidental.
Conversely, a full 
$\nu=\nu_{min}+3$ calculation might resolve the $A_y$ puzzle.

The purely short-range components of the potential can only be determined from
few-nucleon systems \cite{civK1}.
As we have seen in Sect. \ref{greatness}, 
the Pauli principle
leads to a single LEC.
Once this LEC is determined from one $3N$
input (such as the triton binding energy),
all other observables (such as $A_y$) can be predicted
(once the OPE/short-range component has been fixed by data
involving just two nucleons).

It is apparent that the pionful EFT brings new forces into play, 
and that these new elements
might resolve remaining issues in the description of data.
This prospect
calls for a fully consistent $\nu_{min}+3$ calculation
with maximal $NN$ input.

\subsection{Processes with external probes}
\label{pionfulprobes}

The power counting arguments of
Sect. \ref{piladder} 
can be easily generalized to the case where
external probes with momenta $Q\sim M_{NN}$
interact with few-nucleon systems.
The probes deposit an energy $\sim M_{NN}$ onto the nuclear system, so
that, if
we define the kernel $K$ as the sum of irreducible diagrams
to which the probes are attached,
the power counting Eq. (\ref{nu}) applies equally well to $K$.
Interactions among nucleons occurring before or after
scattering can be treated as before: iteration
of the potential gives rise to the wave-function
$|\psi\rangle$ ($|\psi'\rangle$) of the initial (final) nuclear state.
The full scattering amplitude is then 
\begin{equation}
T = \langle \psi'|K|\psi\rangle.
\label{probes}
\end{equation}
The pionful EFT can also handle scattering at smaller $Q$, of course,
but then Eqs. (\ref{nu}, \ref{probes}) have to be modified.
When the deposited energy
is $\sim M_{NN}^2/m_N$ ---for example, when the incoming probes are photons of
momenta $Q\sim M_{NN}^2/m_N$--- there can be intermediate
few-nucleon states that are reducible, and  the breakdown
of $T$ into kernel and wave-functions is more complicated \cite{ourcompton}.
In this situation a perturbative treatment of pions, or even better,
the pionless EFT, should suffice.

In practice, it is frequently desirable to minimize nuclear
wave-function errors by using a high-precision
phenomenological potential.
That this is a good approximation is suggested by
a comparison \cite{taesunpions}
between a simplified version of the EFT potential 
of Ref. \cite{ciOLvK} and the Argonne V18
potential \cite{argonne}, which show agreement
in most aspects of the wave-function.
The cost of this ``hybrid approach'' \cite{inwei5}
is the introduction of an uncontrolled error due to a possible
mismatch between the off-shell extensions of the kernel and the potential.
This error can, on the other hand, be estimated
by the use of several different phenomenological potentials of
similar quality.

As with few-nucleon forces, the factor $-2C$ in Eq. (\ref{nu})
implies that external probes tend to interact 
predominantly with a single nucleon, simultaneous interactions with more than 
one nucleon being suppressed by powers of $(Q/M_{QCD})^2$. Again, this 
generic dominance of the impulse approximation is a 
well-known result that arises naturally here. 
This is of course what allows extraction, to a certain accuracy, of  
one-nucleon parameters from nuclear experiments. 
A valuable by-product of the EFT is to provide a consistent
framework for one- and few-nucleon dynamics, whereby
few-nucleon processes can be used to infer one-nucleon properties.
More interesting from the purely nuclear-dynamics perspective are, however,
those processes where the leading single-nucleon contribution vanishes by 
a particular choice of experimental conditions, 
for example the threshold region. 
In this case, certain two-nucleon contributions, especially in the 
relatively large deuteron, can become important. 
Further examination of the structure of the chiral Lagrangian
reveals that two-nucleon contributions tend to be
dominated by pion exchange.
Indeed, photons and pions couple to four-nucleon operators only
at ${\cal O}(Q/M_{QCD})$ relative to pion-exchange diagrams
constructed out of the leading-order Lagrangian.
Thus power counting justifies the ``chiral filter hypothesis''
that was put forward to summarize some empirical results
on electroweak form factors
\cite{chiralfilter}.
This ``pion dominance'' ensures that
two-nucleon contributions from the EFT in lowest orders
tend to be similar to those in phenomenological models that
include pion-exchange currents.

Many processes have been analysed in the pionful EFT.
Some of those processes are extensions to higher energies
of the same electroweak processes described in Sect. \ref{pionlesstwobod}.
For example, 

\noindent
$\bullet \hspace{.2cm}  e d\rightarrow e d$ 
and deuteron form factors
\cite{ed}

\noindent
$\bullet \hspace{.2cm}  \vec{e} d\rightarrow e NN$ 
and parity violation 
\cite{pved}

\noindent
$\bullet \hspace{.2cm}  n p\rightarrow d \gamma$ 
and meson-exchange currents
\cite{radcap}

\noindent
$\bullet \hspace{.2cm}  \vec{n} p\rightarrow d \gamma$ 
and parity violation
\cite{polradcap}

\noindent
$\bullet \hspace{.2cm} p p\rightarrow d e^+ \nu_e$
and axial currents
\cite{solarfusion}

\noindent
$\bullet \hspace{.2cm} p \; ^3{\rm He}\rightarrow {^4 {\rm He}} \; e^+ \nu_e$
and solar neutrino production
\cite{solarhep}

\noindent
$\bullet \hspace{.2cm} \nu d\rightarrow l NN$ 
and solar neutrino detection
\cite{nud}

\noindent
$\bullet \hspace{.2cm} \mu d\rightarrow \nu_\mu nn$ 
and its measured rate
\cite{mud}

\noindent
$\bullet \hspace{.2cm} \gamma d\rightarrow \gamma d$
and nucleon polarizabilities
\cite{compton,ourcompton}

\noindent
For details, we refer the reader to more extensive reviews 
\cite{vanKolck_review} 
and the original papers.
Here we briefly discuss those processes more germane to the pionful theory,
involving pions in initial and/or final states.

\subsubsection{$\pi d\rightarrow \pi d$}\label{pid}
This is perhaps the most direct way to check the consistency of EFT in
one- and few-nucleon systems.
For simplicity, consider the region near threshold
with the delta integrated out.
There the lowest-order, $\nu=\nu_{min}=-2$ 
contributions to the kernel vanish because 
the pion is in an $s$ wave and the target is isoscalar.
The $\nu=\nu_{min}+1$ term comes from the 
(small) isoscalar pion-nucleon seagull,
related in lowest-order to the pion-nucleon isoscalar amplitude $b^{(0)}$.
$\nu=\nu_{min}+2$ contributions 
come from corrections to $\pi N$ scattering
and two-nucleon diagrams, which involve besides $b^{(0)}$ also the much larger 
isovector amplitude $b^{(1)}$. 
These various contributions to the $\pi d$ scattering length
have been estimated \cite{inwei5,beanenew}.
They were
found in 
agreement with previous, more phenomenological calculations,
which have been used to extract $b^{(0)}$. 
Partial sets of higher-order corrections 
have been evaluated in Ref. \cite{kaiserization} 
(and in Ref. \cite{koltun} for the
related, double-charge-exchange process).
A consistent $\nu=\nu_{min}+3$ calculation of $\pi d$ elastic
scattering is in progress
\cite{bbemp}.
Eventually, a $\nu=\nu_{min}+4$ calculation
might be required in order to determine
the chiral-symmetry breaking LEC
 $C_2^{(qm)}$ discussed further below in connection 
with lattice QCD.
Charge-symmetry-breaking effects were considered in Ref. \cite{rockmore}.
Note that
an alternative approach with perturbative pions \cite{boragriess}
also seems to accommodate the available experimental data.

\subsubsection{$\gamma^{(*)} d\rightarrow \pi^0 d$}
This reaction offers the possibility to test a
prediction arising from a combination of two-nucleon contributions 
and the single-neutron amplitude. 
The differential cross section for a photon of momentum $k$ and
longitudinal polarization $\epsilon_L$ to
produce a pion of momentum $q$ is, at threshold, 
\begin{equation}
\left[ \left( \frac{3k}{8q} \right) 
       \left( \frac{d\sigma}{d\Omega} \right) \right]_{q=0} = 
|E_d|^2 + \epsilon_L |L_d|^2,
\end{equation}
where the electric dipole amplitude 
$E_d(k^2)$ characterizes the transverse response
and $L_d(k^2)$ the longitudinal response.

$E_d(0)$ was studied up to $\nu=\nu_{min}+3$ with the delta integrated out
in Ref. \cite{vk:beane}. 
There are two classes of contributions, according to whether
the external light particles interact with one 
or with both nucleons.
The one-nucleon part of the kernel is given by standard
$A=1$ ChPT, with due account of $P$ waves and 
Fermi motion inside the deuteron.
The neutrality of the outgoing $s$-wave pion ensures
that the leading $\nu=-2=\nu_{min}$ terms vanish. 
The first two-nucleon part of the kernel 
appears at $\nu=\nu_{min}+2$; it comes from
a virtual charged pion photoproduced on one nucleon
which rescatters on the other nucleon with charge exchange.
These contributions are actually numerically larger than indicated
by power counting due to the relatively large deuteron size.
Smaller two-nucleon terms 
appear at $\nu=\nu_{min}+3$ from corrections in either nucleon.  
Results for $E_{d}(0)$ up to $\nu=\nu_{min}+3$ \cite{vk:beane}
are shown in Table \ref{T:vkolck:Ed}.
They correspond to the Argonne V18 potential \cite{argonne} and a cutoff 
$\Lambda=1000$ MeV. Other realistic potentials and cutoffs from 650 to
1500 MeV give the same result within 5\%.
The chiral potentials of Sect. \ref{potfits}
are more cumbersome to use, but it has been verified that
they give 
results that are similar to other potentials.  
Two-nucleon
contributions seem to be converging, although more convincing
evidence would come from next order, where loops appear. 
A model-dependent estimate \cite{vk:wilhelm} of some $\nu=\nu_{min}+4$ terms 
suggests a 10\% or larger error from neglected higher orders 
in the kernel itself. 
The single-scattering contribution depends on the
amplitude for $\gamma n\rightarrow \pi^0 n$, 
$E_{0+}^{(\pi^0 n)}$,
in such a way that $E_{d}(0) \sim -1.79 -0.38(2.13-E_{0+}^{(\pi^0 n)})$ 
in units of $10^{-3}/m_{\pi^+}$.
Thus, sensitivity to $E_{0+}^{(\pi^0 n)}$ survives the large two-nucleon
contribution at $\nu=\nu_{min}+2$. 

\begin{table}[!b]
\caption{Values for $E_{d}(0)$ in units of $10^{-3}/m_{\pi^+}$
from one-nucleon contributions ($1N$) up to $\nu=\nu_{min}+3$,
two-nucleon kernel ($2N$) at $\nu=\nu_{min}+2$ 
and at $\nu=\nu_{min}+3$,
and their sum ($1N+2N$).\label{T:vkolck:Ed}}
\vspace{0.0cm}
\begin{center}
%added \footnotesize
\begin{tabular}{cccc}
$1N$   & \multicolumn{2}{c}{$2N$} &    $1N+2N$        \\
\cline{2-3}
$\nu\le \nu_{min}+3$ & $\nu= \nu_{min}+2$ & $\nu= \nu_{min}+3$   
                                          & $\nu\le \nu_{min}+3$  \\
\hline
 0.36  & $-$1.90 & $-$0.25                  & $-$1.79 \\
\end{tabular}
\end{center}
\end{table}

\begin{figure}[!t]
\centerline{
\epsfysize=8cm \epsfbox{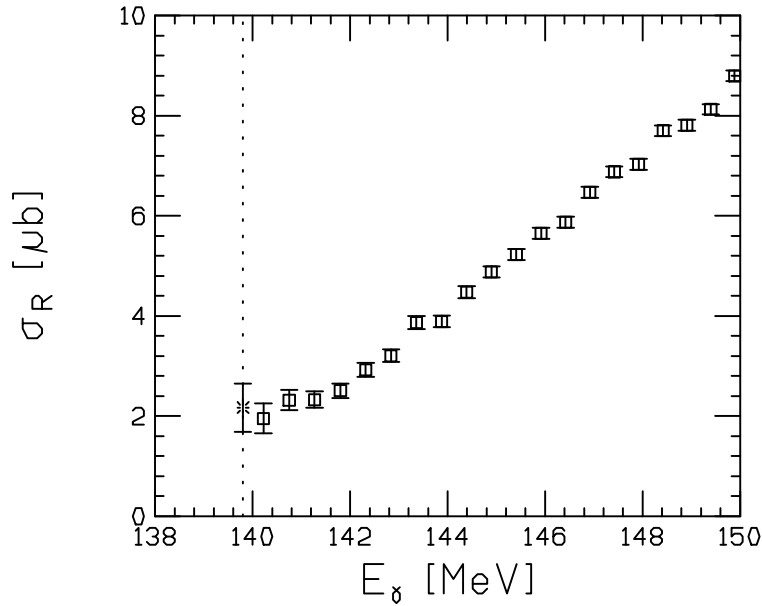}  
}
\vspace{0.05cm}
\caption[ ]{ Reduced cross section  $\sigma_R=(k/q)\sigma$ 
for neutral-pion photoproduction as function of the photon 
energy.
Threshold is marked by a dotted line.
The star at threshold is the 
EFT {\it prediction} \cite{vk:beane},
and
squares are data points \cite{vk:berg}.
Figure courtesy of U.-G. Mei{\ss}ner.}
\label{F:vkolck:piphotodata}
\end{figure}

We see that working within the EFT yields a testable
prediction, $E_d(0)=-(1.8 \pm 0.2)\cdot 10^{-3}/m_{\pi^+}$
\cite{vk:beane}. 
It is remarkable that for this process EFT gives
results that are significantly different from tree-level models of
the type traditionally used in nuclear physics.
For example, the models in Ref. \cite{vk:justusetal} predict the
threshold cross section about twice as large as the EFT.
Most of the difference comes from one-nucleon loop diagrams:
tree-level models tend to differ from ChPT mostly
by having a smaller $E_{0+}^{(\pi^0 n)}$, which increases $|E_d|$.
A test of this prediction is thus an important 
check of our understanding of the role of QCD at low energies.
Such a test was carried out
at Saskatoon \cite{vk:berg}.
The experimental results for 
the pion photoproduction cross section near threshold are shown 
in Fig. (\ref{F:vkolck:piphotodata}),
together with 
the EFT prediction at threshold \cite{vk:beane}.
Inelastic contributions have been estimated
in Refs. \cite{vk:berg,Levchuk:2000hz} and are smaller than 10\% 
throughout the range of energies shown.
At threshold, Ref. \cite{vk:berg} finds 
$E_{d}(0)=-(1.45\pm 0.09)\cdot 10^{-3}/m_{\pi^+}$.
While agreement with the EFT to order $\nu=\nu_{min}+3$ 
is not better than a reasonable estimate of higher-order
terms, it is clearly superior to
tree-level models. This is compelling evidence of chiral loops. 

A further test of the EFT comes from the coherent
neutral-pion electroproduction
on the deuteron.
$E_d(k^2)$ and $L_d(k^2)$ were predicted 
to $\nu=\nu_{min}+2$ (with no new free parameters)
in 
Ref. \cite{Ulfelectroprod},
for transferred momenta in the range $0-0.1$ GeV$^2$.
Because $E_{0+}^{(\pi^0 p)}$ is not well reproduced
at this order, only the $k^2$ dependence can be tested.
This reaction was measured at $k^2=-0.1$ GeV$^2$
in Mainz \cite{vk:bernstein}, and 
values for $|E_d(-0.1)|$
and 
$|L_d(-0.1)|$ were extracted.
If they are
 compared with 
the results from Ref. \cite{Ulfelectroprod}
simply shifted by a $k^2$-independent amount
in order to reproduce the $E_d(0)$ of the $\nu=\nu_{min}+3$
calculation,
then there is good agreement for $|E_d(-0.1)|$
but $|L_d(-0.1)|$ fails by a factor 2 \cite{vk:bernstein}.
Because the calculated $L_d(k^2)$ is not dominated
by a single mechanism to the extent $E_d(k^2)$ is,
it is possible that it suffers from stronger
corrections in next order.
An extension of these calculations to higher
order and beyond threshold is also highly
desirable.

\subsubsection{$NN \rightarrow NN\pi$}
This reaction has attracted a lot of interest because of the failure of 
standard phenomenological mechanisms in reproducing 
the small cross section observed near threshold. 
It involves larger momenta of ${\cal O}(\sqrt{m_\pi m_N})$, so 
the relevant expansion parameter here is not so small,
$(m_\pi/m_N)^{\frac{1}{2}}$.
This process is therefore not a good testing ground for 
the above ideas. 
But $(m_\pi/m_N)^{\frac{1}{2}}$ is still $<1$, so at least in some formal sense
we can perform a low-energy expansion. 
In Refs. \cite{vkolck:cohen,hanhart}
the chiral expansion was adapted to this reaction and the first few 
contributions estimated. 
(Note that ---contrary to what is stated in Ref. \cite{ulfpespelasmauns}---
momenta $\sim \sqrt{m_\pi m_N}$ do not necessarily
imply a breakdown of the non-relativistic
expansion, as $p^4/m_N^3 \sim  (m_\pi/m_N) (p^2/m_N)$
is still small.)

Initial attention concentrated on $pp\rightarrow pp \pi^0$ at threshold.
The lowest-order terms all vanish, and the formally-leading non-vanishing 
terms ---an 
impulse term and 
a similar diagram from the delta isobar--- 
are anomalously small and partly cancel. 
The bulk of the cross section must then arise from contributions that 
are relatively unimportant in other processes:
isoscalar pion rescattering,
TPE, and 
high-order short-range $\pi (\bar{N}N)^2$ terms.
While the first two contributions are calculable,
the third involves LECs that can only be fitted or modeled.
These LECs can be thought of as
originating from heavier-meson exchange: 
pair diagrams with $\sigma$ and $\omega$ exchange, and 
a $\pi\rho\omega$ coupling, among other, smaller terms
\cite{withriska}.    
In Ref. \cite{withriska} it was shown that a large uncertainty
comes from the short-range features of the wave-function,
so a more systematic study has to await the development of
chiral potentials that are accurate at the relevant energies.
Other EFT studies of this channel,
including attempts to compute TPE, can be found in 
Ref. \cite{vkolck:pppppizero}. 
More problematic is the situation with
the threshold cross section of other, not so suppressed channels
$NN \rightarrow d\pi, \rightarrow pn\pi$.
In those channels the Weinberg-Tomozawa $\pi\pi NN$ term,
fixed by chiral symmetry,
dominates.
Wave-function dependence is much smaller, yet
a calculation including leading and some sub-leading contributions
underpredicts the data by a factor of $\sim 2$ \cite{vkolck:carocha}.
A calculation including TPE is badly needed.

Despite these problems, much can be learned
from this reaction in the threshold region.
One example is charge-symmetry breaking.
The nucleon mass splitting comes from both the quark mass difference
and from electromagnetic effects,
$\Delta m_N= \delta m_N + \bar{\delta} m_N$, with
$\delta m_N= {\cal O}(\varepsilon m_{\pi}^{2}/M_{QCD})$
and $\bar{\delta} m_N= {\cal O}(\alpha M_{QCD}/\pi)$.
Determining the two LECs $\delta m_N$ and $\bar{\delta} m_N$ separately
is interesting for several reasons: coupled to a lattice evaluation
of $\delta m_N$,
it can be used to extract quark masses; 
it can test quark models that evaluate $\bar{\delta} m_N$;
and it can constrain a possible time variation of $\alpha$ because
$^4$He nucleosynthesis is sensitive to $\Delta m_N$.
These LECs contribute in combinations other than $\Delta m_N$
to processes involving pions, because the two operators that
generate the nucleon mass difference have different chiral partners,
which involve an even number of pions.
Unfortunately, these LECs are hard to measure directly in $\pi N$ 
scattering.
The forward-backward asymmetry in $np \to d\pi^0$, on the other hand,
is sensitive to the charge-symmetry breaking from
these operators, and it has been calculated
\cite{vanKolck:2000ip}.
Since the asymmetry  is related to a ratio of amplitudes, some
of the uncertainties in the strong-interaction physics
are reduced.
The asymmetry is being measured at TRIUMF
\cite{allena}, at a level that could allow for an observation
of the quark-mass-difference effect.
A related experiment, $dd\rightarrow \alpha \pi^0$,
which can address the same issues but with
different theoretical uncertainties, has been proposed
at IUCF \cite{edandandy}.
%till here

It is possible that some of the problems encountered
at threshold stem from the smallness of pion $s$ waves,
which show poor convergence also in $A=1$ ChPT
---for example, in neutral pion photoproduction on the proton.
Indeed, $p$-wave pion production seems better behaved.
Ref. \cite{hanhart} calculates the first two
orders of the cross section
for the $pp\rightarrow pp\pi^0$ reaction with initial
nucleons in the spin $S=1$ state in the direction of
the incoming center-of-mass momentum, as function
of the outgoing pion momentum in the range $0.5-1$ $m_\pi$.
With no free parameters,
good convergence and reasonable agreement with data
are found.
It was also pointed out that other observables
would, at that order, be sensitive
to a combination of $\pi (\bar{N}N)^2$ LECs
that affects the leading $3N$ force, discussed in Sect. \ref{3and4}.
In particular,
the amplitude for the $^1S_0 \rightarrow (^3S_1-^3D_1) p$
transition, which vanishes in leading order,
is very sensitive to this LEC.
This amplitude, extracted from the $pp\rightarrow pn\pi^+$
data for pion
momenta in the range $0.2-0.5 $ $m_\pi$,
can be fitted quite well with a natural-sized LEC \cite{hanhart}.
This value for the LEC can be used in the $3N$ potential
to improve the predictive power of the chiral potential. 

All calculations of pion-production observables have involved
approximations necessary to match the kernel and wave-functions.
A critical discussions of these approximations
can be found in Ref. \cite{piprodtoy}.
Issues such as the size of the contribution
of the $\pi NN$ cut, not well accounted for in the common approximations,
need to be better understood.
Pion production is clearly wide open for
further development.

\subsection{Connection with lattice QCD}

The holy grail of nuclear physics has for some time been
its derivation from QCD.
As we have seen, 
light nuclei are large objects of size 
$\sim 1/M_{NN} \gg 1/M_{QCD}$, or larger.
Dynamics at this scale can be understood within the EFT, and
all nuclear information is encoded in the parameters of the EFT Lagrangian.
These parameters, in turn, are fixed by the physics of smaller distances.
If the EFT can somehow be matched onto QCD at
some scale not far below $M_{QCD}$, the EFT can be used
to predict all of traditional nuclear physics.
The EFT allows to split the quest for the holy grail
in two stages, according
to the two energy scales.

At present, the best hope for a solution of QCD in the regime
of large coupling constant relevant for nuclear physics
is by explicit numerical solution on the lattice.
However, the 
large size of nuclei make their direct simulation
practically and intellectually unsound.
A more reasonable goal for is to 
match with the EFT, which requires 
lattices of size not much larger than $1/M_{QCD}$.
We are still far from this goal, but a few steps have already been taken. 

One obstacle arises from the difficulty in simulating
small pion masses.
For example, Ref. \cite{fuku} computes the $\si$ and
$\siii$ scattering lengths in quenched QCD with $m_\pi\gaprox 500$ MeV.

\begin{figure}[!t]
\vskip 0.15in
\centerline{{\epsfxsize=2.5in \epsfbox{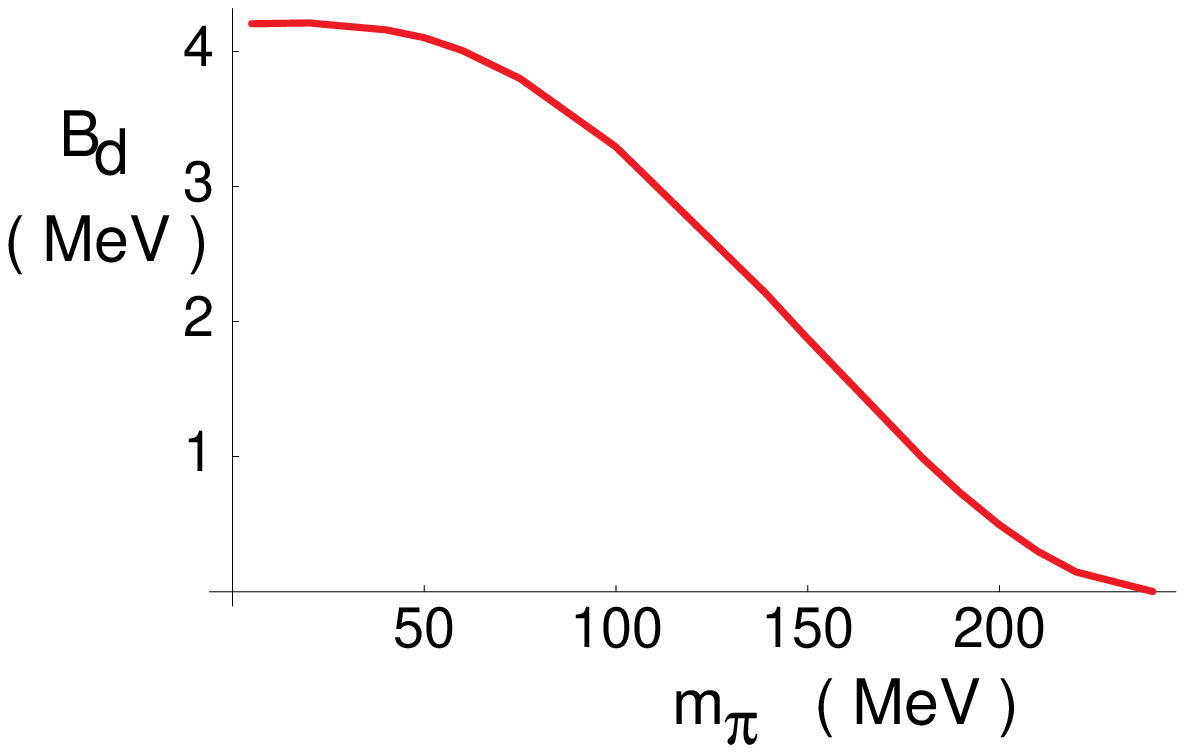}}
{\epsfxsize=2.5in \epsfbox{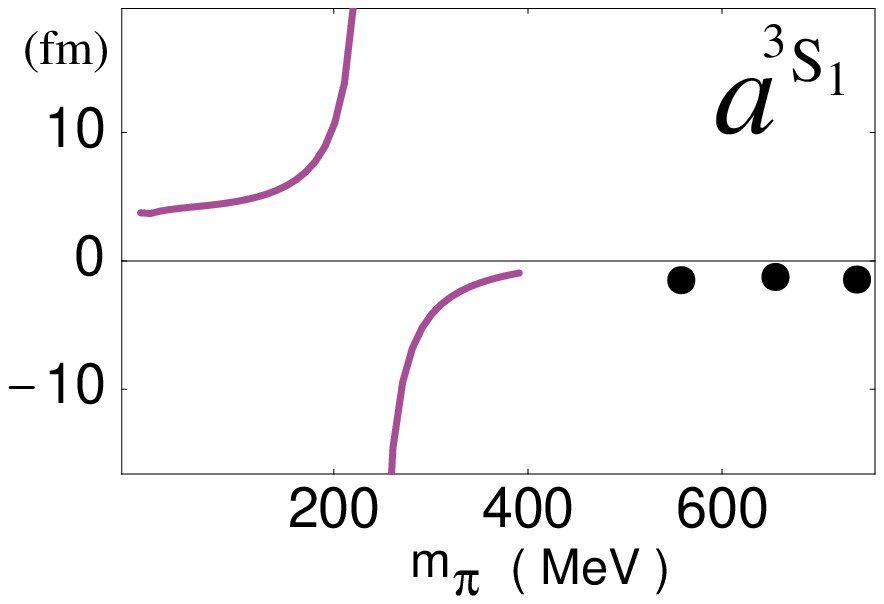}} }
\noindent
\caption{
The deuteron binding energy (left panel) and
the $\siii$ scattering length (right panel)
as functions
of the pion mass that explicitly appears in the OPE potential.
Implicit pion-mass dependence was not calculated, as other 
parameters were set to their physical values for all $m_\pi$.
The dots are quenched lattice
QCD data.
{}From Ref. \cite{towards}.
}
\label{fig:scattpanel}
\vskip .2in
\end{figure}

The $m_\pi$ dependence of nuclear forces
comes in explicitly from pion propagators in pion exchange,
and implicitly from short-range interactions.
For illustration,
in Fig. (\ref{fig:scattpanel}) we exhibit
the deuteron binding energy and the
$\siii$ scattering length 
stemming from the leading {\it explicit} $m_\pi$ dependence 
in the expansion around the chiral limit \cite{towards}.
A higher-order
two-derivative contact interaction was also included
and fitted to the triplet effective range.
For the physical value of the pion mass, one
gets the deuteron binding energy
to reasonable accuracy, $B_d=2.211~{\rm MeV}$ (essentially independent of 
the cutoff $R$).
In the chiral limit the deuteron is bound by 
$B_d^0\simeq 4.2~{\rm MeV}$.  
This value is still somewhat small compared to $f_\pi^2/2 m_N\sim
10~{\rm MeV}$, which one might expect to arise in QCD, 
and therefore one would
conclude that the deuteron is still weakly bound in the chiral limit!  
This 
calculation of the explicit $m_\pi$ dependence
agrees with that
obtained with the AV18 potential \cite{BobW} with $m_\pi=0$,
of $B_d^{0 (AV18)}\simeq 4.1~{\rm MeV}$.
The
lattice data for the triplet scattering length
from Ref.~\cite{fuku} are also shown.

While phenomenological models typically can only vary $m_\pi$
in OPE, all aspects of $m_\pi$ dependence can in principle be determined
in the pionful EFT.
Since the pion mass can be varied  up to
$M_{QCD}$, the EFT can be used to extrapolate
lattice results to realistic values of $m_\pi$.
It was pointed out in Ref. \cite{towards}
that the leading (explicit {\it and implicit})
$m_\pi$ dependence of nuclear forces can be calculated
once the chiral-symmetry-breaking LEC
$C_2^{(qm)}$ is known. That is because the leading
$m_\pi$ dependence in $f_\pi$, $g_A$ and $m_N$ is known
from ChPT.
Unfortunately, determination of $C_2^{(qm)}$ requires calculation
of processes involving external pions ---e.g. $\pi d$
scattering, see Sect. \ref{pid}--- at high orders and, consequently, precise
low-energy data.
Alternatively, one can imagine fitting  $C_2^{(qm)}$
to the lattice data themselves.

Note that in Fig. (\ref{fig:scattpanel}) we illegally compared the 
EFT with quenched QCD. 
Most simulations
cannot yet be done in QCD itself, but only in quenched or, more generally, 
partially-quenched QCD (PQQCD), where different masses are assigned
to valence and sea quarks.
PQQCD has a different symmetry pattern than QCD,
a different low-energy dynamics, and thus a different
dependence on the pion mass.
A proper extrapolation of PQQCD data to smaller pion masses 
requires a partially-quenched EFT.
Implications of PQEFT to the $NN$ interaction are
discussed in Ref. \cite{pqeft}.

\section{OUTLOOK}
\label{outlook}

\subsection{More-nucleon systems}

We have seen that the EFT paradigm has been
extensively applied to systems with $A=2-4$, 
for momenta below and above the
pion mass.
Work remains to be done even for those systems, of course.
For example, the chiral expansion of the pionful EFT
still needs to be better understood 
for $A=2,3$; and $A=4$ has to be studied in the pionless EFT
in order to uncover the role of a $4N$ force that could appear in LO.  

As these issues get settled, 
a natural next step for the EFT program is to increase $A$.
There have, in fact, been attempts
to extend the paradigm to heavier nuclei.

For example, 
EFT methods are being  used to perfect the nuclear shell
model \cite{Haxton:2000bi}. 
The goal here is to take some modern potential model and
simplify the bound-state problem for large nuclei in such a way as
to make a numerical solution of the Schr\"odinger equation
feasible.
This simplification comes about by reducing
the dimensionality of the original Hilbert space of the shell model,
the effect of the highly excited states being included into local operators
acting on a reduced Hilbert space. The reduced problem obtained by 
``integrating out'' the high-energy modes can then be solved 
by standard numerical methods. 

Another approach is to develop an EFT to handle other nuclei
that are, like the deuteron, particularly shallow \cite{Bertulani}.
Examples are halo nuclei, where the separation energy
of one or more nucleons is much smaller than the energies
associated with a core of the remaining nucleons.

There have also been many EFT-inspired studies of
nuclear matter
($A\rightarrow \infty$, $\alpha=0$) and very heavy nuclei.
They have all aimed at identifying the relevant degrees of freedom
and an
expansion parameter that can describe physics for densities
around the saturation density.
See, for example, 
Ref. \cite{Friar:1995dt}.

Whether these approaches prove to be {\it bona fide}
EFTs (in the sense used in this review) or not, the problem
remains of deriving the saturation of nuclear matter from an
EFT adjusted to describe few-nucleon physics.
This is a formidable problem:
the complexity of the necessary resummations
of LO operators increases rapidly with $A$,
becoming high already at $A=5$.
Lacking the identification of a further expansion parameter,
we might, as in QCD itself, have to resort to lattice simulations.
A step in this direction was taken in Ref. \cite{Muller:1999cp},
where a toy model with no- and two-derivative contact two-body interactions
was solved (at zero and at finite temperature)
on a spatial lattice using Monte Carlo techniques,
and the interaction parameters were fitted to saturation properties.
The next step involves using EFT interactions determined from
few-nucleon systems.

These attempts, interesting as they are, fall outside
the scope of this review, and we refer the reader to the 
original literature.

\subsection{Conclusion}

For the last couple of years, the pionless EFT
has been developed and applied to two- and three-nucleon systems.
Although for the $NN$ system in isolation it amounts to nothing more than ERT,
the full power of the field-theory arsenal comes to
fruition when more nucleons and/or external probes are considered.
We have seen that the extension to the $3N$ system is 
full of surprises, such as the appearance of limit-cycle behavior
and of a relevant three-body force.
These surprises have been turned into successes, and 
relatively simple calculations yield results of quality not
inferior to polished potentials models.
Although limited in energy,
this EFT can achieve high precision for reactions
of interest to astrophysics,
such as $np\rightarrow d\gamma$. 
 
The older pionful EFT is less well understood.
There are hints that the expansion has finally been
identified as an expansion around the chiral limit,
but higher orders in the expansion
must be studied.
Among the higher-order terms is a LEC,
$C_2^{(qm)}$, that is the main uncertainty in the extrapolation
to the chiral and heavy-pion limits.
External probes might be able to determine this LEC.
Anyhow,
considerable progress has been achieved in the development
of the EFT $NN$ potential.
Isospin-violating effects are a unique virtue of the pionful EFT because
they are so tightly linked to the pattern of QCD symmetries.
There remain issues regarding the size of short-range
$3N$ forces, but novel longer-range $3N$ forces naturally appear
and can play an important role in nuclear dynamics. 
An assessment of this progress
from the perspective of the historical development of nuclear potentials
can be found in Ref. \cite{kudos}.

Yet, most nuclei still await us.

\vspace{0.5cm}
\noindent
{\it Acknowledgments}

\noindent
We would like to thank our collaborators for teaching us most of what we 
know about effective theories and nuclear physics. 
UvK is grateful to 
the Nuclear Theory Groups at the U of Washington, 
U of South Carolina, Ohio U
and Ohio State U, and to the INT
for hospitality during stages of the writing of this paper.
Thanks to RIKEN, Brookhaven National Laboratory and to the U.S.
Department of Energy [DE-AC02-98CH10886] for providing the facilities
essential for the completion of this work. 
This work was supported  by the Director, Office of Energy Research, 
Office of High Energy and Nuclear Physics, and by the Office of 
Basic Energy Sciences, Division of Nuclear Sciences, 
of the U.S. Department of Energy under Contract No.
DE-AC03-76SF00098 (PFB), 
and by
a DOE Outstanding Junior Investigator Award (UvK).

%%%*** Da' uma melhorada na conclusao...
%%ta otima, vou jantar
%%%*** Preencha seus agradecimentos
%%done

%%%%%%%%%%%%%%%%%%%%%%%%%%%%%%%%%%%%%%%%%%%%%%%%%%%%%%%%%%%%%%%%%%


\begin{thebibliography}{99}


% right format: 
%\bibitem  Jones S. {\it Phys. Rev. D} 68:93939 (2000); Jones S, Pamonha F. hep-th/0000000

\bibitem{Manohar_review}
Manohar AV.
hep-ph/9606222;
Kaplan DB. 
nucl-th/9506035;
Lepage GP.
In {\it From Actions to Answers, TASI'89},
ed. T DeGrand, D Toussaint. Singapore: World Sci. (1990)

\bibitem{Lepage_NRQFT}
Caswell WE, Lepage GP.
{\it Phys. Lett.} B167:437 (1986)

\bibitem{Georgi_heavyquark}
Georgi H.
{\it Phys. Lett.} B240:447 (1990)

\bibitem{Son_mesonmasses}
Son DT, Stephanov MA.
{\it Phys. Rev.} D61:074012 (2000);
Beane SR, Bedaque PF, Savage MJ.
{\it Phys. Lett.} B483:131 (2000);
Barducci A, Casalbuoni R, Pettini G, Gatto R.
{\it Phys. Rev.} D63:074002 (2001)

\bibitem{Georgi_Manohar}
Manohar A, Georgi H.
{\it Nucl. Phys.} B234:189 (1984)

\bibitem{inwei6} 
Weinberg S. 
{\it Phys. Lett.} B251:288 (1990); 
{\it Nucl. Phys.} B363:3 (1991)

\bibitem{ciOvK} 
Ord\'{o}\~{n}ez C, van Kolck U. 
{\it Phys. Lett.} B291:459 (1992)

\bibitem{inwei5} 
Weinberg S.
{\it Phys. Lett.} B295:114 (1992)

\bibitem{ciOLvK} 
Ord\'{o}\~{n}ez C, Ray L, van Kolck U.
{\it Phys. Rev. Lett.} 72:1982 (1994);
{\it Phys. Rev.} C53:2086 (1996)

\bibitem{civK1} 
van Kolck U.
{\it Phys. Rev.} C49:2932 (1994)

\bibitem{invk}
van Kolck U. 
{\it Few-Body Syst. Suppl.} 9:444 (1995);
U. of Texas Ph.D. Dissertation (1993)

\bibitem{Bedaque_vanKolck_quartet}
Bedaque PF, van Kolck U.
{\it Phys. Lett.} B428:221 (1998)

\bibitem{vanKolck:1997ut}
van Kolck U.
In {\em Proceedings of the Workshop on Chiral Dynamics 1997, Theory and 
Experiment\/}, ed. A Bernstein, D Drechsel, T Walcher.
Berlin: Springer-Verlag (1998), hep-ph/9711222.

\bibitem{vanKolck_shortrange}
van Kolck U.
{\it Nucl. Phys.} A645:273 (1999)

\bibitem{seattle_pionless}
Chen JW, Rupak G, Savage MJ.
{\it Nucl. Phys.} A653:386 (1999)

\bibitem{Bethe} 
Bethe HA.
{\it Phys. Rev.} 76:38 (1949)

\bibitem{Efimov_qualitative}
Efimov V.
TPI-MINN-89-31-T.

\bibitem{vanKolck_review}
van Kolck U.
{\it Prog. Part. Nucl. Phys.}  43:337 (1999);
Beane SR, Bedaque PF, Haxton WC, Phillips DR, Savage MJ.
In {\it Boris Ioffe Festschrift}, ed. M Shifman, 
Singapore: World Sci. (2001), nucl-th/0008064

\bibitem{books}
{\it Nuclear Physics with Effective Field Theory},
ed. R Seki, MJ Savage, U van Kolck.
Singapore: World Sci. (1998);
{\it Nuclear Physics with Effective Field Theory II},
ed. PF Bedaque, MJ Savage, R Seki, U van Kolck.
Singapore: World Sci. (2000)

\bibitem{Braaten_dilute}
Braaten E, Nieto A.
hep-th/9609047;
Hammer H-W, Furnstahl RJ.
{\it Nucl. Phys.} A678:277 (2000);
Furnstahl RJ, Hammer H-W, Tirfessa N.
{\it Nucl. Phys.} A689:846 (2001);
Hammer H-W, Furnstahl RJ.
nucl-th/0108069 

\bibitem{stooges_3bosons}
Bedaque PF, Hammer H-W, van Kolck U.
{\it Phys. Rev. Lett.} 82:463 (1999);
{\it Nucl. Phys.} A646:444 (1999).

\bibitem{Bedaque_recombination}
Bedaque PF, Braaten E, Hammer H-W.
{\it Phys. Rev. Lett.} 85:908 (2000);
Braaten E, Hammer H-W.
{\it Phys. Rev. Lett.} 87:160407 (2001);
Braaten E, Hammer H-W, Mehen T.
{\it Phys. Rev. Lett.}  88:040401 (2002)

\bibitem{ksw_1}
Kaplan DB, Savage MJ, Wise MB.
{\it Phys. Lett.} B424:390 (1998);
{\it Nucl. Phys.} B534:329 (1998)

\bibitem{Mehen_Stewart_momsub}
Mehen T, Stewart IW.
{\it Phys. Lett.} B445:378 (1999);
Gegelia J.
nucl-th/9802038

\bibitem{Birse_fixedpoints}
Birse MC, McGovern JA, Richardson KG.
{\it Phys. Lett.} B464:169 (1999)

\bibitem{nijmanal}
Stoks VGJ, Klomp RAM, Rentmeester MCM, de Swart JJ.
{\it Phys. Rev.} C48:792 (1993)

\bibitem{Ravndal}
Kong X, Ravndal F.
{\it Phys. Lett.} B450:320 (1999); 
{\it Nucl. Phys.} A665:137 (2000)

\bibitem{range_large}
Beane SR, Savage MJ.
{\it Nucl. Phys.} A694:511 (2001)

\bibitem{Rupak_Griesshammer_compton}
Grie{\ss}hammer HW, Rupak G.
{\it Phys. Lett.} B529:57 (2002)

\bibitem{Riska_Brown}
Riska DO, Brown GE.
{\it Phys. Lett.} B38:193 (1972)

\bibitem{Rupak_BBN}
Rupak G.
{\it Nucl. Phys.} A678:405 (2000)

\bibitem{data_BBN}
Arenh\"ovel H, Sanzone M.
{\it Photodisintegration of the Deuteron: A Review of  
Theory and Experiment}, 
Berlin: Springer-Verlag (1991)

\bibitem{seattle_suppressed}
Chen JW, Rupak G, Savage MJ.
{\it Phys. Lett.} B464:1 (1999)

\bibitem{Chen_neutrino_1}
Butler M, and Chen JW.
{\it Nucl. Phys.} A675:575 (2000);
Butler M, Chen JW, Kong X.
{\it Phys. Rev.} C63:035501 (2001);
Chen JW.
{\it Nucl. Phys.} A684:484 (2001)

\bibitem{ravndal1}
Kong X, Ravndal F.
{\it Nucl. Phys.} A656:421 (1999);
{\it Phys. Lett.} B470:1 (1999); 
{\it Phys. Rev.} C64:044002 (2001);
Butler M, Chen JW.
{\it Phys. Lett.} B520:87 (2001)


\bibitem{stooges_dependence}
Bedaque PF, Hammer H-W, van Kolck U.
{\it Phys. Rev.} C58:641 (1998)

\bibitem{Bedaque_highL}
Gabbiani F, Bedaque  PF, Grie{\ss}hammer HW.
{\it Nucl. Phys.} A675:601 (2000)

\bibitem{stooges_triton}
Bedaque PF, Hammer H-W, van Kolck U.
{\it Nucl. Phys.} A676:357 (2000)

\bibitem{Kaplan:1996nv}
Kaplan, DB
{\it Nucl. Phys.} B494:471 (1997)

\bibitem{Mehen:1999qs}
Mehen T, Stewart IW, Wise M.
{\it Phys. Rev. Lett.} 83:931 (1999)

\bibitem{Amado_review}
Amado RD.
In {\it Elementary Particle Physics and Scattering Theory, Vol.2},
ed. M Chr\'etien, S Schweber. New York: Gordon and Breach (1970)

\bibitem{skorny}
Skorniakov GV, Ter-Martirosian KA.
{\it Sov. Phys. JETP} 4:648 (1957)

\bibitem{data_quartet_phaseshifts}
van Oers WTH, Seagrave JD.
{\it Phys. Lett.} B24:562 (1967);
Phillips AC, Barton G.
{\it Phys. Lett.} B28:378 (1969)

\bibitem{data_quartet}
Dilg W, Koester L, Nistler W.
{\it Phys. Lett.} B36:208 (1971)

\bibitem{Efimov_range}
Efimov VN, Tkachenko EG.
JINR-E4-8473

\bibitem{Friar_quartet}
Friar JL, H\"uber D, Wita\l a H, Payne GL.
{\it Acta Phys. Polon.} B31:749 (2000)

\bibitem{Rupak_coulomb}
Rupak G, Kong XW.
nucl-th/0108059

\bibitem{Faddeev_Minlos}
Faddeev LD, Minlos RA.
{\it Sov. Phys. JETP} 14:1315 (1962)

\bibitem{Danilov}
Danilov GS.
{\it Sov. Phys. JETP} 16:1010 (1963)

\bibitem{Efimov_effect}
Efimov VN.
{\it Sov. J. Nucl. Phys.} 12:589 (1971);
                          28:546 (1979);
{\it Nucl. Phys.} A210:157 (1973);
Amado RD, Noble JV.
{\it Phys. Rev.} D5:1992 (1972)

\bibitem{conformal}
Mehen T, Stewart IW, Wise MB.
{\it Phys. Lett.} B474:145 (2000)

\bibitem{Hammer:2000nf}
Hammer H-W, Mehen T.
{\it Nucl. Phys.} A690:535 (2001)

\bibitem{wilson1}
Wilson KG. 
{\it Phys. Rev.} D3:1818 (1971);
G\l azek SD, Wilson KG. 
{\it Phys. Rev.} D47:4657 (1993)

\bibitem{wilson2}
G\l azek SD, Wilson KG. 
hep-th/0203088

\bibitem{Hammer_range}
Hammer H-W, Mehen T.
{\it Phys. Lett.} B516:353 (2001)

\bibitem{toappear}
Bedaque PF, Grie{\ss}hammer H, Hammer H-W,  Rupak G.
In preparation  

\bibitem{kievsky}
Kievsky A, Rosati S, Tornow W, Viviani M.
{\it Nucl. Phys.} A607:402 (1996);
Kievsky A. Private communication (2002)

\bibitem{Phillips}
Phillips AC.
{\it Nucl. Phys.} A107:209 (1968)

\bibitem{tkachenko}
Efimov V, Tkachenko EG.
{\it  Few-Body Syst.} 4:71 (1988)

\bibitem{Efimov_phillips}
Efimov V, Tkachenko EG.
{\it Phys. Lett.} B157:108 (1995)

\bibitem{Hammer:2001ng}
Hammer H-W.
nucl-th/0110031

%%%%%

\bibitem{inGSW} 
Goldstone J, Salam A, Weinberg S.  
{\it Phys. Rev.} 127:965 (1962)

\bibitem{inCCWZ} 
Coleman S, Wess J, Zumino B. 
{\it Phys. Rev.} 177:2239 (1969);
Callan CG, Coleman S, Wess J, Zumino B. 
{\it Phys. Rev.} 177:2247 (1969)

\bibitem{bkm} 
Bernard V, Kaiser N, Mei{\ss}ner U-G. 
{\it Int. J. Mod. Phys.} E4:193 (1995) 

\bibitem{inwei2} 
Weinberg S. 
{\it Physica} 96A:327 (1979)

\bibitem{lutz} 
Lutz M. 
In hep-ph/9606301; 
Private communication (1996, 1997)

\bibitem{cohenhansen1}
Cohen TD, Hansen JM.
{\it Phys. Lett.} B440:233 (1998) 

\bibitem{variouspert}
Gegelia J. 
nucl-th/9806028;
Cohen TD, Hansen JM. 
{\it Phys. Rev.} C59:13,304 (1999)

\bibitem{mehenstewart}
Mehen T, Stewart IW.
{\it Phys. Lett.} B445:378 (1999); 
{\it Phys. Rev.} C59:2365 (1999);
{\it Nucl. Phys.} A665:164 (2000)

\bibitem{furnsteele}
Steele JV, Furnstahl RJ. 
{\it Nucl. Phys.} A645:439 (1999)

\bibitem{rupakshoresh} 
Rupak G, Shoresh N. 
{\it Phys. Rev.} C60:054004 (1999) 

\bibitem{fms}
Fleming S, Mehen T, Stewart IW.
{\it Nucl. Phys.} A677:313 (2000); 
{\it Phys. Rev.} C61:044005 (2000)

\bibitem{twokaiserpapers}
Kaiser N, Brockmann R, Weise W.
{\it Nucl. Phys.} A625:758 (1997); 
Kaiser N, Gerstend\"orfer S, Weise W.
{\it Nucl. Phys.} A637:395 (1998);
Ballot JL, Robilotta MR, da Rocha CA.
{\it Phys. Rev.} C57:1574 (1998)

\bibitem{friarcoon} 
Coon SA, Friar JL. 
{\it Phys. Rev.} C34:1060 (1986);
Friar JL.
{\it Czech. J. Phys.} 43:259 (1993) 

\bibitem{nonppions}
Lepage, GP.
nucl-th/9706029;
Gegelia J. 
{\it Phys. Lett.} B463:133 (1999)

\bibitem{taesunpions}
Park T-S, Kubodera K, Min D-P, Rho M.
{\it Phys. Rev.} C58:637 (1998);
{\it Nucl. Phys.} A646:83 (1999);
Hyun C-H, Min D-P, Park T-S.
{\it Phys. Lett.} B473:6 (2000)

\bibitem{FTT}
Frederico T, Tim\'oteo VS, and Tomio L.
{\it Nucl. Phys.} A653:209 (1999)

\bibitem{epelfit}
Epelbaum E, Gl\"ockle W, Mei{\ss}ner U-G.
{\it Nucl. Phys.} A671:295 (2000)

\bibitem{singpotrev}
Frank WM, Land DJ, Spector RM.
{\it Rev. Mod. Phys.} 43:36 (1971);
Perelomov AM, Popov VS.
{\it Teor. i. Mate. Fiz.} 4:48 (1970)

\bibitem{kiddies}
Beane SR, Bedaque PF, Childress L, Kryjevski A, McGuire J, van Kolck U.
{\it Phys. Rev.} A64:042103 (2001)

\bibitem{Camblong:2001zi}
Camblong HE, Ord\'o\~nez CR.
hep-th/0110278

\bibitem{Barford:2001sx}
Barford T, Birse MC.
nucl-th/0108024

\bibitem{towards}
Beane SR, Bedaque PF, Savage MJ, van Kolck U.
{\it Nucl. Phys.} A700:377 (2002) 

\bibitem{Sprung}
Sprung DWL, van Dijk W, Wang E, Zheng DC, Sarriguren P, Martorell J.
{\it Phys. Rev.} C49:2942 (1994)

\bibitem{Kaplan:1999qa}
Kaplan DB and Steele JV.
{\it Phys. Rev.} C60:064002 (1999)

\bibitem{scaldeferri}
Scaldeferri KA, Phillips DR, Kao C-W, Cohen TD. 
{\it Phys. Rev.} C56:679 (1997)

\bibitem{gegeliasoto}
Gegelia J, Japaridze G.
{\it Phys. Lett.} B517:476 (2001); 
Eiras D, Soto J.
nucl-th/0107009

\bibitem{newlutz}
Lutz M. 
{\it Nucl. Phys.} A677:241 (2000) 

\bibitem{jimandsid}
Friar JL, Coon SA. 
{\it Phys. Rev.} C49:1272 (1994);
Epelbaoum E, Gl\"ockle W, Mei{\ss}ner U-G.
{\it Nucl. Phys.} A637:107 (1998);
Friar JL.
{\it Phys. Rev.} C60:034002 (1999)

\bibitem{argonne}
Wiringa RB, Stoks VGJ, Schiavilla R.
{\it Phys. Rev.} C51:38 (1995)

\bibitem{ciBwciSHciSO}
Brueckner KA, Watson KM. 
{\it Phys. Rev.} 92:1023 (1953);
Sugawara M, Okubo S. 
{\it Phys. Rev.} 117:605,611 (1960); 
Sugawara H, von Hippel F. 
{\it Phys. Rev.} 172:1764 (1968)

\bibitem{robetal}
Rentmeester MCM, Timmermans RGE, Friar JL, de Swart JJ.
{\it Phys. Rev. Lett.} 82:4992 (1999); 
Timmermans RGE. 
Private communication (2001)

\bibitem{ulfressat}
Epelbaum E, Mei{\ss}ner U-G, Gl\"ockle W, Elster C.  
nucl-th/0106007

\bibitem{mach1}
Entem DR, Machleidt R. 
{\it Phys. Lett.} B524:93 (2002) 

\bibitem{Bogner:2001gq}
Bogner SK, Kuo TT, Schwenk A, Entem DR, Machleidt R.
nucl-th/0108041

\bibitem{kaiserloops}
Kaiser N.
{\it Phys. Rev.} C61:014003 (2000);
                 C62:024001 (2000); 
                 C63:044010 (2001);
                 C64:057001 (2001);
                 C65:017001 (2002)

\bibitem{mach2}
Entem DR, Machleidt R. 
nucl-th/0202039 

\bibitem{jerry}
Miller GA, Nefkens BMK, Slaus I.
{\it Phys. Rept.} 194:1 (1990)

\bibitem{prlwithnijm}
van Kolck U, Rentmeester MCM, Friar JL, Goldman T, de Swart JJ. 
{\it Phys. Rev. Lett.} 80:4386 (1998)

\bibitem{mejim}
Friar JL, van Kolck U.  
{\it Phys. Rev.} C60:034006 (1999)

\bibitem{niskanenpot}
Niskanen JA, 
{\it Phys. Rev.} C65:037001 (2002)

\bibitem{mejimandterry} 
van Kolck U, Friar JL, Goldman T. 
{\it Phys. Lett.} B371:169 (1996)

\bibitem{coon}
Coon SA, McKellar BHJ, Stoks VGJ.
{\it Phys. Lett.} B385:25 (1996) 

\bibitem{waltz}
Walzl M, Mei{\ss}ner U-G, Epelbaum E.
{\it Nucl. Phys.} A693:663 (2001) 

\bibitem{Epelbaum:1999zn}
Epelbaum E, Mei{\ss}ner U-G.
{\it Phys. Lett.} B461:287 (1999)

\bibitem{Pieper:2001mp}
Pieper SC, Wiringa RB.
{\it Ann. Rev. Nucl. Part. Sci.} 51:53 (2001)

\bibitem{epel34NLO}
Epelbaum E, Kamada H, Nogga A, Wita\l a H, Gl\"ockle W, Mei{\ss}ner U-G. 
{\it Phys. Rev. Lett.} 86:4787 (2001)

\bibitem{AywithchiralNN}
Entem DR, Machleidt R, Wita\l a H.
nucl-th/0111033 

\bibitem{TMbrazil} 
Coon SA, Scadron MD, McNamee PC, Barrett BR, Blatt DWE, McKellar BHJ.
{\it Nucl. Phys.} A317:242 (1979); 
Coelho HT, Das TK, Robilotta MR. 
{\it Phys. Rev.} C28:1812 (1983)

\bibitem{vk:huber} 
Friar JL, H\"uber D, van Kolck U.
{\it Phys. Rev.} C59:53 (1999)



\bibitem{coonrework}
Coon SA, Han HK.
{\it Few-Body Syst.} 30:131 (2001);
Kamada H, H\"uber D, Nogga A.
{\it Few-Body Syst.} 30:121 (2001)

\bibitem{huber2}
H\"uber D, Friar JL, Nogga A, Wita\l a H, van Kolck U. 
{\it Few-Body Syst.} 30:95 (2001) 


\bibitem{newformod}
Ellis RG, Coon SA, McKellar BHJ.
{\it Nucl. Phys.} A438:631 (1985);
Coon SA, Pe\~na MT, Riska DO.
{\it Phys. Rev.} C52:2925 (1995)

\bibitem{stewartepelbaumPC}
Stewart IW. 
Private communication (2000);
Epelbaum E.
Private communication (2001)

\bibitem{ourcompton}
Beane SR, Phillips DR, Malheiro M, van Kolck U.
{\it Nucl. Phys.} A656:367 (1999) 

\bibitem{chiralfilter} 
Rho M. 
{\it Phys. Rev. Lett.} 66:1275 (1991)

\bibitem{ed}
Kaplan DB, Savage MJ, Wise MB.
{\it Phys. Rev.} C59:617 (1999);
Phillips DR, Cohen TD.
{\it Nucl. Phys.} A668:45 (2000); 
Walzl M, Mei{\ss}ner U-G.
{\it Phys. Lett.} B513:37 (2001);
Phillips DR.
In {\it Mesons and Light Nuclei}, ed. J. Adam et al, 
New York: AIP Press (2001), nucl-th/0108070

\bibitem{pved}
Savage MJ, Springer RP.  
{\it Nucl. Phys.} A644:235 (1998), (E) A657:457 (1999);
                  A686:413 (2001);
Diaconescu L, Schiavilla R, van Kolck U.
{\it Phys. Rev.} C63:044007 (2001)

\bibitem{radcap}
Park T-S, Min D-P, Rho M.
{\it Phys. Rev. Lett.} 74:4153 (1995);
{\it Nucl. Phys.} A596:515 (1996);
Savage MJ, Scaldeferri KA, Wise MB.
{\it Nucl. Phys.} A652:273 (1999); 
Park T-S, Kubodera K, Min D-P, Rho M.
{\it Phys. Lett.} B472:232 (2000);
Hyun C-H, Min D-P, Park T-S. 
{\it Phys.Lett.} B473:6 (2000) 

\bibitem{polradcap}
Kaplan DP, Savage MJ, Springer RP, Wise MB. 
{\it Phys. Lett.} B449:1 (1999);
Hyun C-H, Park T-S, Min D-P.
{\it Phys. Lett.} B516:321 (2001)

\bibitem{solarfusion}
Park T-S, Min D-P, Rho M.
{\it Phys. Rep.} 233:341 (1993);
Park T-S, Kubodera K, Min D-P, Rho M.
{\it Astrophys. J.} 507:443 (1998);
Park T-S, Marcucci LE, Schiavilla R, Viviani M, Kievsky A, Rosati S, 
Kubodera K, Min D-P, Rho M.
nucl-th/0106025 

\bibitem{solarhep}
Park T-S, Marcucci LE, Schiavilla R, Viviani M, Kievsky A, Rosati S, 
Kubodera K, Min D-P, Rho M.
nucl-th/0107012 

\bibitem{nud}
Nakamura S, Sato T, Ando S, Park T-S, Myhrer F, Gudkov V, Kubodera K.
nucl-th/0201062 

\bibitem{mud}
Ando S, Park T-S, Kubodera K, Myhrer F.
nucl-th/0109053 

\bibitem{compton}
Chen JW, Grie{\ss}hammer HW, Savage MJ, Springer RP.
{\it Nucl. Phys.} A644:221,245 (1998);
Chen JW. 
{\it Nucl. Phys.} A653:375 (1999) 

\bibitem{beanenew}
Beane SR, Bernard V, Lee T-SH, Mei{\ss}ner U-G.
{\it Phys. Rev.} C57:424 (1998)

\bibitem{kaiserization}
Kaiser N.
nucl-th/0203001

\bibitem{koltun} 
Misra A, Koltun DS. 
{\it Phys. Rev. C} 61:024003 (2000)
 
\bibitem{bbemp}
Beane SR, Bernard V, Epelbaum E, Mei{\ss}ner U-G, Phillips DR.
In preparation

\bibitem{rockmore}
Rockmore RM.
{\it Phys. Lett.} B356:153 (1995)

\bibitem{boragriess}
Borasoy B, Griesshammer HW. 
nucl-th/0105048 

\bibitem{vk:beane}
Beane SR, Lee CY, van Kolck U.
{\it Phys. Rev.} C52:2914 (1995);
Beane SR, Bernard V, Lee T-SH, Mei{\ss}ner U-G, van Kolck U.
{\it Nucl. Phys.} A618:381 (1997)        

\bibitem{vk:wilhelm} 
Wilhelm P. 
{\it Phys. Rev.} C56:1215 (1997) 

\bibitem{vk:justusetal}
Koch JH, Woloshyn RM. 
{\it Phys. Rev.} C16:1986 (1997);
Laget JM. 
{\it Phys. Rep.} 69:1 (1981)

\bibitem{vk:berg}
Bergstrom JC, et al. 
{\it Phys. Rev.} C57:3203 (1998)

\bibitem{Levchuk:2000hz}
Levchuk MI, Schumacher M, Wissmann F.
{\it Nucl. Phys.} A675:621 (2000)

\bibitem{Ulfelectroprod}
Bernard V, Krebs H, Mei{\ss}ner U-G.
{\it Phys. Rev.} C61:058201 (2000)
 
\bibitem{vk:bernstein}
Ewald I, et al (A1 Collaboration).
{\it Phys. Lett.} B499:238 (2001)

\bibitem{vkolck:cohen} 
Cohen TD, Friar JL, Miller GA, van Kolck U.
{\it Phys. Rev.} C53:2661 (1996)

\bibitem{hanhart}
Hanhart C, van Kolck U, Miller GA.
{\it Phys. Rev. Lett.} 85:2905 (2000)

\bibitem{ulfpespelasmauns}
Bernard V, Kaiser, Mei{\ss}ner U-G.
{\it Eur. Phys. J.} A4:259 (1999) 

\bibitem{withriska}
van Kolck U, Miller GA, Riska DO.
{\it Phys. Lett.} B388:679 (1996);
Pe\~na MT, Riska DO, Stadler A.
{\it Phys. Rev.} C60:045201 (1999)

\bibitem{vkolck:pppppizero}
Park BY, Myhrer F, Morones JR, Meissner T, Kubodera K.
{\it Phys. Rev.} C53:1519 (1996); 
Sato T, Lee T-SH, Myhrer F, Kubodera K.
{\it Phys. Rev.} C56:1246 (1997);
Hanhart C, Haidenbauer J, Hoffmann M, Mei{\ss}ner U-G, Speth J. 
{\it Phys. Lett.} B424:8 (1998);
Gedalin E, Moalem A, Razdolskaya L. 
{\it Phys.Rev.} C60:031001 (1999);
Dmitrasinovic V, Kubodera K, Myhrer F, Sato T.
{\it Phys. Lett.} B465:43 (1999); 
Ando S, Park T-S, Min D-P.
{\it Phys. Lett.} B509:253 (2001)


\bibitem{vkolck:carocha} 
da Rocha CA, Miller GA, van Kolck U.
{\it Phys. Rev.} C61:034613 (2000)

\bibitem{vanKolck:2000ip}
van Kolck U, Niskanen JA, Miller GA.
{\it Phys. Lett.} B493:65 (2000)

\bibitem{allena}
Opper AK, Korkmaz E (spokespersons). 
TRIUMF E-704 Proposal

\bibitem{edandandy}
Bacher AD, Stephenson EJ (spokespersons). 
IUCF CE-82 Proposal

\bibitem{piprodtoy}
Hanhart C, Miller GA, Myhrer F, Sato T, van Kolck  U.
{\it Phys. Rev.} C63:044002 (2001)

\bibitem{fuku}
Fukugita M, Kuramashi Y, Okawa M, Mino H, Ukawa A.
{\it Phys. Rev.} D52:3003 (1995)

\bibitem{BobW} 
Wiringa R. 
Private communication (2001)

\bibitem{pqeft}
Beane SR, Savage MJ. 
hep-lat/0202013.

\bibitem{Haxton:2000bi}
Zheng DC, Barrett BR, Jaqua L, Vary JP, McCarthy RJ.
{\it Phys. Rev.} C48:1083 (1993);
Haxton WC, Song CL.
{\it Phys. Rev. Lett.} 84:5484 (2000);
Haxton WC, Luu T.
{\it Nucl. Phys.} A690:15 (2001);
Fayache MS, Vary JP, Barrett BR, Navratil P, Aroua S.
nucl-th/0112066

\bibitem{Bertulani} 
van Kolck U.
In hep-ph/0201266;
Bertulani CA, Hammer H-W, van Kolck U.
In preparation

\bibitem{Friar:1995dt}
Friar JL, Madland DG, Lynn BW.
{\it Phys. Rev.} C53:3085 (1996);
Furnstahl RJ and Serot BD.
{\it Nucl. Phys.} A663:513 (2000);
Steele JV.
nucl-th/0010066;
Brown GE, Rho M.
hep-ph/0103102

\bibitem{Muller:1999cp}
M\"uller H-M, Koonin SE, Seki R, van Kolck U.
{\it Phys. Rev.} C61:044320 (2000);
M\"uller H-M, Seki R. 
In Ref. \cite{books}.

\bibitem{kudos}
Machleidt R, Slaus I.
{\it J. Phys.} G27:R69 (2001);
Friar JL.  
{\it Nucl. Phys.} A684:200 (2001);
Coon SA. 
nucl-th/9903033 





\end{thebibliography}
\end{document}